\documentclass[11pt]{article}
\pdfoutput = 1

\usepackage[utf8]{inputenc}
\usepackage{color,graphicx}
\usepackage{amsmath}
\usepackage{verbatim}
\usepackage{bm}
\usepackage{amsthm}    
\usepackage{booktabs} 
\usepackage{physics}
\usepackage{amsfonts}
\usepackage{array}
\usepackage{setspace}
\usepackage{float}
\usepackage{url}
\usepackage{standalone}
\usepackage{leftindex}
\usepackage{tikz}
\usepackage{mathrsfs}
\usepackage{fancybox}
\usepackage{tensor}
\usepackage{caption}
\usepackage{subcaption}
\usepackage{dsfont}
\usepackage{jheppub}
\usepackage{relsize}
\definecolor{pinkk}{rgb}{1,0,1}

\theoremstyle{remark}

\theoremstyle{plain}

\theoremstyle{definition}

\hypersetup{
    colorlinks,
    citecolor=blue,
    filecolor=black,
    linkcolor=blue,
    urlcolor=blue,
    linktocpage=true
}

\numberwithin{equation}{section}

\title{On the Positivity of de Sitter Group Averaging}

\author[a]{Donald Marolf,}
\author[a]{Xuyang Yu}
\affiliation[a]{Department of Physics, University of California, Santa Barbara, CA 93106, USA}
\emailAdd{marolf@ucsb.edu}
\emailAdd{xuyangyu@ucsb.edu}

\abstract{%
Perturbative quantum gravity about a global de Sitter (dS$_d$) background involves constraints that restrict the state of the linearized theory but which appear only at higher orders in perturbation theory.  In particular,  for familiar gravitational theories with $d>2$ these constraints require all charges defined by continuous de Sitter isometries to vanish in the linearized state.  While standard Fock spaces have the property that the only normalizable state satisfying these constraints is the vacuum, Higuchi suggested that this lack of allowed states should be remedied by instead using the so-called group-averaging inner product.  This inner product satisfies important uniqueness conditions, but its positivity has to date remained an open question.  For $d > 2$, we establish this positivity here for all such states with $N$ particles (including particles of arbitrary spin) in standard global dS${}_d$ modes  where the particle-weights $\Delta_i$  satisfy the condition ${\rm Re} \, \sum_{i=1}^N \Delta_i > d-1$, which is the criterion for group averaging to converge absolutely for arbitrary finite linear combinations of such states.   We also treat the case of bosonic and fermionic  (integer and half-integer spin) particles in dS$_2$ as a toy model for the more general case.

However,  we can identify certain classes of states for which group averaging continues to converge absolutely for smaller weights.  Interestingly, in some cases these smaller-weight states turn out to have {\it negative} group-averaging norms.  We thus also discuss the implications of this observation for perturbative de Sitter quantum gravity.   
}

\begin{document}

\maketitle

\section{Introduction}
\label{sec:intro}

Gravitational perturbation theory around global de Sitter space has long been known \cite{Moncrief:1975,Moncrief:1976un} to feature special constraints that arise only at 2nd order in perturbation theory.  The physical content of these constraints is that they render generic solutions of the linearized theory spurious, as only solutions that satisfy the constraints can source consistent gravitational back-reaction.  While they are known in the gravitational literature as linearization-stability constraints, they may be seen as gravitational examples of what string theorists call tadpole constraints (or tadpole cancellation constraints) \cite{FischlerSusskind1986,FischlerSusskind1986b,Sagnotti1988,GiddingsKachruPolchinski2002}. Indeed, they are associated with a gravitational Gauss law which simply states that, as with the electromagnetic analogue, the appropriate total charge must vanish in a closed universe.

Let us suppose that our theory admits a solution in which the metric is given by global de Sitter space (dS$_d$) in $d$ spacetime dimensions and where all other fields vanish.  This is the case for familiar theories with a positive cosmological constant for $d>2$, though for $d=2$ interesting theories typically require a non-vanishing dilaton that violates the above assumption.  We will refer to this solution as ``pure global de Sitter space'' to indicate the lack of any non-vanishing non-metric fields.

In perturbation theory around pure global dS$_d$, the gravitational linearization-stability constraints turn out to impose the vanishing of all charges associated with continuous de Sitter isometries.
At the quantum level, they thus require the linearized states to be invariant under the connected component SO$_0(d,1)$ of the de Sitter group \cite{Moncrief_1978}.  But since the vacuum is the only normalizable state with this property in a standard Fock space \cite{Higuchi_1991}, basing the quantum theory on a perturbatively-corrected Fock inner product defines a much smaller space of states than is suggested by the classical theory.  While one can also consider linearized theories that include either zero-modes or exponentially-growing modes (and which are thus not described by a standard Fock space), we will focus on the Fock-space case below. 

Higuchi \cite{Higuchi_1991_2} suggested long ago that a physically-useful larger Hilbert space could be constructed by renormalizing the Fock inner product in a specific way.\footnote{See e.g. \cite{Kaplan:2024xyk} for a recent discussion of how to extract useful physics from the resulting Hilbert space.}  From a general Fock state $|\psi\rangle$, one may formally construct a `group-averaging' state $|\Psi\rangle$ that satisfies the desired constraints by writing
\begin{equation}
\label{eq:DSIGA}
|\Psi\rangle := \int_G \, dg\, U(g) |\psi\rangle,
\end{equation}
where $G$ is the (connected component of the) $d$-dimensional de Sitter group ($G= {\rm SO}_0(d,1)$), $U(g)$ denotes the unitary action of $g\in G$ on the Fock space, and $dg$ is the associated Haar measure.  Due to the invariance of the Haar measure under group multiplication, the Fock inner product of two such states may be formally written
\begin{equation}
\label{eq:divIP}
\langle \Psi_1|\Psi_2\rangle_{\rm Fock} := {\rm Vol}(G) \int_G\, dg\,  \langle \psi_1|U(g)|\psi_2\rangle,
\end{equation}
where ${\rm Vol}(G)$ is the (infinite) volume of the de Sitter group.  Higuchi thus suggested that the inner product on the space of such invariant states be instead defined by dropping the factor of ${\rm Vol}(G)$ to write
\begin{equation}
\label{eq:GAIP}
\langle \Psi_1|\Psi_2\rangle^{\rm GA} :=  \int_G dg\,  \langle \psi_1|U(g)|\psi_2\rangle,
\end{equation}
where the label ${\rm GA}$ indicates that this is now to be called the group-averaging inner product.  This construction was further formalized and connected to other problems involving gravitational constraints in \cite{Marolf:1994wh,Marolf:1994ae,Ashtekar:1995zh,Giulini:1998rk,Marolf:2000iq}, which emphasized the importance of restricting the allowed `seed' states $|\psi\rangle$ to a suitable subspace $\Phi$ of the Fock space on which
\eqref{eq:GAIP} converges absolutely and then defining the group-averaging Hilbert space ${\cal H}^{\rm GA}$ by the usual process of taking a quotient by null states followed by Cauchy completion.  In this sense, \eqref{eq:GAIP} is what has come to be called a coinvariant construction \cite{Chandrasekaran:2022cip,Penington:2023dql}, meaning that any Fock state of the form $|\psi\rangle = [1-U(g_0)]|\phi\rangle$ defines a null state.  See also the independent line of work \cite{Landsman1995Rieffel,
      Landsman1998Constrained,
      LandsmanWiedemann1995,
      WiedemannLandsman1996,
      LandsmanWren1997,
      Landsman1995WDW,
      Landsman1998Book} which gives an equivalent formulation in terms of Rieffel induction of representations of $C^*$-algebras from the trivial representation of the constraint group. 

The group-averaging inner product can then be further justified by appealing to a uniqueness result \cite{Giulini_1999} stating that, when it converges absolutely, it is the only possible inner product consistent with providing a $*$-representation of a certain algebra of gauge-invariant operators.  Interestingly, however, despite its motivation above as a renormalization of \eqref{eq:divIP} after dividing by a positive divergent ${\rm Vol}(G)$, and aside from a few exactly-solvable special cases studied in \cite{Higuchi_1991_2}, none of the above arguments actually identified a domain $\Phi$ in the perturbative dS Fock space on which \eqref{eq:GAIP} could be shown to be positive semi-definite.\footnote{Though the cases studied in \cite{Higuchi_1991_2} include the important setting of perturbative gravity about dS$_4$ in which the linearized state contains only gravitons.  Group averaging can also be shown to be positive for various other theories and models, see e.g. \cite{Marolf:1994wh,Marolf:1994ae,Ashtekar:1995zh,GomberoffMarolf1999,Louko:1999tj,Marolf:2008it}.}

Our work here fills this lacuna by establishing positivity for group averaging in global dS$_d$ with $d \ge 2$ in states with a sufficient number of sufficiently heavy particles, with a caveat for the special case $d=2$ explained below.  In particular, our main positivity result applies to states with $N$ particles defined by standard mode-functions on global dS$_d$ in which the weights $\Delta_i$ of the particles satisfy 
\begin{equation}
\label{eq:gaconv}
\Sigma: = {\rm Re} \, \sum_{i=1}^N \Delta_i > d-1.
\end{equation}  The definition of such $\Delta_i$ will be reviewed below, but for particles with masses above the dS scale it is worth noting that \eqref{eq:gaconv} is equivalent to choosing $N\ge 3$ (while more particles can be required for smaller masses).  

For the case $d>2$ in which pure de Sitter solutions are expected, we emphasize that our argument allows all possible types of free particles.  By this we mean that we allow the particles to transform in arbitrary representations of the universal cover Spin$_0(d,1)$ of the connected component SO$_0(d,1)$ of the dS$_d$ isometry group.   We also consider the case $d=2$, which we include largely as a toy model and as a warm-up to the higher-dimensional case.  Einstein-Hilbert gravity is of course trivial in $d=2$ dimensions.  Furthermore, as noted above, dilaton gravity theories tend to admit dS$_2$ solutions only with accompanying dilaton profiles that break the SO$_0(2,1)$ isometry group to a one-dimensional subgroup. For such cases group averaging can be shown to be positive using the arguments of \cite{Marolf:1994wh,Marolf:1994ae} so that the arguments below are not needed.  We therefore restrict discussion of $d=2$ to bosons and fermions (which provide useful toy models for the case $d>2$), commenting only briefly in section \ref{sec:disc} on the more general anyonic case that can arise for $d=2$.

As noted in \cite{Marolf_2009}, the criterion \eqref{eq:gaconv} is sufficient to make the group-averaging integral absolutely convergent for standard Fock basis states defined by global dS mode functions.\footnote{While ref. \cite{Marolf_2009} discussed only scalar, vector, and symmetric-tensor fields, its observations relating this absolute convergence to the exponential decay of matrix elements extend directly to spinning representations as well.}  Interestingly, while the group-averaging norms of some states diverge as expected at $\Sigma=d-1$,  we also identify various classes of states for which group averaging continues to converge even when the naive convergence criterion of \cite{Marolf_2009}  is violated; i.e., when \eqref{eq:gaconv} fails to hold.
In a $d=2$ case we find that the group-averaging norm continues to be positive for all $\Sigma > d-2$, and we expect this to continue to hold for related cases with $d>2$.  However, there are also cases where, despite the absolute convergence of group averaging for those states, and despite the heuristic argument for positivity given above, we find negative group-averaging norms for $\Sigma < d-1$.

Our work here was motivated by the analyses in \cite{KolchmeyerLiu2024}, which emphasized two points.  The first was that there is only one formal de Sitter invariant state in the 2-particle sector of the Fock space of any free field, and that its explicit form is known; these statements also follow from an argument in \cite{Higuchi_1991_2} (see also \cite{Giddings:2026fhw} for an explicit construction for a massive scalar in general $d$).  From this expression a certain `distributional' positivity in the 2-particle sector is easily established (though we will use a simpler, more abstract argument below).  This argument was initially used for particular examples in \cite{Higuchi_1991_2}, but it turns out to generalize readily. 

The second motivating point emphasized in \cite{KolchmeyerLiu2024} was that $N$-particle states may be usefully written in terms of states containing a smaller number of `effective particles' by decomposing tensor product representations of $G$ into a sum of irreducibles.   Putting these two facts together is sufficient to establish the desired positivity result for both principal-series and discrete-series particles when $d=2$ (i.e., for $M^2 \ge \frac{1}{4\ell_{dS}^{2}}$ and for other special cases), and for the analogous so-called `tempered representations' in higher dimensions. For $d=2$ some further manipulation then also establishes positivity for light (complementary-series) fields that satisfy \eqref{eq:gaconv}.  The three-particle states play a key role in this analysis as, outside the so-called discrete series,  $N=3$ is the smallest number of particles that can satisfy \eqref{eq:gaconv}. 
Similar arguments then turn out to hold for all $d>2$, though presenting a unified treatment requires a more abstract approach that draws heavily from the mathematics literature.

To set the stage, and to motivate the finite (and often small) number of 3-particle states with vanishing dS charge that arise in any quantum free-field Fock space,  we begin in section \ref{sec:classical} with a discussion of classical configurations of either 2 or 3 free spinless positive-mass point particles in dS$_d$ that are carefully tuned to set the  total de Sitter charges of the configuration  to zero.  In the 2-particle case, there is a vanishing charge configuration only when the two particles have the same mass.  In that case, the configuration is unique (up to dS isometries) and the particles necessarily follow antipodal geodesics. In contrast, the 3-particle case allows arbitrary masses, though for given generic values of $m_1,m_2,m_3$ the configuration is again unique up to dS isometries.   It is then unsurprising that 2- and 3-particle states built from scalar fields and satisfying the desired constraints have analogous properties in the quantum case.

Our main quantum argument then appears in sections \ref{sec:principal-discrete-positivity}--\ref{sec:higherD}.  In these sections we restrict discussion to bosonic  (integer spin) particles, after which it will turn out to be essentially trivial to include fermions as well.  Equivalently, in the main text we take our particles to be unitary irreducible representations of SO$_0(d,1)$ and consider the covering group Spin$_0(d,1)$ only in section \ref{sec:disc}.

We first provide a very concrete proof in section \ref{sec:principal-discrete-positivity} of positivity for principal- and discrete-series bosonic (integer spin) fields in dS$_2$. This also allows us to introduce some basic machinery and notation.   A similarly concrete argument for dS$_2$ complementary-series fields then follows in section \ref{sec:complementary-positivity}.  The higher-dimensional bosonic argument is described more abstractly in section \ref{sec:higherD}, with the critical 3-particle complementary-series case analyzed separately in section \ref{subsec:3pgend}.  We close with a brief summary and discussion of remaining issues in section \ref{sec:disc}.  This in particular includes describing the trivial extension to states with fermions (particles with half-integer spin), as well as commenting on the implications of our convergent-but-negative group-averaging norms.

\section{Classical two- and three-particle states in \texorpdfstring{dS$_d$}{dS\_d}}
\label{sec:classical}

As noted above, in order for a configuration of classical matter to be a consistent source of perturbative gravitational back-reaction, all of the de Sitter charges associated with the matter fields must vanish.  Here we consider matter consisting of spinless  classical free particles with positive mass.  Such particles necessarily travel on geodesics and, should they happen to collide, simply pass through each other without deflection.

It is useful to begin with a brief reminder of the de Sitter symmetries and the associated charges.  Recall that dS$_d$ with de Sitter length $\ell_{dS}$ may be defined as the hyperboloid
\begin{equation}
\ell_{dS}^2 = -T^2 + \sum_{i=1}^{d} (X^i)^2
\label{eq:dShyp}
\end{equation}
in $(d+1)$-dimensional Minkowski space $M^{d+1}$.  The dS isometries are then simply the (origin-preserving) Lorentz transformations in $M^{d+1}$.  Given a particular choice of time coordinate $T$, it is then natural to divide the generators into rotations (which preserve every constant-$T$ surface) and boosts.

\begin{figure}
    \centering
    \includegraphics[width=0.45\linewidth]{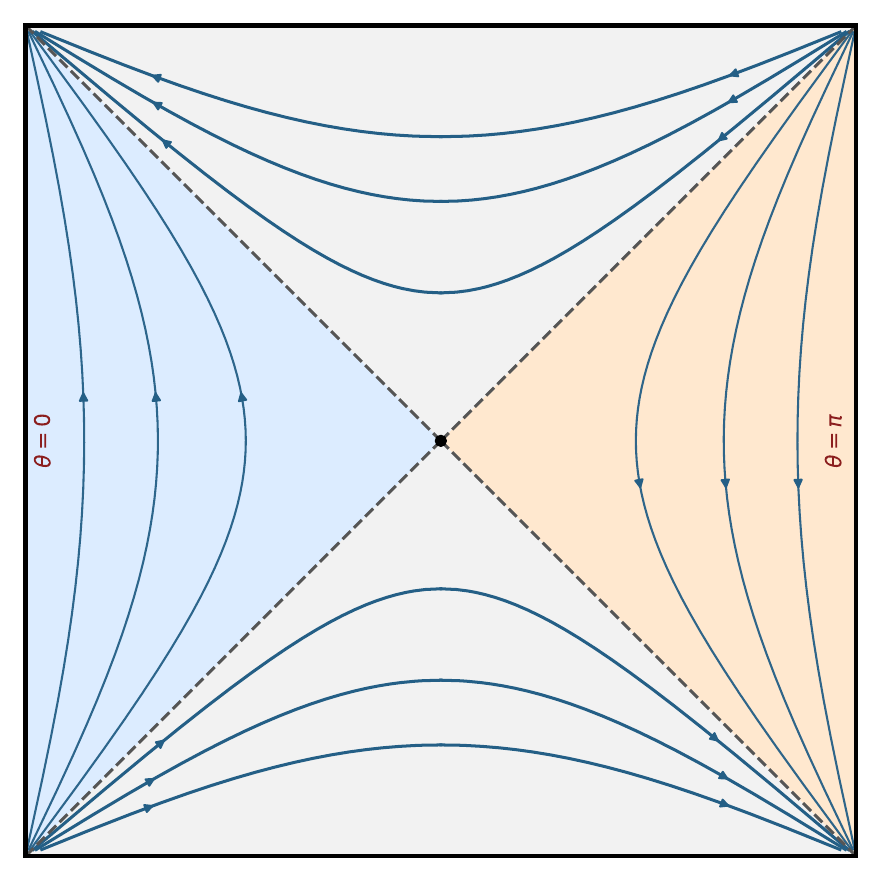}
    \caption{
    A conformal diagram of $\mathrm{dS}_d$ that suppresses an $S^{d-2}$ (away from the poles of the $S^{d-1}$ at $\theta=0,\pi$).  The left and right boundaries are the pair of antipodal timelike geodesics at these poles. The dashed
    lines are the Killing horizons of the boost in the $X^d$ direction, and
    the corresponding left and right static patches are shaded blue and orange.
    The dark blue curves (with arrows) show the flow generated by the boost along $X^d$.
    }
    \label{fig:confdiag}
\end{figure}

A useful conformal diagram for dS$_d$ is shown in figure \ref{fig:confdiag}.  Each point on the conformal diagram represents a full $S^{d-2}$ (so that for $d=2$ it in fact represents a pair of points), associated with an O$(d-1)$ subgroup of the de Sitter group. To be definite, we can take this to be the $S^{d-2}$ defined by the intersection of the hyperboloid \eqref{eq:dShyp} with the $(d-1)$-plane of constant $T,X^d$. This sphere of course degenerates at the poles defined by the two curves $X^d = \pm \sqrt{T^2+\ell_{dS}^2}$, which we may call $\theta=0, \pi$, and which form the left and right boundaries of the conformal diagram.

Figure \ref{fig:confdiag} also shows the flows generated by the boost in the $X^d$ direction.  This isometry clearly preserves the two curves $X^d = \pm \sqrt{T^2+\ell_{dS}^2}$ which in fact define an antipodal pair of geodesics in dS$_d$.  Since our boost is hypersurface-orthogonal, it may be said to define a notion of time evolution under which dS$_d$ is static.  But since the boost becomes null on its Killing horizons, this sense applies only in the two regions where the boost is timelike.  These are the so-called `static patches' associated with the geodesics at $\theta=0,\pi$.

Let us now turn to our discussion of classical massive free particle configurations with vanishing dS charges.
In the two-particle case, it is well known that this configuration is unique.  One way to see this is to note that, without loss of generality, we may place one particle on the $\theta=0$ geodesic and take it to define a pair of static patches as in figure \ref{fig:confdiag}.  Since the angular momentum of this particle on any constant-$T$ slice clearly vanishes, the vanishing of the total angular momentum then requires the same to be true of the second particle as well.  Particle 2 must therefore also follow a geodesic at some constant value of $\theta$.

However, we must also impose the vanishing of  the total boost charge $B_i$ of the configuration along each direction $X^i$.  Note that $B_i$ is odd under the reflection $X^i \rightarrow - X^i$.  Recall also that particle 1 travels on a geodesic that satisfies $X^i=0$ for $i \neq d$.  This means that it is invariant under the above reflections for $i\neq d$, and thus that it must have $B_i=0$.

Vanishing of the total $B_i$ then also requires $B_i=0$ for particle 2.  As a result, the constant-$\theta$ geodesic followed by particle 2 must lie at either $\theta=0$ or $\theta=\pi$.  Placing it at $\theta=0$ gives a total $X^d$ boost charge $B_d = (m_1+m_2)\ell_{dS}$, which of course cannot vanish for positive $m_1,m_2$.  Placing particle 2 at $\theta=\pi$ instead gives  $B_d = (m_1-m_2)\ell_{dS}$, which vanishes precisely when $m_1=m_2$. This is thus the unique solution, and we see that it exists only for $m_1=m_2$.   There are simply no other positive-mass 2-particle cases that admit consistent gravitational back-reaction.

Let us now consider the more interesting case of 3 free particles.   We can again place particle 1 on the $\theta=0$ geodesic on the left boundary of figure \ref{fig:confdiag}.  Since this geodesic has no angular momentum, the angular momenta of particles 2 and 3 must be equal and opposite.  But this requires particles 2 and 3 to be confined to a common great circle on each $S^{d-1}$ at constant $T$.  Since our $\theta=0$ worldline for particle 1 preserves an SO$(d-1)$ rotational symmetry, we may use that symmetry to place the great circle associated with particles 2 and 3 in the subspace associated with $X^1,X^2,X^d$.  I.e., the problem has now been reduced to studying the cases of dS$_3$ and dS$_2$.

Let us first consider the case $d=3$.  We may use rotational symmetry in the $X^1,X^2$ plane to set $X^2=0$ for particle 2
at $T=0$. At $T=0$ we thus find both particles 1 and 2 to lie on the Killing horizon for the boost along $X^2$ so that their contributions to the $X^2$ boost charge ($B_2$) must vanish.  Vanishing of the total boost charge $B_2$ thus also requires  particle 3 to be at such a Killing horizon at $T=0$; i.e., it also has $X^2=0$ at $T=0$.

Since particles 2 and 3 move on a common great circle containing their $T=0$ positions, this great circle lies in the plane $X^2=0$ unless those positions are antipodal or coincide (and if particles 2 and 3 have vanishing angular momentum, each simply remains at its $T=0$ angular position). In the antipodal or coincident cases, vanishing of $B_1$ and $B_d$ at $T=0$ requires particles 2 and 3 to lie at $\theta=0,\pi$, and a rotation about the $X^d$ axis then brings their great circle into the plane $X^2=0$. It follows that all particles lie in a dS$_2$ submanifold, so we have in fact reduced the general dS$_d$ case to a study of dS$_2$ (though, to preserve the above conventions, we associate it with the directions $T,X^1,X^d$ in the embedding space).  In particular, we can introduce a standard dS$_2$-like angular coordinate $\tilde \theta$ by taking $\tilde \theta \in[0,2\pi)$ and defining $X^d =\sqrt{\ell^2_{dS}+T^2}  \cos \tilde \theta,$
$X^1 = \sqrt{\ell^2_{dS}+T^2} \sin \tilde \theta,$ so that $\tilde \theta$ combines information from $\theta$ with the sign of the embedding coordinate $X^1$.   To provide a uniform treatment, we will also use $\tilde \theta$ to denote the angular coordinate in dS$_2$.

The attentive reader will note that we have yet to make full use of the freedom to act with boosts along $X^d$.  Before doing so, let us observe that,
since particle 1 has vanishing charge under boosts along $X^1$, we can assign particles 2 and 3 boost charges $\pm b_1$ for some $b_1 \in {\mathbb R}$.  The total boost charge for the system is then $B_1 = b_1 - b_1=0$.  Let us also call their angular momenta $\pm J$.  Acting with a boost of rapidity $\eta$ along $X^d$ then gives new such charges
\begin{align}
J' &= J\, \cosh \eta  - b_1 \, \sinh \eta,  \\
b_1'&= b_1 \, \cosh \eta - J \, \sinh \eta.
\end{align}
For $|J|\neq |b_1|$ we may choose $\eta$ to minimize $|J|$ or, what one may check is equivalent, to minimize $|b_1|$.  There are now two cases.  For $|b_1|> |J|$ the minimum is
\begin{align}
J' &= 0, \\
b_1'&= {\rm sign}(b_1) \sqrt{b_1^2 -J^2}\neq0,
\end{align}
while for $|J|> |b_1|$ the minimum is
\begin{align}
J' &= {\rm sign}(J) \sqrt{J^2 -b_1^2}\neq 0, \\
b_1'&= 0.
\end{align}
For $|J|= |b_1|$, the pair $(J,b_1)$ transforms like a null vector.  By taking $\eta$ large (and of the appropriate sign) we can make $(J,b_1)$ arbitrarily small.  But they cannot be set to zero unless we start with $J=b_1=0$.

In the first case we have found a global dS reference frame in which all particles are at rest at angular coordinates $\tilde \theta_1=0, \tilde \theta_2, \tilde \theta_3$.  Setting the total charges to zero for boosts along $X^1$ and $X^d$ then yields
\begin{align}
0&= m_2  \sin \tilde \theta_2 + m_3 \sin \tilde \theta_3,\\
0 &= m_1 + m_2 \cos \tilde \theta_2 + m_3 \cos \tilde \theta_3.
\end{align}
This pair of equations can then be solved for $\tilde \theta_2,\tilde \theta_3$.  Real solutions $\tilde \theta_2,\tilde \theta_3$ exist when the masses satisfy the so-called triangle inequalities
\begin{align}
\label{eq:mless}
m_1 &< m_2+m_3, \\
m_2 &< m_3+m_1, \\
m_3 &< m_1+m_2, \label{eq:mless3}
\end{align}
and the solution is unique up to a reflection $X^1 \rightarrow -X^1$. (For $d\ge3$ the two solutions may equally well be related by a $\pi$ rotation in the $X^1X^2$ plane.)

In the second case, particles 2 and 3 must pass through the bifurcation surface of the Killing horizon for boosts along $X^1$.  I.e., they must have $\tilde \theta=0,\pi$ at $T=0$. Setting the total charge to zero for the boost along $X^d$ thus requires
\begin{equation}
\label{eq:Bcase}
m_1 \pm \sqrt{m_2^2+(J')^2/\ell_{dS}^2}\pm \sqrt{m_3^2+(J')^2/\ell_{dS}^2}=0,
\end{equation}
where the terms for particles 2 and 3 enter with $(+)$ signs for particles with $\tilde \theta=0$ at $T=0$ and they enter with $(-)$ signs for particles with $\tilde \theta=\pi$ at $T=0$. While it is clear that \eqref{eq:Bcase} cannot be satisfied for real $J'$ when both $\pm$ signs are positive, each of the other three choices of the two signs gives a unique real solution for $|J'|$ for masses in the ranges shown below for the allowed choices of $\tilde \theta_2,\tilde \theta_3$:
\begin{align}
\label{eq:mgreat}
m_1 &> m_2+m_3, \ \ \ \tilde \theta_2=\tilde \theta_3=\pi,\\
m_2 &> m_3+m_1, \ \ \ \tilde \theta_2=\pi,\tilde \theta_3=0,\\
m_3 &> m_1+m_2,  \ \ \ \tilde \theta_2=0,\tilde \theta_3=\pi. \label{eq:mgreat3}
\end{align}
Again, the two choices of sign for $J'$ are related by the reflection $X^1\rightarrow -X^1$, or by a rotation in the $X^1X^2$ plane for $d\ge3$.

The generic null-vector case $|J|=|b_1|\neq 0$ is clearly intermediate between the above two cases.  It thus occurs precisely when one of the inequalities \eqref{eq:mless}--\eqref{eq:mless3} (or, equivalently, \eqref{eq:mgreat}--\eqref{eq:mgreat3}) is saturated.  In this case some pair of particles meets at a point on the past or future boundary, but there are no intersections at finite time.

Now, when one of the inequalities \eqref{eq:mgreat}--\eqref{eq:mgreat3} is saturated, there is also the additional possibility $J=b_1=0$.  This case is exceptional in several ways.    First, up to isometry it corresponds to having two particles at rest along the geodesic at $\theta=0$ and one at $\theta=\pi$ (or vice versa).  It thus resembles the 2-particle configuration and is invariant under boosts along $X^d$.  Second, when discussed in a generic reference frame it clearly requires an extra fine-tuning to make both $J$ and $b_1$ vanish.

In summary, if we ignore the finely-tuned cases $J=b_1=0$ or $m_1+m_2=m_3$ (and permutations thereof), for each choice of $m_1,m_2,m_3> 0$ in dS$_d$ with $d\ge 3$, up to isometries in the connected component SO$_0(d,1)$ there is precisely one set of geodesics for which the total de Sitter charges all vanish.  For $d=2$, for each $m_1,m_2,m_3$ there are precisely two such sets of geodesics that are related by a reflection.  In both cases we will see that de Sitter invariant quantum states are similarly unique for $d>2$ (and that they are unique up to a reflection for $d=2$), as are the 3-point functions of spinless operators in $(d-1)$-dimensional conformal field theories.  Since $J$ and $b_1$ do not commute, it is natural to expect that there is no quantum analogue of the finely-tuned case $J=b_1=0$.

\section{Principal- and discrete-series group-averaging positivity for \texorpdfstring{dS$_2$}{dS\_2}}
\label{sec:principal-discrete-positivity}

We now turn to the positivity of the group-averaging inner product on free-field Fock states defined by bosonic particles in which the state of each particle is given by a finite superposition of standard global dS modes.  Following standard discussions of Fock states in Minkowski quantum field theory (see e.g. \cite{Weinberg:1995mt}), we take our Fock space to be defined by $N$-particle states in which each particle transforms in a unitary irreducible representation of the de Sitter isometry group, and for which the full $N$-particle state then transforms in the tensor product representation.   This section considers the simple cases  corresponding to fields  whose one-particle states provide the so-called principal- and discrete-series representations of the dS${}_2$ isometry group SO$_0(2,1)$.  This also provides an opportunity to introduce notation, conventions, and basic properties of 
unitary irreducible representations (UIRs) of SO$_0(2,1)$ that will continue to be of use in higher dimensions as well and, in addition, to describe the relation between the principal- and complementary-series representations for $d=2$.  We will do both in section \ref{subsec:PD-conventions-domain} below.

In such cases, the fact that the de Sitter invariant two-particle state requires $m_1=m_2$, and that it is then unique, allows us to quickly show in section \ref{subsec:PD-two-particle-base} that group averaging is positive in the two-particle sector.   Positivity on states with higher numbers of particles then follows in section \ref{subsec:PD-CG-decomposition} by using Clebsch-Gordan decomposition of tensor products to write such states in terms of a smaller number of effective particles.  This will set the stage for our study of the remaining (complementary-series) representations in section \ref{sec:complementary-positivity}.

\subsection{Notation, conventions, and preliminaries}
\label{subsec:PD-conventions-domain}

We now provide a very explicit review of various useful results  related to group averaging over the connected component SO$_0(2,1)$ of the dS$_2$ isometry group. 
We begin with basic properties of de Sitter group averaging itself (in section \ref{sec:dSGArev}), followed by an overview in section \ref{sec:dS2UIRs} of the unitary irreducible representations of SO$_0(2,1)$ and some final comments about notation in section \ref{subsubsec:Notation}.

\subsubsection{Group averaging for \texorpdfstring{SO$_0(2,1)$}{SO\_0(2,1)}}
\label{sec:dSGArev}

Recall that we are interested in the group-averaging inner product
\begin{align}
        \braket{\phi}{\psi}^{\rm GA}=\int_G dg\braket{\phi}{U(g)|\psi},
        \label{eq:PD-GA-definition}
\end{align}
where in this section
\(G={\rm SO}_0(2,1)\).
As described in e.g. \cite{Marolf_2009},   one can write the Haar measure $dg$ in the explicit form
\begin{align}
dg=d\theta_L\,d\theta_R\,\sinh\lambda\,d\lambda
        \label{eq:PD-Haar-measure}
\end{align}
in terms of the Cartan decomposition
\begin{align}
        g=k_La(\lambda)k_R,\qquad a(\lambda)=e^{i\lambda B_1},\qquad \lambda\in[0,\infty)
        \label{eq:PD-Cartan-decomposition}
\end{align}
with $k_L = e^{i\theta_L J}$ and $k_R = e^{i\theta_R J}$ for $0\le \theta_L, \theta_R < 2\pi$ and $J$ the rotation generator of SO$_0(2,1)$.

The integrals over $\theta_L, \theta_R$ each give $2\pi$ times the projector onto rotation-invariant states.   The group-average of a general state thus depends only on its component with vanishing total angular momentum. On the other hand, if \(\phi\) and \(\psi\) already satisfy $J\phi=J\psi=0$, the two compact integrations each give $2\pi$ and we find
\begin{align}
        \braket{\phi}{\psi}^{\rm GA}=4\pi^2\int_0^\infty d\lambda\,\sinh\lambda\,\langle \phi | e^{i\lambda B_1} | \psi \rangle.
        \label{eq:PD-GA-boost-reduced}
\end{align}
For future reference, we mention that group averaging for dS$_d$ is of course much the same.  In particular, for rotation-invariant states one again finds an expression of the form \eqref{eq:PD-GA-boost-reduced} but with a measure proportional to $\sinh^{d-1} \lambda\, d \lambda$; see e.g. \cite{Marolf_2009}.

\subsubsection{UIRs of \texorpdfstring{SO$_0(2,1)$}{SO\_0(2,1)}}
\label{sec:dS2UIRs}

It will clearly be important to understand the decay of the matrix elements $\langle \phi | e^{i\lambda B_1}| \psi \rangle$ at large $\lambda$.  This decay will depend on the SO$_0(2,1)$ representation to which the states $\phi, \psi$ belong, as well as the particular choice of states $|\phi\rangle, |\psi\rangle$ themselves.  

As stated above, 
we are most interested in states that arise in linear quantum field theories on dS$_d$, and we focus in this work on theories (or sectors of theories) that give rise to Fock-type Hilbert spaces with a well-defined vacuum and with a well-defined notion of particle number.  In any such theory, the one-particle states must transform in a unitary representation of SO$_0(d,1)$.  It is then natural to decompose the one-particle sector into a sum of irreducible such representations, and to then think of different representations appearing in such a decomposition as corresponding to distinct species of particle. 

As a result, it will be useful to introduce some basic terminology and notation for unitary irreducible representations (UIRs) of
SO$_0(2,1)$.  These representations were first studied by Bargmann \cite{Bargmann1947}, who found that they could be labeled by values of the quadratic Casimir and that the UIRs could then be sorted into three types that are now known as the principal-, discrete-, and complementary-series UIRs \cite{Knapp2001Semisimple}.  When such representations are interpreted as one-particle states of free quantum scalar fields\footnote{In two spacetime dimensions there are no independent higher-spin linear bosonic fields.} on dS$_2$, this Casimir is just the mass-squared of the quantum field, though it is both useful and conventional to discuss such representations in terms of the so-called weight $\Delta$ that satisfies
\begin{equation}
\label{eq:mDelta}
M^2 = \Delta(1-\Delta)/\ell_{dS}^2.
\end{equation}

The quadratic relationship \eqref{eq:mDelta} generically assigns two values of $\Delta$ to each UIR, with the two values being related by the so-called shadow transformation $\Delta\leftrightarrow 1-\Delta$, and with one of the allowed values of $\Delta$ sometimes being called the `shadow' of the other; see e.g. \cite{Sun:2021thf}.  

Particles with $M^2 >0$ turn out to correspond to the UIRs of the principal and complementary series.  In particular, the principal-series UIRs occur for $M^2 \ge \frac{1}{4\ell_{dS}^2}$, which yields complex weights $\Delta = \frac{1}{2} +is$ for real $s$.  For such representations, the shadow weight $1-\Delta$ is just the complex conjugate $\Delta^*$, and the shadow transformation is trivial at $\Delta = \frac{1}{2}$ (which gives $s=0$).  It will be useful to denote the principal-series representation with weight $\Delta = \frac{1}{2}+is$ using the symbol $P_s$ (or, equivalently, by $P_\Delta$ for $\Delta = \frac{1}{2} + is$) and to take $s\ge 0$ by convention.

The cases $0<M^2 < \frac{1}{4\ell_{dS}^2}$ correspond to complementary-series UIRs.  In this case $\Delta$ is real, and we will find it useful to break the shadow symmetry $\Delta \leftrightarrow 1-\Delta$ by labeling our complementary-series UIRs with the smaller of the two allowed values; i.e., in the complementary series we take $0 < \Delta < \frac{1}{2}$.  

The remaining discrete series has integer $\Delta\ge 1$, and there are in fact two such representations for each such $\Delta$ that are distinguished by a choice of sign $\epsilon \in\{+,-\}$.  We will denote the corresponding UIRs by $D^\pm_\Delta$. The representations 
$D^+_\Delta$ turn out to contain only states with positive angular momentum quantum numbers $m \ge \Delta$, while the $D^-_\Delta$ turn out to contain only states with negative angular momentum quantum numbers $m \le -\Delta$.
Note that (for $\Delta \neq 1$) \eqref{eq:mDelta} then gives $M^2<0$ so that such UIRs can only describe one-particle states of tachyonic fields.  Here it is useful to recall that, while the long-wavelength modes (those with $|m|<\Delta$) of tachyonic fields are unstable and do not admit a well-defined notion of particle number,\footnote{Instead, they behave like a Newtonian particle in an inverted harmonic potential and thus have continuous (not quantized) energy spectra.} for fixed $M^2 <0$ the modes with sufficiently short wavelength are in fact stable and do lead to a Fock space with well-defined particle number.  
Together, the two representations $D^\pm_\Delta$ for given $\Delta$ may be thought of as describing one-particle excitations of the Fock space defined by such short-wavelength modes.

In all three series, the angular momentum eigenstates define a useful orthogonal basis for each UIR.  We denote the normalized elements of this basis by $|m\rangle_s$ or $|m\rangle_\Delta$ (with $\Delta = \frac{1}{2}+is$ in the principal series), and the Hilbert-space-adjoint bra-vectors by ${}_\Delta \langle m|$.  In all cases we will use the conventions of appendix E of \cite{KolchmeyerLiu2024} for these states; note that these conventions differ from those of the main text of \cite{KolchmeyerLiu2024} and from those used in 
\cite{Sun:2021thf}.  Note also that there is no need to explicitly indicate the sign $\epsilon \in \{+, -\}$ for a discrete-series representation as this is determined by the sign of the angular momentum quantum number $m$.

The weights $\Delta$ will play a central role in our discussion of group averaging.  This stems from the fact that, as emphasized by e.g. \cite{Marolf_2009}, the weight of a representation controls the leading fall-off of $U(g)$ matrix elements at large $g$ and thus the convergence properties of \eqref{eq:PD-GA-boost-reduced}. To make this explicit, let us recall that the  closed-form expressions for arbitrary matrix elements of any $U(g)$ in any SO$_0(2,1)$ UIR were worked out long ago by Bargmann \cite{Bargmann1947} (see also \cite{BarutWilson1976,Rajabov2018,VK}).  Since we will eventually be interested in the expressions in all three series, it is also useful to recall that matrix elements ${}_\Delta\langle m|U(g)|m'\rangle_{\Delta}$ are {\it not} strictly analytic functions of $\Delta$.  More specifically, using the conventions of appendix E of \cite{KolchmeyerLiu2024}, the matrix elements of the principal series {\it do} define a meromorphic function of $\Delta$, but the analytic continuation of this function to the complementary series is not just 
${}_\Delta\langle m|U(g)|m'\rangle_{\Delta}$. Instead, it  gives\footnote{\label{foot:hdisc}A related formula also holds in the discrete series though, with our conventions, that formula involves $h_{1-\Delta}(m)$.} 
$\sqrt{\frac{h_\Delta(m')}{h_\Delta(m)}}{}_\Delta\langle m|U(g)|m'\rangle_{\Delta}$ where, following \cite{VK}, \begin{equation}
\label{eq:hDef}
h_\Delta(m) =  \frac{\Gamma(m+\Delta)}{\Gamma(m+1-\Delta)}. 
\end{equation}
Noting that $h_\Delta(m)$ is real and positive in the complementary series, while it is a pure phase in the principal series, one may also simply say that the quantities 
\begin{equation}
\label{eq:hme}
\sqrt{\Bigg|\frac{h_\Delta(m')}{h_\Delta(m)}\Bigg|}\ {}_\Delta\langle m|U(g)|m'\rangle_{\Delta}
\end{equation}
{\it do} define meromorphic functions of $\Delta$.  By this we mean that, while they
are a priori defined only when $\Delta$ labels a UIR (since their definition involves a Hilbert space inner product), we may consider \eqref{eq:hme} in the principal series (where $h_\Delta$ is just a phase) and note that it defines an analytic function for such $\Delta$. We may then analytically extend this function to obtain a meromorphic function on the complex $\Delta$-plane.  The statement is then that, when evaluated in the complementary series (where $h_\Delta >0$ so that the absolute value operation has no effect), this extended function again coincides with \eqref{eq:hme}.  Here we take the positive branch of the square root in both the principal and complementary series.

Before we leave the discussion of \eqref{eq:hme},  it is worth noting that the additional $h_\Delta$ factors in \eqref{eq:hme} cancel explicitly for $m=m'$ and so are relevant only for off-diagonal matrix elements.  Furthermore, using the identity $\Gamma(z)\Gamma(1-z) = \frac{\pi}{\sin \pi z}$ twice and noting that $m\in {\mathbb Z}$ implies $\sin[\pi(m+1-\Delta)] = \sin[\pi(\Delta+m)]$ yields $h_\Delta(m) = h_\Delta(-m)$.  As a result, the $h_\Delta$ factors in 
\eqref{eq:hme} also cancel  for $m=-m'$.

We may now describe the explicit matrix elements of $U(g)$ in a useful form.
Following the conventions of appendix E of \cite{KolchmeyerLiu2024}, taking $U(g)$ to be a boost in the $X^1$ direction (with $B_1 =\frac{i}{2}(L_1-L_{-1})$ in the conventions of \cite{KolchmeyerLiu2024}), defining $r=m-m'$, and taking $m \ge m'$, for both the principal and complementary series the explicit results of \cite{Bargmann1947}  may be written
\begin{equation}
\label{eq:SO21explicit}
\sqrt{\Bigg|\frac{h_\Delta(m')}{h_\Delta(m)}\Bigg|}{}_\Delta\langle m|e^{i\lambda B_1}|m'\rangle_\Delta
=
\frac{(it)^r}{r!}\,
c^{-2\Delta}\,
(m'+1-\Delta)_r\,
{}_2F_1\!\left(\Delta-m',\,\Delta+m;\,r+1;\,t^2\right),
\end{equation}
where 
\begin{equation}
t := \tanh\frac{\lambda}{2},
\qquad
c := \cosh\frac{\lambda}{2},
\end{equation}
and the Pochhammer symbol is
\begin{equation}(m'+1-\Delta)_r
=
(m'+1-\Delta)(m'+1-\Delta+1)\cdots(m'+1-\Delta+r-1).
\end{equation}
The case $m' \ge m$ then follows from unitarity $\left({}_\Delta\langle m|e^{i\lambda B_1}|m'\rangle_\Delta \right)^*={}_\Delta\langle m'|e^{-i\lambda B_1}|m\rangle_\Delta $.

At large $\lambda$ these expressions can be expanded in exponentials and the leading terms are always of the form $e^{-\lambda \Delta}$ in the discrete and complementary series, while in the principal series one finds a pair of co-leading terms proportional to $e^{-\lambda \Delta}$ and $e^{-\lambda \Delta^*}$.  As a result, as noted in \cite{Marolf_2009}, since $U(g)$ acts diagonally on multiparticle states (so that matrix elements of $U(g)$ in tensor product states are obtained by multiplying together the matrix elements of $U(g)$ in the individual factors), if $|\phi\rangle, |\psi\rangle$ are both $N$-particle states with particle weights $\Delta_i$ then, for finite linear combinations of tensor products of the above angular momentum basis states,\footnote{This set of states is not itself de Sitter invariant.  However, it can be promoted to a de Sitter invariant linear space by acting with the group algebra $L^1(G)$ as in section IIIA of \cite{Giulini_1999}.  We will implicitly assume that this has been done throughout the rest of this work.} our SO$_0(2,1)$ group averaging \eqref{eq:PD-GA-boost-reduced} converges absolutely for
\begin{equation}
\Sigma:= {\rm Re} \, \sum_{i=1}^N \Delta_i > 1.
\label{eq:D>1}
\end{equation}
While we will also investigate special cases in section \ref{sec:complementary-positivity} below where group averaging continues to converge absolutely for even smaller values of $\Sigma$, it will be useful to refer to \eqref{eq:D>1} (and its higher-dimensional analogues) as the naive absolute convergence bound of \cite{Marolf_2009}.  

Importantly, states with $N<3$ particles can satisfy \eqref{eq:D>1} only if at least one of those particles lies in a discrete-series representation.  The case of a 1-particle state in a discrete representation with $\Delta >1$ is rather trivial as, since it was mentioned above that there are no rotation-invariant states in any discrete representation, group-averaging any such Fock state gives a null state in ${\cal H}^{\rm GA}$.  The two-particle case will be discussed in detail in section \ref{subsec:PD-two-particle-base}.  However, we emphasize that two principal-series particles always have ${\rm Re} \left(\Delta_1+\Delta_2\right)=1$.  As a result, the group-averaging integral \eqref{eq:PD-GA-boost-reduced} for inner products is oscillatory.  In particular, it does not converge absolutely.  Instead, it converges only in a distributional sense that will be discussed in section \ref{subsec:PD-two-particle-base} below.  

For studies of the complementary series, it is often useful to note that the above explicit form of the matrix elements \eqref{eq:SO21explicit} simplifies greatly for low angular momenta.  In particular, for $m, m' = 0$ in either the principal or complementary series  we find 
\begin{align}
 \phi_\Delta(\lambda) & :=
{}_\Delta\langle0|e^{i\lambda B_1}|0\rangle_\Delta
=
c^{-2\Delta}\,
{}_2F_1\!\left(
\Delta,\Delta;1;t^2
\right) \nonumber \\ &
=P_{\Delta-1}(\cosh\lambda) \nonumber \\
\label{eq:0002d}
&=\frac1{2\pi}\int_0^{2\pi}d\vartheta\,
   \left(\cosh \lambda +\sinh \lambda\cos\vartheta\right)^{\Delta-1},
\end{align}
where $P_{\Delta-1}$ is the Legendre function with index $\Delta-1$.
In particular, the final line is manifestly positive (and convergent) for complementary-series weights (and, indeed, for all positive real $\Delta$).  Together with the fact that, in such cases, the leading decay of \eqref{eq:0002d} is precisely $e^{-\lambda \Delta}$, the positivity of this function makes explicit the fact that the group-averaging norm of a state with two complementary-series particles diverges (since it is given by a divergent integral of a positive quantity).

For later use we also note that for $m,m' = \pm 1$ we find
\begin{align}
{}_\Delta\langle 1|e^{i\lambda B_1}|1\rangle_\Delta\,
&=
c^{-2\Delta}\,
{}_2F_1\!\left(
\Delta-1,\Delta+1;1;t^2
\right)\nonumber \\ 
&=
\phi_\Delta(\lambda)
-\frac{\tanh \frac{\lambda}{2}}{\Delta(\Delta-1)}
\frac{d\phi_\Delta}{d\lambda},
\label{eq:11}
\end{align}
and, since $h_\Delta(1)=h_\Delta(-1)$, we also obtain
\begin{align}
{}_\Delta\langle1|e^{i\lambda B_1}|{-1}\rangle_\Delta
&=
-\frac{\Delta(\Delta-1)}{2}\,
t^2 c^{-2\Delta}\,
{}_2F_1\!\left(
\Delta+1,\Delta+1;3;t^2
\right) \nonumber
\\
&=
-\phi_\Delta(\lambda)
+\frac{\coth \frac{\lambda}{2}}{\Delta(\Delta-1)}
\frac{d\phi_\Delta}{d\lambda}.
\label{eq:1-1}\end{align}
The final lines of both \eqref{eq:11} and \eqref{eq:1-1} follow from standard  hypergeometric identities. 

\subsubsection{Notation and conventions for integrals over principal-series representations}
\label{subsubsec:Notation}

We noted above that the group-averaging inner product of 2-particle states is generally oscillatory when both particles lie in the principal series.  As a result, such integrals generally converge only in a distributional sense.  This suggests that it will be mathematically useful to consider direct integrals of principal-series representations, even if the physical relevance of such a construction is not yet clear.\footnote{Indeed, a quantum field theory that actually involved a continuum of scalar fields would have a divergent density of states and would have ill-defined thermodynamics.}

Let us therefore pursue a brief aside to introduce notation relevant to integrating over $s$ so that we  define a single principal-series Hilbert space
\begin{equation}
{\cal H}_P := \int_{{\mathbb R}^+}^\oplus ds \, P_s.
    \end{equation}
Note that we may still discuss simultaneous eigenstates of angular momentum and the quadratic Casimir in  ${\cal H}_P$ (i.e., states with definite values of both $m$ and $s$), but that such states are only delta-function normalizable in $s$.    

We denote these delta-function normalizable states by $|m,s\rangle$ to distinguish them from the unit-normalized states $|m\rangle_s$ introduced above.  The precise normalization of $|m,s\rangle$ will be fixed below.  Before doing so, however, we comment that we will more generally use the convention that states are delta-function normalizable in any continuous representation labels ($s$-labels) written inside the bra or ket, though there are no inner-product delta-functions associated with labels that appear as subscripts on the bra or ket.  For example, the state
\begin{equation}
|m_1, m_2\rangle_{s_1,s_2} : =
\ket{m_1}_{s_1}\otimes\ket{m_2}_{s_2} \in P_{s_1}\otimes P_{s_2}
        \label{eq:PD-fixed-PP-state}
\end{equation}
has unit norm, while the states
\begin{align}
\ket{m,s;n}_{;s_0}:=\ket{m,s}\otimes\ket{n}_{s_0} \in {\cal H}_P \otimes P_{s_0}
        \label{eq:PD-mixed-PP-state}
\end{align}
satisfy
\begin{align}
        {}_{;s_0}\braket{m',s';n'}{m,s;n}_{;s_0}=\frac{K_P(s)}{\pi^3}\delta(s'-s)\delta_{m'm}\delta_{n'n}.
        \label{eq:PD-mixed-PP-inner-product}
\end{align}
Here we have introduced the factor 
\begin{equation}
        \label{eq:PD-KP}
\frac{K_P(s)}{\pi^3}:=    \frac{4}{\pi s \tanh \pi s}
\end{equation} to match the conventions for the states $|m,r\rangle^x$ (introduced in \eqref{eq:PD-PP-CG} below) in appendix E of \cite{KolchmeyerLiu2024} (which were there called $|n\rangle_s^x$ and whose normalizations follow from the normalizations of their Clebsch-Gordan symbols stated in their (E.21) and (E.22)), as we will later make significant use of that appendix.
To reduce clutter, we will often write \eqref{eq:PD-mixed-PP-inner-product} in the form
\begin{align}
        \braket{m',s';n'}{m,s;n}_{;s_0}=\frac{K_P(s)}{\pi^3}\delta(s'-s)\delta_{m'm}\delta_{n'n},
        \label{eq:PD-mixed-PP-inner-product2}
\end{align}
suppressing the subscript $s_0$ on the bra-state (since the inner product can be non-trivial only when it is associated with the same representation as the ket-state whose subscript is kept explicit).

In contrast,
a fully delta-normalized principal-principal two-particle state is  written
\begin{align}
        \ket{m_1,s_1;m_2,s_2}:=\ket{m_1,s_1}\otimes\ket{m_2,s_2} \in {\cal H}_P \otimes {\cal H}_P,
        \label{eq:PD-full-PP-state}
\end{align}
and has inner product
\begin{align}
        \braket{m_3,s_3;m_4,s_4}{m_1,s_1;m_2,s_2}=\frac{K_P(s_1)}{\pi^3}\frac{K_P(s_2)}{\pi^3}\delta_{m_3,m_1}\delta(s_3-s_1)\delta_{m_4,m_2}\delta(s_4-s_2).
        \label{eq:PD-full-PP-inner-product}
\end{align}

It is worth mentioning that, in the discrete and complementary series, we will consider only two-particle states in some tensor product of UIRs (e.g. $D_{\Delta_1}^{\epsilon_1} \otimes D_{\Delta_2}^{\epsilon_2}$),  without integrating over $\Delta_1,\Delta_2$.   
Two-particle states in such representations are again written $|m_1,m_2\rangle_{\Delta_1,\Delta_2}$, where as usual in the discrete series the sign of $m_i$ determines the corresponding sign $\epsilon_i$ so that additional $\epsilon_i$ labels are unnecessary.

\subsection{The two-particle sector}
\label{subsec:PD-two-particle-base}

It will be convenient to first prove positivity of group averaging on the two-particle Hilbert space ${\cal H}_P \otimes {\cal H}_P.$  Due to the associated integrals over the parameter $s$, this Hilbert space has an infinite number of states in any small window associated with natural notions of 2-particle energy.  As a result,  ${\cal H}_P \otimes {\cal H}_P$ does not form the 2-particle sector of a physically interesting quantum field theory.  However, as discussed previously, for 2-particle states in QFTs with a finite density of states (so that the two-particle Hilbert space involves only direct sums over the $s_1,s_2$ labels of the two particles and not direct integrals), the group-averaging norms  are divergent; the space of group-averaging normalizable states in such theories must instead have $N \ge 3$ particles.  This is related to the fact that 2-particle states with both particles in the principal series always have exactly $\Sigma =1$ and so cannot satisfy $\Sigma > 1$. Nevertheless, the study of
${\cal H}_P \otimes {\cal H}_P$ will be a useful mathematical crutch that will make the discussion of states with $N\ge 3$ particles particularly clear and concrete.\footnote{The space ${\cal H}_P \otimes {\cal H}_P$ was also implicitly used to discuss 2-particle states in \cite{KolchmeyerLiu2024}.}

For states $|\phi\rangle, |\psi\rangle \in {\cal H}_P \otimes {\cal H}_P$, the group-averaging inner product takes the form \eqref{eq:PD-GA-definition} as usual.  However, we again emphasize that the relevant representation $U(g)$ acts diagonally on ${\cal H}_P \otimes {\cal H}_P$.  As a result, if we choose (distributional) product states $|\phi\rangle = |m_1,s_1;m_2,s_2\rangle$ and $|\psi\rangle = |n_1,s_1{}';n_2,s_2{}'\rangle$, we find
\begin{align}
\label{eq:PPGAIP}
\langle m_1,s_1;m_2,s_2|n_1,s_1{}';n_2,s_2{}'\rangle^{\rm GA} &= \int_G dg\, \langle m_1,s_1|U(g)|n_1,s_1{}'\rangle \, \langle m_2,s_2|U(g)| n_2,s_2{}'\rangle \nonumber \\
&= \int_G dg\,  \langle m_1|U(g)| n_1\rangle_{s_1} \, \langle m_2|U(g)| n_2\rangle_{s_2} \, \nonumber \\ &\quad \times \frac{K_P(s_1)}{\pi^3}\frac{K_P(s_2)}{\pi^3} \delta(s_1-s_1{}') \delta(s_2-s_2{}'),
\end{align}
where the Dirac delta-functions arise from the fact that $U(g)$ preserves the representation label $s$.

Now, the right-hand-side of \eqref{eq:PPGAIP} is closely related to a property of principal-series representations that is an analogue of Schur orthogonality for compact groups.   In particular, Bargmann \cite{Bargmann1947} and Harish-Chandra \cite{HarishChandra1952} showed that principal-series representations satisfy the relation
\begin{align}
        \int_G dg\,\bra{m_1}U(g)\ket{n_1}_{s_1} \, \bra{n_2}U(g^{-1})\ket{m_2}_{s_2}=K_P(s_1)\delta(s_1-s_2)\delta_{m_1,m_2}\delta_{n_1,n_2},
        \label{eq:PD-Kitaev-P}
\end{align} where we have used the notation and normalizations of \cite{Kitaev2018}.  The attentive reader will note that expression \eqref{eq:PPGAIP} involves $U(g)$ in the matrix element for $s_2$ while expression \eqref{eq:PD-Kitaev-P} involves $U(g^{-1}) = U^\dagger(g)$.  However, one may readily
translate between the two using the fact that principal-series representations are real (or, in the mathematical terminology, self-dual, or self-contragredient\footnote{In the language of \cite{Sun:2021thf}, one may equivalently say that the principal-series representation of weight $\Delta$ is isomorphic to the shadow representation of weight $\bar \Delta$.}), so that the complex-conjugate representation defined by $P_s$  is isomorphic to the original $P_s$.  In particular, using the conventions of appendix E of \cite{KolchmeyerLiu2024} for the phases of the
states $\ket{m}_{s}$ with $s >0$, we have
\begin{equation}
\label{eq:Pshadow}
\bra{n}U(g^{-1})\ket{m}_{s}
= (-1)^{n+m}e^{i(\phi_{n,s} - \phi_{m,s})}
\bra{-m}U(g)\ket{-n}_{s},
\end{equation}
with
\begin{equation}
  e^{i\phi_{n,s}} = i(-1)^n \frac{\Gamma(\Delta) \, \Gamma(n+1-\Delta)}{\Gamma(1-\Delta) \,  \Gamma(n + \Delta)},  
\end{equation}
where this is a pure phase in the principal series since $\Gamma(z^*) = \left[\Gamma(z)\right]^*$.
With these conventions we may thus use \eqref{eq:PD-Kitaev-P} and \eqref{eq:Pshadow} to compute the desired group-averaging inner product \eqref{eq:PPGAIP}:
\begin{align}
\label{eq:PPGAIP2}
\langle m_1,s_1;m_2,s_2|n_1,s_1{}';n_2,s_2{}'\rangle^{\rm GA} &= \int_G dg\, \langle m_1|U(g)| n_1\rangle_{s_1} \, \bra{-n_2}U(g^{-1})\ket{-m_2}_{s_2}
\, \nonumber\\ &\times (-1)^{n_2+m_2}e^{i(\phi_{m_2,s_2}-\phi_{n_2,s_2})}\frac{K_P(s_1)}{\pi^3}\frac{K_P(s_2)}{\pi^3}\delta(s_1-s_1{}') \delta(s_2-s_2{}') \nonumber
 \\&=\frac{K_P^3(s_1)}{\pi^6}
\delta(s_1-s_2) \delta(s_1-s_1{}') \delta(s_2-s_2{}') \nonumber \\&
\times (-1)^{n_2+m_2}e^{i(\phi_{m_2,s_2}-\phi_{n_2,s_2})}\delta_{n_1+n_2=0}\, \delta_{m_1+m_2=0}
\, ,
\end{align}
where $\delta_{a+b=0}$ denotes the Kronecker delta $\delta_{a+b,0}$ that sets $a+b=0$.  This form displays the expected result that group averaging should annihilate any state with definite total angular momentum $m_1+m_2\neq 0$.  For two-particle states in some ${\cal H}_P \otimes P_s$, we analogously find
\begin{align}
\label{eq:PPGAIPfixs}
\langle m_1,s_1;m_2|n_1,s_1{}';n_2\rangle_{;s_2}^{\rm GA} &=\frac{K_P^2(s_1)}{\pi^3}
\delta(s_1-s_2) \delta(s_1-s_1{}') \nonumber \\ &\times \delta_{n_1+n_2=0}\, \delta_{m_1+m_2=0}
\, (-1)^{n_2+m_2} e^{i(\phi_{m_2,s_2}-\phi_{n_2,s_2})}.
\end{align}

The key point is now that the right-hand-sides of \eqref{eq:PPGAIP2} and \eqref{eq:PPGAIPfixs} factorize into  products of functions that depend separately on $n_1,n_2$ and on $m_1,m_2$.  This means that the associated Gram matrices have rank one.  As a result, after performing the quotient by null vectors, the Hilbert space defined by \eqref{eq:PPGAIP2} or \eqref{eq:PPGAIPfixs} has at most one state for each pair $s_1,s_2$. The fact that the inner products \eqref{eq:PPGAIP2} and \eqref{eq:PPGAIPfixs} are clearly diagonal in $s_1,s_2$ then means that these computations decouple, and that we need only test positivity for each such state separately.  Choosing representatives of the associated equivalence classes for which   $m_1=n_1,m_2=n_2$, the Gamma-functions in \eqref{eq:PPGAIP2}  cancel among themselves and  the right-hand-side becomes manifestly non-negative as desired.  We have thus established positivity of group averaging on the linear span of the states $|m_1,s_1;m_2,s_2\rangle$ in  ${\cal H}_P \otimes {\cal H}_P$ and the states $|m_1,s_1;m_2\rangle_{;s}$ in ${\cal H}_P\otimes P_s$.

Before leaving the principal series, we note for future reference that,
while one may choose to simply take \eqref{eq:PD-Kitaev-P} as a result from the mathematics literature, this relation in fact follows in a straightforward way from considering the SO$_0(2,1)$ Casimir ${\bf C} = J^2 - B_1^2 - B_2^2$
as a self-adjoint operator on the Hilbert space $L^2(G; dg)$ of functions on the group $G$ that are square-integrable with respect to the Haar measure $dg$. The action of ${\bf C}$ is defined by taking ${\bf C}f(g):= f({\bf C}g) =f(g{\bf C})$, where the final equality uses the fact that ${\bf C}$ commutes with all $g\in G$, and where the inclusion of infinitesimal group generators in the argument of $f$ is interpreted as an appropriate derivative; i.e.,  ${\bf C}$ acts as a covariant Laplacian $\nabla^2$ on $G$.  Since ${\bf C}$ acts as a $c$-number on any irreducible representation, it is thus easy to see that
each matrix element ${}_s\bra{j}U(g)\ket{m}_s$ defines a formal eigenstate of ${\bf C}$ with eigenvalue $-\left( s^2 + \frac{1}{4}\right)$.  Straightforward asymptotic analyses (see e.g. \cite{Marolf_2009}) then show that such functions are
delta-function normalizable and  that they are delta-function orthogonal for distinct values of $s$.  The Kronecker deltas on the right-hand-side of \eqref{eq:PD-Kitaev-P} are then obtained by noting that such states also diagonalize the operators $J_{L,R}$ defined by $J_L f(g) = f(Jg)$ and $J_Rf(g) = f(gJ)$ and that $J_L, J_R,{\bf C}$  are mutually commuting operators on $L^2(G)$.  Finally, the normalization factors \eqref{eq:PD-KP} can be determined by analyzing the coefficients of the above-mentioned asymptotic expansion.

We will also need the analogous results for the so-called discrete-series representations.  Since two-particle states for that case have $\Sigma >1$, group averaging should converge absolutely for such basis states.
Indeed, a useful observation is that matrix elements of the discrete series define normalizable states in $L^2(G)$ which are thus associated with the discrete spectrum of the Casimir operator ${\bf C}$. Orthogonality of eigenstates with distinct eigenvalues then requires any two-particle group-averaging inner product  to vanish when the states involve a mixture of principal- and discrete-series representations; i.e., any two-particle state involving one principal-series particle and one discrete-series particle is a null state under the group-averaging inner product.\footnote{\label{foot:CD}This argument in fact generalizes to 2-particle states with one particle in the complementary series and one particle in the discrete series.  Group averaging again converges absolutely for such 2-particle states, and the matrix elements of $U(g)$ decay rapidly enough to allow the integrations by parts required to show that the fact that the two representations necessarily have distinct eigenvalues for the Casimir implies the vanishing of the group-averaged inner product.  Indeed, as we will discuss in section \ref{subsec:3c}, group averaging also converges absolutely for rotation-invariant odd-parity 2-particle states with either one complementary-series particle and one principal-series particle, or with two complementary-series particles of weights $\Delta_1,\Delta_2$.  The above argument thus again shows the state to be null in either such a rotation-invariant parity-odd principal-complementary case, or in such a rotation-invariant parity-odd complementary-complementary case with $\Delta_1 \neq \Delta_2$.  That the remaining rotation-invariant parity-odd complementary-complementary case with $\Delta_1 =\Delta_2$ also yields a null state then follows by taking the limit $\Delta_1 \rightarrow \Delta_2$ and using the continuity implied by absolute convergence.}

From the corresponding formula for the universal covering group in \cite{Kitaev2018}, it follows that the analogue of \eqref{eq:PD-Kitaev-P} for the discrete series  is
\begin{align}
        \int_G dg\,\langle m_1|U(g)|n_1\rangle{}_{\Delta_1, \epsilon_1}\ & \langle n_2|U(g^{-1})|m_2\rangle{}_{\Delta_2, \epsilon_2} \nonumber \\ &=K_D(\Delta_1)\delta_{\Delta_1, \Delta_2}\delta_{m_1,m_2}\delta_{n_1,n_2},        \label{eq:PD-Kitaev-D}
\end{align}
where
\begin{align}
        K_D(\Delta)=\frac{8\pi^2}{2\Delta-1}
        \label{eq:PD-KD}.
\end{align}
Since
the discrete-series representation $D^+_\Delta$ contains only states of positive angular momentum while
$D^-_\Delta$ contains only negative angular momentum states, the Kronecker deltas involving angular momenta also enforce $\epsilon_1=\epsilon_2$ when the right-hand-side is non-zero.   Note that \eqref{eq:PD-KD} is positive since $\Delta \ge 1$.

Again using the conventions of \cite{KolchmeyerLiu2024} appendix E, one finds from \cite{Sun:2021thf} that the discrete-series shadow transform (the analogue of \eqref{eq:Pshadow}) takes the trivial-looking form\footnote{Discrete-series representations are not self-contragredient but, instead, the distinct representations $D_\Delta^\pm$ form a contragredient pair.}
\begin{equation}
\langle n| U(g^{-1}) |m\rangle_{\Delta, \pm}
= \langle -m| U(g) |{-n}\rangle_{\Delta, \mp}.
\label{eq:3Pip}
\end{equation}
Combining these two results yields
the discrete-series analogue of \eqref{eq:PPGAIP}:
\begin{align}
\label{eq:gen3Pip}
\langle m_1,m_2|n_1,n_2\rangle^{\rm GA}_{\Delta_1,\epsilon_1, \Delta_2, \epsilon_2} &= \int_G dg  \langle m_1|U(g)| n_1\rangle_{\Delta_1,\epsilon_1} \, \bra{-n_2}U(g^{-1})\ket{-m_2}_{\Delta_2,-\epsilon_2}
\nonumber \\&=
K_D(\Delta_1) \delta_{\Delta_1,\Delta_2}\,
\delta_{n_1+n_2=0}\, \delta_{m_1+m_2=0},
\end{align}
where, since the $D^\pm_\Delta$ each contain only angular momenta with a definite sign, the right-hand-side can be non-zero only for $\epsilon_1 = - \epsilon_2$.  Just as in the principal series, noting that the right-hand-side factorizes into a part that is independent of $n_1,n_2$ and a part that is independent of $m_1,m_2$, and also noting that the right-hand-side of \eqref{eq:gen3Pip} is positive,
establishes positivity of group averaging for states with two particles in the discrete series.

\subsection{\texorpdfstring{$N\ge 3$}{N ≥ 3} particles and Clebsch-Gordan reduction}

\label{subsec:PD-CG-decomposition}

We now review how the Clebsch-Gordan decomposition can be used to write states with $N$ principal- and discrete-series particles in terms of states with smaller numbers of ``effective particles'' that again lie in the principal and discrete series.  This will then immediately reduce the desired positivity result for group averaging of states with $N>2$ such particles in the standard Fock spaces to the $N=2$ results of section \ref{subsec:PD-two-particle-base}.

The essential point is that it has been known for some time \cite{Pukanszky1961,Repka1976,Repka1978,Molchanov1980} that decomposing a tensor product of two principal- and discrete-series representations of SO$_0(2,1)$ into irreducibles yields a sum/integral over discrete- and principal-series representations.  In particular, if the two representations are taken from the principal and discrete series, then the decomposition involves {\it only} representations that again lie in the principal and discrete series.  Furthermore, the principal-series representations can arise with multiplicity at most 2, while the discrete-series representations arise with multiplicity at most 1. In particular, in the notation of the previous section, for all $s_1,s_2 >0$ we have
 \begin{equation}
 \label{eq:Pprod}
 P_{s_1}\otimes P_{s_2} \cong_G {\cal H}_P \oplus {\cal H}_P \bigoplus_{\epsilon = \pm,\, \Delta \in {\mathbb Z}_{\ge 1}} D^\epsilon_\Delta,
 \end{equation}
 where $\cong_G$ denotes unitary equivalence of $G$-representations. 
 Furthermore, the product representations $P_s \otimes D^\pm_\Delta$ and $D^{\epsilon_1}_{\Delta_1} \otimes D^{\epsilon_2}_{\Delta_2}$ are also contained in the right-hand-side of  \eqref{eq:Pprod}, though they contain at most one copy of ${\cal H}_P$ and the product of two discrete series of the same sign contains no principal-series representations at all.  
 
 One may say that \eqref{eq:Pprod} and its analogues involving discrete-series particles replace two `physical' particles by a sum over states of a single `effective particle' which comes equipped with an `internal' representation label that can take values in both the principal and discrete series, as well as an additional degeneracy label that distinguishes the two ${\cal H}_P$ terms on the right-hand-side of \eqref{eq:Pprod}.  Concretely, the equivalence in \eqref{eq:Pprod} means that for $m=m_1+m_2$ we may write  
\begin{align}       \ket{m_1}_{s_1}\ket{m_2}_{s_2}&=\sum_{x\in\{S,A\}}\int_0^\infty dr\,\rho_P(r)\left(\widetilde C^{P,x}_{PP}\right)^{r,m\,*}_{s_1,m_1|s_2,m_2}\ket{m,r}^{x}\nonumber\\
        &\quad+\sum_{\Delta=1}^{|m|}\left(\widetilde C^{D,\operatorname{sgn}(m)}_{PP}\right)^{\Delta,|m|\,*}_{s_1,m_1|s_2,m_2}\ket{m}_{\Delta,\operatorname{sgn}(m)},\qquad m\neq 0,
        \label{eq:PD-PP-CG}
\end{align}
where there is no discrete-series sum for $m=0$.  In \eqref{eq:PD-PP-CG} we have used the notation and conventions of  \cite{KolchmeyerLiu2024} which involve 
a complex conjugation ($*$) due to defining the unadorned Clebsch-Gordan coefficients to give the embedding of $P_{s_3}$ in $P_{s_1}\otimes P_{s_2}$, 
the association of the degeneracy index $x$ (which labels the two copies of ${\cal H}_P$ in \eqref{eq:Pprod}) with certain symmetry ($S$) and antisymmetry ($A$) properties,\footnote{In particular, the above Clebsch-Gordan symbols are respectively symmetric and antisymmetric under $ (s_1,m_1) \leftrightarrow (s_2,m_2)$.} and
the use of the Plancherel measure
\begin{align}
        \rho_P(r)=\frac{\pi r\tanh(\pi r)}{4} = \frac{\pi^3}{K_P(r)}>0
        \label{eq:PD-rhoP}.
\end{align}

We will use the term Clebsch-Gordan coefficients below to refer to the symbols
\begin{equation*}
\left(\widetilde C^{P,x}_{PP}\right)^{r,m}_{s_1,m_1|s_2,m_2} \quad \text{and} \quad \left(\widetilde C^{D,\operatorname{sgn}(m)}_{PP}\right)^{\Delta,|m|}_{s_1,m_1|s_2,m_2}
\end{equation*}
defined by \eqref{eq:PD-PP-CG} above. 
Importantly, for all $s_1,s_2,s_3$ the $S$-type Clebsch-Gordan symbols are parity-even while the $A$-type Clebsch-Gordan symbols are parity-odd.  We may then also write corresponding formulae for principal-discrete and discrete-discrete two-particle states.

Before proceeding to use the above result, it is worth remarking that
much of it again follows from an understanding of the spectrum of the Casimir operator ${\bf C}$ on $L^2(G)$.  The key point is that
the matrix elements of principal- and discrete-series representations give a complete set of ${\bf C}$-eigenstates.\footnote{This follows from a straightforward calculation.  As a manifold with metric, the group $G={\rm SO}_0(2,1)$ is a ${\mathbb Z}_2$ quotient of AdS$_3$.  Furthermore, ${\bf C}$ acts as the Laplacian, for which the resulting mode functions are well-studied and can be identified with the matrix elements in the principal and discrete series discussed in section \ref{sec:dS2UIRs} (see \eqref{eq:SO21explicit} and footnote \ref{foot:hdisc}).}  After all, matrix elements in any two-particle representation again define smooth functions of $g$ and, for principal- and discrete-series particles, the above-quoted decay rates imply that the two-particle matrix elements are square-integrable\footnote{This is the same analysis which implies that their squares, which are essentially 4-particle matrix elements, define functions whose integrals over $G$ converge absolutely.} and so lie in $L^2(G)$.  They may thus be written as superpositions of Casimir-eigenstates.
This shows that the decomposition of such tensor-products into irreducibles involves only principal- and discrete-series representations, though more work is required to determine the multiplicities with which each representation appears.

In any case, positivity of group averaging with $N \ge 3$ particles taken from any combination of  principal- or discrete-series representations now follows immediately from \eqref{eq:Pprod} and its analogues involving discrete-series particles. The relevant point is that such a multi-particle state lives in the tensor product of its one-particle representations.  We may thus use a single application of \eqref{eq:Pprod} and its analogues involving discrete representations to reduce the number of particles by one, replacing two physical particles by a single effective particle, finding that any principal-series representations appearing as part of the internal states of the effective particle in fact come as part of an ${\cal H}_P$.   For any $N\ge 3$ we may thus reduce the states to a sum of terms of the form studied in section \ref{subsec:PD-two-particle-base} and for which positivity has already been established.  Here we also use the fact that cross-terms in the group-averaging inner product, associated with distinct terms in the final sum over two-particle states, will always vanish since the action of $U(g)$ preserves any representation or degeneracy labels.

In particular, for three principal-series particles we find   \begin{align}
\langle m_1,m_2,m_3|& m'_1,m'_2,m'_3\rangle_{s_1,s_2,s_3}^{\rm GA} \nonumber \\ ={}&
\pi^3 \sum_{x\in\{S,A\}}    
\, \left(\widetilde C^{P,x}_{PP}\right)^{s_3,-m_3}_{s_1,m_1|s_2,m_2}\left(\widetilde C^{P,x}_{PP}\right)^{s_3,-m'_3\,*}_{s_1,m'_1|s_2,m'_2} (-1)^{m_3+m'_3} e^{i(\phi_{m_3,s_3}-\phi_{m'_3,s_3})}.
\label{eq:PPPip}
\end{align}
The positivity of this expression is then made manifest by considering states
\begin{equation*}
|\psi\rangle = \sum_{m_1,m_2,m_3} \psi_{m_1,m_2,m_3} \, |m_1,m_2,m_3\rangle_{s_1,s_2,s_3}
\end{equation*}
and using \eqref{eq:PPPip} to write
\begin{align}
\langle \psi |\psi \rangle^{\rm GA} &=
\pi^3 \Bigl(\left|\sum_{m_i} (-1)^{m_3}e^{-i\phi_{m_3,s_3}}\psi_{m_1,m_2,m_3} \left(\widetilde C^{P,S}_{PP}\right)^{s_3,-m_3\,*}_{s_1,m_1|s_2,m_2} \right|^2 \nonumber \\ &+ \left|\sum_{m_i} (-1)^{m_3}e^{-i\phi_{m_3,s_3}}\psi_{m_1,m_2,m_3} \left(\widetilde C^{P,A}_{PP}\right)^{s_3,-m_3\,*}_{s_1,m_1|s_2,m_2} \right|^2 \Bigr),
\label{eq:PPPip2}
\end{align}
where $\sum_{m_i}$ denotes a sum over all $m_1,m_2,m_3.$  Due to the parity properties of the Clebsch-Gordan coefficient, the $A$ term in \eqref{eq:PPPip2} vanishes for all states $|\psi\rangle$ that are even under the parity operator ${\cal P}$ (defined by ${\cal P}|m\rangle_s = |{-m}\rangle_s$) while the $S$ term vanishes for all parity-odd states. Corresponding positive-definite expressions also hold when some of the particles in our 3-particle state lie in the discrete series.

Though this concludes the main argument, there are three technical points that are worthy of further brief remarks. The first is simply a reminder that, for some states, the group-averaging inner product can in fact be larger than the original Fock-space inner product.  Indeed, the ratio can be unbounded from above\footnote{Interestingly, however, exact results of \cite{Higuchi_1991_2} and numerical results in \cite{Marolf_2009} suggest that this effect is not associated with growth of the group-averaging norm at large parameters for states defined by standard global-dS modes.  We expect that it instead arises from the fact that there are infinitely many orthogonal Fock basis states that group averaging maps to the same state in ${\cal H}^{\rm GA}$, so normalizable Fock states built from  superpositions of such states can have large norms in ${\cal H}^{\rm GA}$.} so that, even in the principal and discrete series, group averaging converges absolutely only on some dense domain within the space of 3-particle states.  The above argument uses the fact that the decay bounds of \cite{Marolf_2009} show that states defined by standard global-dS mode functions lie in the relevant domain.  The same is thus true for finite linear combinations.  Positive semi-definiteness of group averaging on this space thus leads to a group-averaged Hilbert space ${\cal H}^{\rm GA}$ via the usual quotient by null states and Cauchy completion.

The second comment is that, while a physicist would usually say that the computation of e.g. \eqref{eq:PPPip2} 
used the fact that group-averaging 2-particle states in the principal series yields two delta-functions (shown on the right-hand-side of \eqref{eq:PPGAIPfixs}) to evaluate the integrals in \eqref{eq:PD-PP-CG}, and that it thus group-averages the effective two-particle states before performing the integrals to reconstruct the original 3-particle states, this exchange of orders of integration is actually immediate and requires no justification.  In particular, recall that the delta-functions in \eqref{eq:PPGAIPfixs} have meaning only when they are integrated against suitable test functions; in fact, \eqref{eq:PPGAIPfixs} is shorthand for the statement that, for smooth compactly-supported $f_1, f_2$, we have
\begin{align}
\label{eq:PPGAIPfixs2}
\int ds_1 ds'_1 &\, f^*_1(s_1) f_2(s'_1)\langle m_1,s_1;m_2|n_1,s_1{}';n_2\rangle_{;s_2}^{\rm GA}  \nonumber \\ &= (-1)^{m_2+n_2}
e^{i(\phi_{m_2,s_2}-\phi_{n_2,s_2})} \frac{K^2_P(s_2)}{\pi^3}f^*_1(s_2) f_2(s_2) 
\delta_{n_1+n_2=0}\, \delta_{m_1+m_2=0}
\, ,
\end{align}
when the $s_1,s_1'$ integrals on the left-hand side are to be performed before computing the group-averaging inner product.  
In the desired computation we may thus obtain \eqref{eq:PPPip} by directly applying \eqref{eq:PPGAIPfixs2} with $f_1,f_2$ being appropriate Clebsch-Gordan symbols.

On the other hand, the third point is that
our positivity argument for states with $N \ge 4$ {\it does} involve a meaningful exchange of certain orders of integration (and summation).  In particular, the  argument proceeds by writing the $N\ge 4$ particle states as integrals and sums over states with smaller numbers of effective particles, evaluating the group-averaging inner products of those smaller-$N$ states, and then performing the integrals and sums over the parameters of the effective particles.  This involves exchanging the order of the integration over the group $G$ with that of the integration and sum over the space ${\cal S}$ of parameters of the effective particles.  

Such an exchange is justified when the full integral over $G\times {\cal S}$ converges absolutely.  That this is the case here follows from the observations that:  i) as noted previously, group averaging of states with 3 or more principal- and discrete-series particles converges absolutely and, as shown by item (2) of theorem 2.1 in \cite{Shalom2000Rigidity} (with $p=2$), does so due to a bound that is uniform on ${\cal S}$, ii) sums over the discrete-series weight $\Delta$ always truncate because the representation $D^+_\Delta$ ($D^-_\Delta$) contains only states with angular momentum at least $\Delta$ (at most $-\Delta$), iii) sums over angular momentum labels $m$ are similarly controlled by angular momentum conservation, and iv) our Clebsch-Gordan symbols decay faster than any polynomial in the principal-series representation label $s$ due to the corresponding behavior of Fourier transforms of smooth functions.\footnote{Recall the Clebsch-Gordan coefficients can be studied by performing an integral over $G$ of a function built by multiplying together matrix elements of $U(g)$ in the three relevant representations.}  These comments complete the proof of positivity of group averaging for states with $N\ge 3$ principal- and discrete-series particles in dS$_2$.

\section{Complementary-series positivity in \texorpdfstring{SO$_0(2,1)$}{SO\_0(2,1)}}
\label{sec:complementary-positivity}

This section addresses positivity of group averaging for states involving complementary-series bosonic particles in dS$_2$.
In this low-dimensional context, it is straightforward to discuss the complementary series by building on our arguments and results from sections \ref{subsec:PD-two-particle-base} and \ref{subsec:PD-CG-decomposition}. We first outline the argument for  3-particle states in section \ref{subsec:3c}.  We then provide final details in section \ref{subsec:Aproof} before proceeding to states with $N\ge 4$ particles in section \ref{subsec:4d2}.  For any $N$ we will establish positivity for the case 
\begin{equation}
\label{eq:dS2conv}
{\rm Re}  \,  \sum_{i=1}^N  \Delta_i >1
\end{equation} in which the naive absolute convergence bound of \cite{Marolf_2009} is satisfied.  However, we will also find cases with smaller weights where group averaging converges absolutely and, in some of those cases, negative group-averaging norms will in fact arise.

\subsection{Three-particle cases}
\label{subsec:3c}

We begin with the 3-particle states.  There are three cases, depending on whether any of these particles lie in the principal or discrete series, or whether all 3 lie in the complementary series.

If there are indeed any principal-series particles, then we may proceed as in section \ref{subsec:PD-two-particle-base}. Suppose for example that particle 3 is in the principal series, so that $\Delta_3 = \frac{1}{2}+is$ for $s\in{\mathbb R}.$ Then since $\sum_{i=1}^3 \, {\rm Re} \, \Delta_i > 1$, we must have 
\begin{equation}
\label{eq:D>half}
{\rm Re} \, (\Delta_1+\Delta_2) > 1/2.
\end{equation}
But it has long been known \cite{Repka1976,Repka1978} that, for any representations with weights $\Delta_1, \Delta_2$ satisfying \eqref{eq:D>half}, the tensor product of the associated two representations can be decomposed into a sum/integral of principal- and discrete-series representations in direct analogy with \eqref{eq:Pprod}.  In fact, this again follows by the argument described in the second paragraph below \eqref{eq:PD-rhoP}.  We may thus replace particles 1 and 2 by a sum/integral of principal- and discrete-series particles and then follow the argument of section \ref{subsec:PD-two-particle-base} verbatim.

Let us now instead suppose that at least one of the particles lies in the discrete series.   In this case we again use the fact that the tensor product of the other two representations admits a known sum/integral decomposition into UIRs (whose details can again be found in \cite{Repka1976,Repka1978}).  Furthermore, as described in footnote \ref{foot:CD}, in taking the $L^2(G)$ inner product with any discrete-series matrix element of $U(g)$, we can obtain contributions only from terms in this sum involving the contragredient discrete-series representation.   But just as in section \ref{subsec:PD-CG-decomposition}, the coefficient of this term will involve the norm-squared of a Clebsch-Gordan symbol associated with the above decomposition into UIRs.  It will thus again be manifestly positive in precisely the manner described in section \ref{subsec:PD-CG-decomposition} (see \eqref{eq:PPPip2}).

This leaves us only with the case where all three particles lie in complementary-series representations.  
Since $\Sigma > 1$, one might again note that any pair of particles still satisfies $\Delta_1 +\Delta_2 = \Sigma - \Delta_3 > \frac{1}{2}$, so the tensor product of those representations again decomposes into tempered representations. However, since an effective principal-series particle has ${\rm Re} \, \Delta_{\rm eff} = \frac{1}{2}$, the fact that $\Delta_3 < \frac{1}{2}$ means that the total weight of our effective two-particle state would then be strictly less than $1$.  Thus group-averaging would not generally converge, leaving us unable to complete the desired argument.

Nevertheless, in this case we may proceed by working directly with the 3-particle complementary-series states and using the results
\eqref{eq:PPPip} and \eqref{eq:PPPip2} from the principal series and analytically continuing in $\Delta_1,\Delta_2,\Delta_3$.  To begin, let us recall that explicit results for the complementary-series Clebsch-Gordan coefficients were given in appendix E of \cite{KolchmeyerLiu2024} in terms of $\Gamma$-functions.  While these expressions are rather cumbersome, they make manifest that any given Clebsch-Gordan coefficient is a meromorphic function of $\Delta_1,\Delta_2,\Delta_3$.

Now, the reader may recall that \eqref{eq:PPPip} and \eqref{eq:PPPip2} above were not presented as explicitly analytic functions of $\Delta_1, \Delta_2, \Delta_3$ since the right-hand-side of \eqref{eq:PPPip} was written in terms of both Clebsch-Gordan symbols and their complex-conjugates.  The right-hand-sides of \eqref{eq:PPPip} and \eqref{eq:PPPip2} thus involve both $\Delta$ and $\Delta^*$.  However, those expressions were valid only in the principal series, where we have $\Delta = \frac{1}{2} + is$ with $s\in {\mathbb R}$.  Thus $\Delta^* = 1-\Delta$ everywhere in the domain of validity of \eqref{eq:PPPip} and \eqref{eq:PPPip2}.  

Making the replacement\footnote{\label{foot:CGchi}One may think of this replacement as being done either using the explicit formulae for Clebsch-Gordan coefficients given in e.g. \cite{KolchmeyerLiu2024} or, alternatively, by using the fact discussed above that Clebsch-Gordan coefficients are associated with integrals over SO$_0(2,1)$ of products of matrix elements of $U(g)$ to see that the behavior of Clebsch-Gordan coefficients under complex conjugation is inherited from the corresponding properties \eqref{eq:Pshadow} of the matrix elements. As we will discuss further in section \ref{subsec:PDgend} (see \eqref{eq:CGbetterconj}), the latter route allows one to express complex-conjugates of Clebsch-Gordan coefficients in terms of the unconjugated coefficients (which thus involve $\Delta$ instead of $\Delta^*$).  These are two distinct routes that lead to the same final result for the group-averaged inner product.}
 $\Delta^* \rightarrow 1-\Delta$ then allows us to write the right-hand-sides of \eqref{eq:PPPip} and \eqref{eq:PPPip2} in a form that is manifestly analytic in $\Delta,$ though it is no longer manifestly positive definite.  Instead, for each  $x\in\{S,A\}$ we find the product of a Clebsch-Gordan coefficient evaluated at $\Delta_1,\Delta_2,\Delta_3,$ with a 
Clebsch-Gordan-like coefficient evaluated at $1-\Delta_1,1-\Delta_2,1-\Delta_3$. 

With enough work, one can use the various formulae in \cite{KolchmeyerLiu2024} (and especially (E.23)) to show that the analytic continuation of the right-hand-side of \eqref{eq:PPPip2} to weights in the complementary-series (with $\Sigma = \Delta_1+\Delta_2+\Delta_3 >1$) gives a group-averaging inner product that remains positive-definite (and, in fact, to give an explicit formula for the result that we include below in footnote \ref{foot:hole}).  However, without introducing new notation that argument is rather cumbersome.  We will therefore proceed in the main text by arguing below that the comments above imply that for complementary-series $\Delta_1,\Delta_2,\Delta_3$  the right-hand-side of \eqref{eq:PPPip} remains a sum of two terms, each of which factorizes.\footnote{And which, by the discussion around \eqref{eq:hDef}, in fact gives, in the complementary series, the result of evaluating $\langle m_1,m_2,m_3|m'_1,m'_2,m'_3\rangle^{\rm GA}_{\Delta_1,\Delta_2,\Delta_3} \prod_{i=1}^3 \sqrt{\frac{h_{\Delta_i}(m'_i)}{h_{\Delta_i}(m_i)}}$.}  As a result, at any $\Delta_i$ the analytic continuation of the right-hand-side will define at most a two-dimensional group-averaged inner-product space.  The corresponding inner product will thus be positive definite if we can find two group-averaged-orthogonal states whose norms are both positive, which will in fact follow by using \eqref{eq:0002d}, \eqref{eq:11}, and \eqref{eq:1-1}.  

In particular, it is straightforward to show that the left-hand-side is also analytic in the $\Delta_i$ for $\Sigma > 1$.  To see this, note that the left-hand-side is the integral of a product of matrix elements of $U(g)$ and, as shown explicitly in \eqref{eq:SO21explicit}, each individual matrix element can be expressed as an  analytic function of $\Delta_1$, $\Delta_2$, or $\Delta_3$. We then recall that,  as discussed in section \ref{sec:dS2UIRs}, for $\Sigma >1$ group averaging always converges absolutely. Furthermore, it is straightforward to check that $\Delta$-derivatives of the matrix elements \eqref{eq:SO21explicit} retain the same exponential fall-offs as the undifferentiated matrix elements. This means that the group-averaging integral continues to converge absolutely when we replace any matrix element by its $\Delta$-derivative.  As a result, for $\Sigma >1 $ we may take $\Delta$-derivatives before performing the group-averaging integral and use analyticity of the integrand (and the fact that it thus satisfies the Cauchy-Riemann equations) to show that the integrated group-averaging inner product satisfies the Cauchy-Riemann equations as well.

It therefore follows that the space of group-averaged 3-particle states continues to be at most 2-dimensional even when all particles are in the complementary series.  We may thus show that the group-averaged Hilbert space of such states has a positive-definite inner product by simply finding two group-averaging-orthogonal states with positive group-averaging norms.  Furthermore, since we have already noted that the $x=A$ term on the right-hand-side of \eqref{eq:PPPip2} vanishes for even-parity states while the $x=S$ term vanishes for odd-parity states, it follows that even- and odd-parity states are always group-averaging orthogonal.  It thus suffices to find a single state of each parity with positive group-averaging norm.

In the even-parity sector it is useful to focus on the 3-particle state in which  all three angular momenta vanish. The group-averaging norm is manifestly positive since this norm is the integral of the product of 3 copies of \eqref{eq:0002d} which, as is clear from the final line of \eqref{eq:0002d}, are positive at all $\lambda$ for all real values of  $\Delta_i $.   For future reference, we also note that it is straightforward to check analytically that the integral has a simple pole at $\Delta_1+\Delta_2+\Delta_3=1$ with non-zero residue.  This makes it clear that the result cannot be extended to $\Delta_1+\Delta_2+\Delta_3 <1$.  In particular, analytic continuation to such a regime would yield a negative inner product.

It thus remains only to study odd-parity states.  A particularly simple example is given by the choice\footnote{Since we consider only bosons here, this is a non-trivial QFT state only when particles 2 and 3 are not identical.}
\begin{equation}
\label{eq:Astate3C}
|\psi\rangle_{\Delta_1,\Delta_2,\Delta_3} = |0,1,-1\rangle_{\Delta_1,\Delta_2,\Delta_3} -     |0,-1,1\rangle_{\Delta_1,\Delta_2,\Delta_3}.
\end{equation}
Three-particle positivity for $\Sigma > 1$ will then follow if we can show that the group-averaging norm of \eqref{eq:Astate3C} is positive for complementary-series values of $\Delta_1,\Delta_2,\Delta_3$.  Since this argument is slightly more involved than the simple manifest-positivity of the naive group-averaging integrand for the above even-parity state, we will defer the details to section \ref{subsec:Aproof}.

Interestingly, we will see in section \ref{subsec:Aproof} that (with the proper understanding) the bound \eqref{eq:dS2conv} is not in fact required for this positivity to hold.  Instead, group averaging in the odd-parity 3-particle sector will be positive for arbitrary choices of UIRs (see also the discussion at the end of this section).  This fact is associated with an improved domain of convergence for group averaging of rotation-invariant parity-odd (RIPO) states.  Indeed,  expanding the single-particle-state hypergeometric functions 
\eqref{eq:SO21explicit} in exponentials  at large $\lambda$, one finds that the coefficients of the leading exponentials $e^{-\Delta \lambda}$ and $e^{(\Delta-1)\lambda}$ are both parity-even, with $e^{-(\Delta+1)\lambda}$ being the first exponential whose coefficient is parity-odd.  As a result, at large boosts the generically-leading exponentials in the matrix element of $U(g)$ will then also have parity-even coefficients and, in particular, these coefficients must vanish for all RIPO states.  The convergence domain for RIPO states is thus shifted in $\Delta$ by $1$ so that group averaging converges absolutely for RIPO states whenever ${\rm  Re} \, \sum_{i=1}^N \, \Delta_i >0$; i.e., since ${\rm  Re} \, \Delta_i >0$ for all UIRs, group averaging always converges absolutely for RIPO states. For parity-odd states that are not invariant under rotations, we can of course obtain similar results by taking the group-averaging integral to integrate over rotations and then to act on the resulting RIPO state as in \eqref{eq:PD-GA-boost-reduced}. We will also show positivity for all $\Delta_i$ below.

This observation for parity-odd states then raises the question of whether there are also parity-even states in which the above slowly decaying exponential tails cancel so as to enjoy an enlarged domain of absolute group-averaging convergence.  Working out the coefficients by expanding the results \eqref{eq:SO21explicit}, one finds that this is indeed possible, and that it requires tuning only a single parameter.  Moreover, in computing an inner product $\langle \psi|\phi\rangle^{\rm GA}$, the relevant coefficients factorize into a linear function of the state $|\phi\rangle$ times an anti-linear function of the state $|\psi\rangle$.  There is thus a natural linear space of even-parity states with an enlarged domain of group-averaging absolute convergence that includes some open set with $\Sigma <  1$.

When $\Sigma = \Delta_1+\Delta_2+\Delta_3=1$, the condition to remove these exponential tails becomes the same as the condition that the $S$-channel contribution vanish in the group-averaging norm \eqref{eq:PPPip2}; i.e., at such $\Delta_i$  all even-parity states for which group averaging converges absolutely become null states in ${\cal H}^{\rm GA}$.
However, these conditions become distinct for $\Sigma <1$.  So in that case one can find even-parity states for which group averaging converges absolutely and gives non-trivial results.  

On the other hand, this absolute convergence means that we can again analytically continue 
\eqref{eq:PPPip} to the desired regime. After doing so, and recalling that it is in fact only quantities of the form \eqref{eq:hme} that can be analytically continued from the principal series to the complementary series, we find
\begin{align}
\langle \psi |\psi \rangle^{\rm GA} &=
\pi^3 {\cal N}_S(\Delta_1,\Delta_2,\Delta_3) \nonumber \\
&\times \left|\sum_{m_i} (-1)^{m_3}  e^{i\phi_{m_3,s_3}} \psi^*_{m_1,m_2,m_3} \sqrt{h_{\Delta_1}(m_1) h_{\Delta_2}(m_2) h_{\Delta_3}(m_3) }\left(\widetilde C^{P,S}_{PP}\right)^{\Delta_3,-m_3}_{\Delta_1,m_1|\Delta_2,m_2} \right|^2 ,
\label{eq:PPPip3} 
\end{align}
where ${\cal N}_S(\Delta_1,\Delta_2,\Delta_3)$
is a real normalization coefficient and we have abused the notation somewhat by defining\footnote{\label{foot:ReIm}It is worth noting that, using the conventions of appendix E of \cite{KolchmeyerLiu2024}, these $S$-channel coefficients become real for real $\Delta_1,\Delta_2,\Delta_3$ while the corresponding $A$-channel coefficients become purely imaginary.}
\begin{equation}
\left(\widetilde{C}^{P,S}_{PP}\right)^{\Delta_3,-m_3}_{\Delta_1,m_1|\Delta_2,m_2} 
:= \left(\widetilde{C}^{P,S}_{PP}\right)^{s_3,-m_3}_{s_1,m_1|s_2,m_2}.
\end{equation}
In the complementary series
the  $s_1,s_2,s_3$ are now imaginary and are defined by $is=\Delta-\frac{1}{2}$.  The coefficients $\psi_{m_1,m_2,m_3}$ are defined by writing 
\begin{equation}
|\psi \rangle_{\Delta_1, \Delta_2, \Delta_3} = \sum_{m_1,m_2,m_3} \psi_{m_1,m_2,m_3}
|m_1,m_2,m_3\rangle_{\Delta_1,\Delta_2, \Delta_3}.   
\end{equation}     The non-trivial factor ${\cal N}_S$ arises from \eqref{eq:PPPip2} after using $\Delta^*=1-\Delta$ in the principal series, analytically continuing to the complementary series, and then re-expressing the result in terms of a modulus-squared.  

As can be seen from the explicit formulae in \cite{KolchmeyerLiu2024}, the $S$-channel Clebsch-Gordan coefficients are meromorphic in the $\Delta_i$ and have a simple pole at $\Sigma:= \Delta_1+\Delta_2+\Delta_3=1$. As a result, unless this pole is canceled by a zero from the $\psi_{m_1,m_2,m_3}$ (as these coefficients implicitly depend on the $\Delta_i$), the factor of ${\cal N}_S$ in \eqref{eq:PPPip3} appears multiplied by a double pole at $\Sigma=1$.
    Analyticity of the left-hand-side of \eqref{eq:PPPip3} for our state thus requires ${\cal N}_S$ to be meromorphic in $\Delta_i$ as well. As a result, we may use the fact (noted above in the context of the state $|0,0,0\rangle_{\Delta_1, \Delta_2, \Delta_3}$) that the group-averaging norm of general even-parity states has a {\it simple} pole at $\Sigma=1$ to see that ${\cal N}_S$ must have a simple zero.  In particular, \eqref{eq:PPPip3} then shows the group-averaging norm of our special even-parity states to be {\it negative} for some open set\footnote{\label{foot:hole}As mentioned above, direct calculations of ${\cal N}_S$ and the corresponding ${\cal N}_A$ can be performed using the formulae in \cite{KolchmeyerLiu2024}.  The key ingredients are their (E.23) and the corresponding result for the $A$-channel (which is not given in \cite{KolchmeyerLiu2024}, but which can be derived from their explicit formula for the $A$-channel coefficients in the same way as their (E.23)).  One should also use the observation in footnote \ref{foot:ReIm}.  Tracing through the relevant conventions, identities, and charge-conjugations is cumbersome, but yields
\begin{align}
{\cal N}_S(\Delta_1,\Delta_2,\Delta_3)
&=
\frac{\Gamma\!\left(1-\frac{\Sigma}{2}\right)}
     {\Gamma\!\left(\frac{\Sigma-1}{2}\right)} 
\left[\frac{\Gamma(1-\Delta_3)}{\Gamma(\Delta_3)} \right]^2\frac     {\Gamma\!\left(\frac{1-\delta_3}
{2}\right)}{\Gamma\!\left(\frac{\delta_3}{2}\right)}\prod_{i=1}^{2}
\frac     {\Gamma\!\left(\frac{\delta_i}{2}\right)}{\Gamma\!\left(\frac{1-\delta_i}{2}\right)}
,
     \qquad {\rm and} \nonumber \\ {\cal N}_A(\Delta_1,\Delta_2,\Delta_3)
&=
\frac{\Gamma\!\left(\frac{3-\Sigma}{2}\right)}
     {\Gamma\!\left(\frac{\Sigma}{2}\right)}\left[\frac{\Gamma(1-\Delta_3)}{\Gamma(\Delta_3)} \right]^2
\frac{\Gamma\!\left(1-\frac{\delta_3}{2}\right)}{\Gamma\!\left(\frac{1+\delta_3}{2}\right)}
     \prod_{i=1}^{2}
\frac
     {\Gamma\!\left(\frac{1+\delta_i}{2}\right)}{\Gamma\!\left(1-\frac{\delta_i}{2}\right)}, \nonumber
\end{align}
where $\delta_1 := \Delta_2+\Delta_3-\Delta_1$ and $\delta_2, \delta_3$ are defined analogously but with cyclically permuted indices $1,2,3$, and where we have also checked the above result numerically. Thus, in accordance with the results of section \ref{subsec:Aproof} below, ${\cal N}_A>0$ everywhere in the complementary series, but ${\cal N}_S<0$ for $\Sigma < 1$ with all $\delta_i$ positive.} of $\Delta_i$ with $\Sigma< 1$ even though for such states group averaging remains absolutely convergent in that regime. 

The observation that RIPO states enjoy absolute convergence for all $\Sigma$ suggests that we should also consider a few additional 2- and 3-particle cases. However, the 2-particle cases turn out to be trivial.  In particular, as noted in footnote \ref{foot:CD}, for either principal-complementary or complementary-complementary RIPO two-particle states, absolute convergence of group averaging implies that the group-average is a null state. Group averaging also converges absolutely for RIPO principal-principal 2-particle states and, again, the group-averaged state is null as can be seen explicitly from e.g. \eqref{eq:PPGAIPfixs}. 

However, it remains to consider 3-particle RIPO states with one principal-series particle (say, particle 1) and two complementary-series particles (particles 2 and 3) with $\Sigma \le 1$.  While the $\Sigma > 1$ case was shown to be positive at the beginning of this section, we have not yet discussed the case $\Sigma\le 1$ (which requires $\Delta_2 + \Delta_3 \le \frac{1}{2}$).  Here we may proceed by replacing particles 2 and 3 by a single effective particle using the fact \cite{Repka1976,Repka1978} that, using ${\mathscr C}_\Delta$ to denote the complementary-series representation of weight $\Delta$, for 
$\Delta_2 + \Delta_3 <\frac{1}{2}$ we have
\begin{equation}
{\mathscr C}_{\Delta_2} \otimes
{\mathscr C}_{\Delta_3} \cong_G {\cal H}_P \oplus {\cal H}_P \oplus {\mathscr C}_{\Delta_2 +\Delta_3} \bigoplus_{\epsilon =\pm,\, \Delta \in {\mathbb Z}_{\ge 1}} D^\epsilon_\Delta.
\end{equation}
For the marginal case $\Delta_2 + \Delta_3 =\frac{1}{2}$ the result is just
\begin{equation}
{\mathscr C}_{\Delta_2} \otimes
{\mathscr C}_{\Delta_3} \cong_G {\cal H}_P \oplus {\cal H}_P  \bigoplus_{\epsilon =\pm,\, \Delta \in {\mathbb Z}_{\ge 1}} D^\epsilon_\Delta.
\end{equation}

Terms that give principal- or discrete-series contributions yield 2-particle states of the form already addressed in section \ref{subsec:PD-two-particle-base}, so we need only consider the term ${\mathscr C}_{\Delta_2 +\Delta_3}$ that arises for $\Delta_2 + \Delta_3 < \frac{1}{2}$.  The associated channel turns out to preserve parity,\footnote{Expanding \eqref{eq:SO21explicit} at large $\lambda$ shows that, for appropriate 
$\kappa(n_2,n_3) \propto \sqrt{h_{\Delta_2}(n_2)h_{\Delta_3}(n_3)/h_{\Delta_2+\Delta_3}(n_2+n_3)}$, the function
${}_{\Delta_2}\langle n_2 |U(g)|m_2\rangle_{\Delta_2}\, {}_{\Delta_3}\langle n_3 |U(g)|m_3\rangle_{\Delta_3} - \kappa(n_2,n_3)\,\kappa(m_2,m_3)\, {}_{\Delta_2+\Delta_3}\langle n_2+n_3 |U(g)|m_2+m_3\rangle_{\Delta_2+\Delta_3}$ lies in $L^2(G)$, and so is a superposition of principal- and discrete-series matrix elements.  Up to normalization, the $\kappa(n_2,n_3)$ are thus the Clebsch-Gordan coefficients for the above ${\mathscr C}_{\Delta_2+\Delta_3}$ channel.  The fact that $\kappa(-n_2,-n_3) = \kappa(n_2,n_3)$ then shows this channel to be parity-even.} so that the effective 2-particle state has the same (odd) parity as the original 3-particle state.  But this means that it is again null by footnote \ref{foot:CD}.  We thus see that RIPO principal-complementary-complementary 3-particle states continue to have non-negative group-averaging norm regardless of the value of $\Sigma$.

\subsection{Positivity of odd-parity group averaging for 3 complementary-series particles}
\label{subsec:Aproof}

This subsection provides a general argument  that the SO$_0(2,1)$ group-averaging inner product is positive semi-definite for all RIPO 3-particle states with each particle lying in the complementary series.    
Since the $S$ contribution to \eqref{eq:PPPip2} vanishes for such states, there is only one coefficient to check for each $\Delta_1,\Delta_2,\Delta_3$.  Verifying positivity of the group-averaging norm of \eqref{eq:Astate3C} will thus imply positivity of the $A$ term in \eqref{eq:PPPip2} for all states.  Furthermore, as noted at the end of section \ref{subsec:3c}, for RIPO states group averaging converges absolutely for all $\Delta_i >0$.  We will see this explicitly for the state \eqref{eq:Astate3C} below.

The key observation\footnote{We thank ChatGPT (GPT 5.6) for assistance with assembling this argument.}  is that expressions \eqref{eq:0002d}, \eqref{eq:11}, and \eqref{eq:1-1} for the matrix elements of $e^{i\lambda B_1}$ in UIRs allow us to write the group-averaged norm of the state $|\psi\rangle_{\Delta_1,\Delta_2,\Delta_3}$ in
a manifestly positive form.  
We begin by observing that using the identity $\coth \frac{\lambda}{2}-\tanh \frac{\lambda}{2} = \frac{2}{\sinh \lambda}$ in computing the sum of  \eqref{eq:11} and \eqref{eq:1-1}  yields
\begin{equation}
\frac{\langle1|e^{i\lambda B_1}|1\rangle + \langle1|e^{i\lambda B_1}|{-1}\rangle}{\langle1|1\rangle}
= -\frac{2}{\Delta(1-\Delta) \,\sinh \lambda}\frac{d\phi_\Delta}{d\lambda}
.
\end{equation}
Let us also note that $\phi_\Delta$ satisfies the hypergeometric equation
\begin{align}
 \phi_\Delta''(\lambda)+\coth \lambda\,\phi_\Delta'(\lambda)
 +\Delta(1-\Delta)\phi_\Delta(\lambda)&=0.
        \label{eq:comp-spherical-ODE}
\end{align}
As a result, using the identity $\tanh \frac{\lambda}{2}+ \coth \frac{\lambda}{2} = 2 \coth \lambda$, we may subtract \eqref{eq:1-1} from \eqref{eq:11} to obtain 
\begin{align}
\frac{\langle1|e^{i\lambda B_1}|1\rangle - \langle1|e^{i\lambda B_1}|{-1}\rangle}{\langle1|1\rangle}
&=
2\phi_\Delta(\lambda)
+2\frac{\coth \lambda}{\Delta(1-\Delta)}
\frac{d\phi_\Delta}{d\lambda} \nonumber \\
&= -\frac{2}{\Delta(1-\Delta)}\frac{d^2 \phi_\Delta}{d\lambda^2}.
\end{align}
The normalized expectation value of $e^{i\lambda B_1}$ in the state 
$|\psi \rangle_{\Delta_1,\Delta_2,\Delta_3}$
thus becomes
\begin{align}
\frac{\langle \psi |e^{i\lambda B_1} |\psi \rangle_{\Delta_1,\Delta_2,\Delta_3}}{\langle \psi | \psi \rangle_{\Delta_1,\Delta_2,\Delta_3}}
 &=\phi_{\Delta_1}(\lambda) \nonumber \\
&\times\frac{\langle1|e^{i\lambda B_1}|1\rangle{}_{\Delta_2} \, \langle1|e^{i\lambda B_1}|1\rangle{}_{\Delta_3}- \langle1|e^{i\lambda B_1}|{-1}\rangle{}_{\Delta_2} \, \langle1|e^{i\lambda B_1}|{-1}\rangle{}_{\Delta_3} }{\langle1|1\rangle{}_{\Delta_2} \langle1|1\rangle{}_{\Delta_3}}\nonumber\\
 &=\frac{2\phi_{\Delta_1}(\lambda)}
{ \Delta_2(1-\Delta_2)\Delta_3(1-\Delta_3)\sinh \lambda}\
 \left(
\phi_{\Delta_2}''(\lambda)\phi_{\Delta_3}'(\lambda)+ \phi_{\Delta_2}'(\lambda)\phi_{\Delta_3}''(\lambda)\right)
\nonumber\\
 &=\frac{2\phi_{\Delta_1}(\lambda)}
{ \Delta_2(1-\Delta_2)\Delta_3(1-\Delta_3)\sinh \lambda}\
 \frac{d}{d\lambda}\left(
\phi_{\Delta_2}'(\lambda)\phi_{\Delta_3}'(\lambda)\right),
        \label{eq:comp-odd-radial-coefficient}
\end{align}
where the first line uses the parity relation \begin{equation}
\label{eq:Prel}
{}_\Delta\langle -m|e^{i\lambda B_1}|{-n}\rangle_{\Delta} = {}_\Delta\langle m|e^{-i\lambda B_1}|n\rangle_{\Delta}= (-1)^{m-n}\,{}_\Delta\langle m|e^{i\lambda B_1}|n\rangle_{\Delta}.
\end{equation}

Since our state $|\psi\rangle_{\Delta_1,\Delta_2,\Delta_3}$  is invariant under rotations, we may now compute the group-averaged norm using \eqref{eq:PD-GA-boost-reduced}.  The $\sinh \lambda$ measure factor in \eqref{eq:PD-GA-boost-reduced} then cancels with the corresponding factor in the denominator of \eqref{eq:comp-odd-radial-coefficient} to yield
\begin{align}
\frac{\langle \psi |\psi \rangle^{\rm GA}_{\Delta_1,\Delta_2,\Delta_3}}{\langle \psi | \psi \rangle_{\Delta_1,\Delta_2,\Delta_3}}
 &=\frac{8\pi^2 }
 {\Delta_2(1-\Delta_2)\Delta_3(1-\Delta_3)}
 \int_0^\infty d\lambda \, \phi_{\Delta_1}(\lambda) \frac{d}{d\lambda}\left(
\phi_{\Delta_2}'(\lambda)\phi_{\Delta_3}'(\lambda)\right)
        \nonumber\\
 &=-\frac{8\pi^2}{\Delta_2(1-\Delta_2)\Delta_3(1-\Delta_3)}
   \int_0^\infty d\lambda \,
   \phi_{\Delta_1}'(\lambda)
   \phi_{\Delta_2}'(\lambda)
   \phi_{\Delta_3}'(\lambda),
        \label{eq:comp-odd-positive-integral}
\end{align}
where the boundary terms vanish since $\phi'_\Delta$  decays rapidly at infinity and vanishes at $\lambda=0$ (due to its symmetry under $\lambda \rightarrow -\lambda$ which follows from the rotational invariance of $|0\rangle_\Delta$).

Note that the fall-off of $\phi'_\Delta(\lambda)$ at large $\lambda$ must involve the same exponential as that of $\phi_\Delta(\lambda)$.  Since the final integral in \eqref{eq:comp-odd-positive-integral} has no factor of $\sinh(\lambda)$, the domain of manifest convergence has been improved to ${\rm Re}\, (\Delta_1+\Delta_2+\Delta_3) >0$. As noted above, this extended convergence domain in fact applies to all RIPO states as can be shown by studying the leading exponential behavior of the matrix elements of $e^{i\lambda B_1}$.

We emphasize that the denominator on the left-hand-side of \eqref{eq:comp-odd-positive-integral} has no ${\rm GA}$ superscript and thus is the {\it kinematic}  norm of $|\psi \rangle_{\Delta_1,\Delta_2,\Delta_3}$ (i.e., the norm in the original Fock space) which is manifestly positive.
We may thus show \eqref{eq:comp-odd-positive-integral} to be positive by showing $\phi'_{\Delta_i} < 0$ for all $\lambda >0$ and all complementary-series $\Delta_i$. This is to be expected since i) unitarity of $e^{i\lambda B_1}$ requires $\phi_{\Delta}(\lambda=0)=1$,  ii) $\phi_{\Delta}(\lambda)$ vanishes rapidly at infinity, and iii) $\phi_{\Delta}(\lambda)$ should have `no nodes' since it comes from an expectation value in the state with vanishing angular momentum.  This negativity is readily established by writing \eqref{eq:comp-spherical-ODE} in the form
\begin{align} 
 \left(\sinh \lambda\,\phi_\Delta'(\lambda)\right)'
 =-\Delta(1-\Delta)\sinh \lambda\,\phi_\Delta(\lambda)<0,
        \label{eq:comp-spherical-monotonicity}
\end{align}
where the final inequality uses positivity of $\phi_\Delta$ and the fact that we are in the complementary series (so that $0< \Delta < 1$).   Using the fact that  $\phi_\Delta'$ vanishes at $\lambda =0$ and integrating the above inequality gives $\sinh \lambda \, \phi'_\Delta <0$ for positive $\lambda$. It then follows that $\phi'_\Delta$ is everywhere negative on ${\mathbb R}^+$ as desired.  This completes the proof of positivity of the group-averaging inner product for 3-particle states in the complementary series with weights $\Delta_1 + \Delta_2 + \Delta_3 >1$,  though for RIPO states  we see that the result in fact holds for all complementary-series representations whether or not this constraint is imposed.

\subsection{States with \texorpdfstring{$N\ge 4$}{N ≥ 4} particles}
\label{subsec:4d2}

Just as for the discussion of principal- and discrete-series particles in section \ref{subsec:PD-CG-decomposition}, the tensor product of two arbitrary UIRs can be written as a sum/integral over UIRs.    Here the key observation is that the weights $\Delta$ of any complementary-series UIRs appearing in that sum always have\footnote{This result is consistent with the statements in section \ref{subsec:3c} that no complementary-series representations can appear in the decomposition when one or both of the original representations lies in the principal or discrete series.} 
\begin{equation}
\label{eq:Deltagrow}
\Delta \ge {\rm Re}\, \left( \Delta_1 + \Delta_2\right),
\end{equation}
where  as usual  we use the convention that  any complementary-series weights $\Delta, \Delta_1, \Delta_2$ lie in $(0,\frac{1}{2})$; 
see e.g. theorem 2.1 of \cite{Shalom2000Rigidity} using the above exponential bounds on simple matrix elements in the $\Delta_1,\Delta_2$ representation (or, equivalently, using H\"older's inequality).   As a result,   if we take particles 1 and 2 to be the particles with the smallest values of ${\rm Re} \, \Delta_i$ then, when the initial state satisfies $\Sigma = {\rm Re} \, \sum_{i=1}^N \Delta_i >1$, for complementary-series weights $\Delta$ we clearly  also find
\begin{equation}
\label{eq:sumDeltagrow}
    {\rm Re}\, \Delta + \sum_{i=3}^N \, {\rm Re } \, \Delta_i >1.
\end{equation} 

That \eqref{eq:sumDeltagrow} also holds for discrete-series and principal-series weights $\Delta$ follows from the fact that, since particles 1 and 2 have the smallest ${\rm Re} \, \Delta_i$, we must have 
$ \sum_{i=3}^N \, {\rm Re } \, \Delta_i \ge {\rm Re}\left( \Delta_1 + \Delta_2\right)$. But we also clearly have $ \sum_{i=3}^N \, {\rm Re } \, \Delta_i > 1-{\rm Re}\left(\Delta_1 + \Delta_2\right)$.  Taking the average of these two inequalities yields $ \sum_{i=3}^N \, {\rm Re } \, \Delta_i > \frac{1}{2}$, which implies \eqref{eq:sumDeltagrow} since the discrete series has only real $\Delta \ge 1$ and the principal series has ${\rm Re}\, \Delta = \frac{1}{2}$.

We have thus reduced the analysis of our $N$-particle state to that of  an $(N-1)$-particle state for which again the sum of the new  ${\rm Re} \, \Delta_i$ exceeds 1.  Iterating this argument down to the case $N=3$ (where positivity for $\Sigma>1$ has already been shown) then establishes positivity for all $N$ (so long as $\Sigma >1$). As in section \ref{subsec:PD-CG-decomposition}, this argument involves exchanging the order in which one integrates over $G$ with that of the integrals and sums over the parameters of the effective particles, but this is again justified by the reasoning that was used at the end of section \ref{subsec:PD-CG-decomposition} (and again explicitly using $\Sigma >1$).

However, this result can still be strengthened further.  In particular, as discussed in sections \ref{sec:dS2UIRs} and \ref{subsec:Aproof}, the condition \eqref{eq:dS2conv} suffices to show that the slowest possible (exponential) fall-off of matrix elements of $e^{i\lambda B_1}$ at large $\lambda$ yields an absolutely convergent group-averaging integral.  But, as discussed in sections \ref{subsec:3c} and \ref{subsec:Aproof}, for certain states the coefficient of that naively-leading exponential simply vanishes.  In cases where this follows from a parity argument, the group-averaging integral will in fact converge so long as $\sum_i {\rm Re}\, \Delta_i >0$, which always holds since every UIR has ${\rm Re}\, \Delta_i >0$ separately.\footnote{The fact that this holds for all RIPO 3-particle states is consistent with the calculation of ${\cal N}_A$ briefly reported in footnote \ref{foot:hole} as this ${\cal N}_A$ remains finite at $\Sigma=1$.}  In other (finely-tuned) cases, this convergence also requires $\Sigma- 2\Delta_i >0$ for all $\Delta_i$. 

In particular, we saw near the end of section \ref{subsec:Aproof} that this was the case for all RIPO states, and that the group-averaging norm of such states is again positive in the 3-particle sector. Our iterative argument then shows that this will again be the case for any $N$-particle state whose Clebsch-Gordan reduction to a state of 3 effective particles is again RIPO.\footnote{Justifying the exchange of orders of integration now requires a slight improvement of the bound used previously.  This improved bound is derived in appendix \ref{app:RIPOconv}.}  For a fixed choice of which (effective) particles are successively Clebsch-Gordan reduced, we may call the space of such $N$-particle states ${\cal H}^N_{\text{3EP-RIPO}}$ (this space in general depends on that choice).  Unfortunately, the fact that 
Clebsch-Gordan reduction in the $A$-channel flips the parity of a state makes it non-trivial to give a simple explicit characterization of 
${\cal H}^N_{\text{3EP-RIPO}}$ at general $N$ for general\footnote{When two $\Delta_i$ coincide, some progress can be made using the fact that $A$-channel ($S$-channel) Clebsch-Gordan reduction on these particles annihilates states that are symmetric (antisymmetric) under interchange of these particles.  Thus, for example, rotation-invariant odd-parity (even-parity) 4-particle states with weights $\Delta_1, \Delta_2, \Delta_3=\Delta_4=\Delta$ lie in ${\cal H}^4_{\text{3EP-RIPO}}$ (defined using a reduction that first combines particles 3 and 4) when they are symmetric (antisymmetric) under interchange of particles $3$ and $4$.  However, this observation is of limited use since Clebsch-Gordan reduction generally leads to superpositions over the resulting effective-particle weights $\Delta$.} $\Delta_i$.  In particular, this ${\cal H}^N_{\text{3EP-RIPO}}$ will generally include states of both even and odd $N$-particle parity, while $N$-particle states of odd parity that are orthogonal (in the Fock inner product) to ${\cal H}^N_{\text{3EP-RIPO}}$ need not have positive group-averaging norms (their reduction yields even-parity states of three effective particles, which can be finely tuned as in section~\ref{subsec:3c}).  We therefore simply restrict group averaging of states violating \eqref{eq:dS2conv} to states in ${\cal H}^N_{\text{3EP-RIPO}}$, though we leave further study of this space to the future.

\section{Positivity in \texorpdfstring{dS$_d$}{dS\_d} for \texorpdfstring{$d >2$}{d > 2} with \texorpdfstring{$\Sigma > d-1$}{Σ > d-1}}
\label{sec:higherD}

Having thoroughly studied positivity of group averaging for Fock states with bosonic particles in dS$_2$, it is natural to ask for corresponding results in higher dimensions. We thus now address the case of Fock states with bosonic particles for general dS$_d$ with $d>2$.   Many of the above $d=2$ arguments in fact carry over rather directly to general dS$_d$, and others can be extended with more work.  The goal of this section is thus to lay out a framework for analyzing general $d >2$ with $\Sigma = {\rm Re} \, \sum_{i=1}^N \Delta_i > d-1$.
Below, we use $G_d$ to denote the connected component SO$_0(d,1)$ of the isometry group of dS$_d$.

Positivity for states built from $N\ge 2$ principal- and discrete-series particles\footnote{As in section \ref{sec:principal-discrete-positivity}, for 2-particle states with both particles in the principal series  we require the state  to belong to ${\cal H}_P \otimes {\cal H}_P$ (with ${\cal H}_P$ defined as the direct integral over all principal-series UIRs) so that the group-averaging norm is finite. And as was also the case for $d=2$, 1-particle states with $\Sigma >d-1$ occur only in representations which lack rotation-invariant states, whence the absolute convergence of group averaging for $\Sigma > d-1$ implies that all states group-average to null states.} is established in section \ref{subsec:PDgend}.  We then show in section \ref{subsec:othergend} that, even when other types of particles are present, positivity of group averaging will follow for all $N$-particle states whose weights $\Delta_i$ (defined in section \ref{subsec:PDgend})  satisfy
\begin{equation}
\label{eq:ddimSigmag1}
\Sigma:= {\rm Re} \, \sum_{i=1}^N \Delta_i > d-1,
\end{equation}
so long as positivity can be shown 
for the associated  3-particle states.  We also explicitly show positivity for 3-particle states with $\Sigma > d-1$ in which all particles are scalars.  This is enough to establish the desired positivity for arbitrary $\Sigma > d-1$ Fock states when $d=3$.  Positivity for the relevant 3-particle states for $d\ge4$ will be established separately via a more abstract approach in section \ref{subsec:3pgend}.

In \eqref{eq:ddimSigmag1}, we again use the convention that so-called complementary-series representations (and also exceptional-series representations) have $\Delta < \frac{d-1}{2}$ while discrete-series representations  have $\Delta > \frac{d-1}{2}$.   Using the comment below \eqref{eq:PD-GA-boost-reduced} regarding the Haar measure on SO$_0(d,1)$, we see that group averaging converges absolutely whenever \eqref{eq:ddimSigmag1} holds (see \cite{Marolf_2009} for scalars, massless vectors and gravitons; for general UIRs this follows from the fall-offs of matrix elements described below, see e.g. \cite{VK,Knapp2001Semisimple}).
 As for the rotationally-invariant parity-odd (RIPO) states for $d=2$ discussed in section \ref{subsec:Aproof}, the positivity result also extends to even smaller $\Delta_i$ for certain classes of states in which group averaging remains absolutely convergent due to symmetries.  But in cases not protected by symmetries, and where the divergence at $\Sigma=d-1$ is removed by direct fine-tuning, we again expect negative group-averaging norms in an open set of $\Delta_i$ with $\Sigma$ just below $d-1$.

\subsection{Positivity for \texorpdfstring{$\Sigma > d-1$}{Σ > d-1} principal- and discrete-series states with \texorpdfstring{$d>2$}{d > 2}}
\label{subsec:PDgend}

The most important observation for treating $d>2$ stems from the fact that, as discussed for dS$_2$ in section \ref{subsec:PD-two-particle-base}, any matrix element $\langle \phi_1| U(g) |\phi_2\rangle$ between two states  $|\phi_1\rangle, |\phi_2\rangle$ in a UIR of $G_d$ defines an eigenfunction of the natural Laplacian on $G_d$ (i.e., an eigenfunction of the quadratic Casimir).  Since the spectrum of this Laplacian cannot be empty, there must be UIRs for which these eigenfunctions are either normalizable or delta-function normalizable in $L^2(G_d)$.  In either of these cases, the corresponding UIRs are called {\it tempered representations} (see \cite{KnappZuckerman1976}, following \cite{HarishChandra1966}), as the eigenfunctions are then analogous to tempered distributions on ${\mathbb R}^n$.  

One thus finds tempered representations for every eigenvalue of the group Laplacian interpreted as an operator acting on $L^2(G_d)$.  In particular, letting $\sigma$ range over all UIRs of the little group SO$(d-1)$, and writing the quadratic SO$_0(d,1)$ Casimir as\footnote{At least in the principal and complementary series, the mathematical literature would say that we choose this form of the Casimir for representations of SO$_0(d,1)$ induced from the representation $\sigma$ of SO$(d-1)$. A related statement is true in the discrete series.  For dS$_4$ we have $d=4$ and thus $d-1=3$, in which case $\sigma$ is a representation of the familiar group SO$(3)$ and is labeled by a single integer $j$ representing its total angular momentum.} $\Delta(d-1-\Delta)-C_2(\sigma)$ for the usual quadratic Casimir $C_2$ on SO$(d-1)$, one finds the leading fall-offs of matrix elements of $e^{i\lambda B_1}$ to be $e^{-\Delta\lambda}$ and/or $e^{-(d-1-\Delta)\lambda}$ where, with our conventions, the second fall-off is absent for so-called discrete-series representations; see e.g. \cite{VK,Knapp2001Semisimple}. For all $\sigma$, the continuous spectrum of the quadratic Casimir on $L^2(G_d)$ is given by the principal-series representations (for which we can take $\Delta = \frac{d-1}{2} +is$ with $s\ge 0$).  These are the only tempered UIRs for odd $d$, though for even $d$ there are also discrete-series representations that define normalizable Casimir eigenstates in $L^2(G_d)$ and for which $\Delta$ ranges over $\frac{d}{2},\frac{d}{2}+1, \frac{d}{2}+2, \dots$ with the maximum value of $\Delta$ set by the associated little-group representation $\sigma$.  Together, these principal- and discrete-series representations provide all tempered representations of $G_d$ \cite{KnappZuckerman1976,Knapp2001Semisimple}.  As in section \ref{sec:principal-discrete-positivity}, we will use the symbol ${\cal H}_P$ to denote the direct integral/sum of all principal-series representations over all allowed pairs $(\Delta,\sigma)$ of weights $\Delta$ and inducing representations $\sigma$ of SO$(d-1)$.

The main arguments of section \ref{sec:principal-discrete-positivity} for principal- and discrete-series representations can thus be applied verbatim for general dimension $d$.  As in section \ref{subsec:PD-two-particle-base}, the group-averaging inner product of a pair of (perhaps distributional) two-particle states associated with representations $\rho_1, \rho_2$ can be written as an inner product in  $L^2(G_d)$ of two functions defined by matrix elements of $U(g)$ in states associated with representations $\rho_1, \rho_2^\vee$, where $\rho_2^\vee$ is the contragredient (dual) representation of $\rho_2$.  Positivity of the $L^2(G_d)$ inner product then implies positivity of group averaging on ${\cal H}_P\otimes {\cal H}_P$, and similarly for two-particle states that involve discrete-series representations.   Note that the only two-particle states that satisfy $\Sigma > d-1$ are those that involve at least one discrete-series particle.  As in the $d=2$ case, our study of group averaging on ${\cal H}_P\otimes {\cal H}_P$ is not directly physical but, instead, serves only as a mathematical crutch to simplify the analysis of $N$-particle states with $N\ge 3$.

In particular, since the product of two $U(g)$ matrix elements associated with tempered representations is again in $L^2(G_d)$, decomposing it into eigenstates of the group Laplacian (each of which is associated with a tempered representation)  shows that the tensor product of tempered representations is a direct sum/integral of tempered representations (perhaps with non-trivial multiplicity) in direct analogy with the $d=2$ case.\footnote{\label{foot:CwithT}The measure for the group-averaging integral of rotation-invariant states involves $\sinh^{d-1} \lambda$. Since tempered representations have matrix elements that decay at least as $e^{-\frac{d-1}{2}\lambda}$ (up to polynomial corrections), and since matrix elements decay in all relevant UIRs, multiplying a tempered-representation matrix element by a matrix element of {\it any} non-trivial UIR gives a member of $L^2(G_d)$.  As a result, the tensor product of a tempered representation with an arbitrary UIR can also be written as a sum/integral of tempered representations. This is again in direct analogy with the case $d=2.$}    Clebsch-Gordan reduction thus establishes positivity of group averaging for all states with $N\ge 3$ particles so long as each particle lies in a tempered representation.  

While this concludes our argument for tempered representations, we of course wish to set the stage for discussion of complementary-series representations in section \ref{subsec:othergend}.  As mentioned above, any complementary-series representation is also induced from a choice of SO$(d-1)$ representation $\sigma$. However, it turns out that complementary-series representations of SO$_0(d,1)$ occur only when $\sigma$ is self-dual, meaning that $\sigma$ admits a charge-conjugation operator as described below.\footnote{The precise criterion is that $\sigma$ is invariant under Weyl-reflections; see e.g. \cite{ArendsBangJensenFrahm2026}, which cites 
\cite{Hirai1962a,BaldoniSilvaBarbasch1983} (see also \cite{Ottoson1968}).  Standard facts from e.g. \cite{FultonHarris1991,Knapp2002}
then show that every reflection-invariant representation is also self-dual.
}  As a result, any induced complementary-series representation is also self-dual and, given any orthonormal basis $|i\rangle_{\Delta, \sigma}$ for this SO$_0(d,1)$ representation, there is a 
unitary charge conjugation operator  $C_{\Delta, \sigma}$  that satisfies  
\begin{equation}
\left( {}_{\Delta,\sigma} \langle i|U_\Delta(g) |i'\rangle _{\Delta,\sigma}\right)^*
 = {}_{\Delta,\sigma} \langle i'|U_\Delta(g^{-1}) |i\rangle_{\Delta,\sigma}
= {}_{\Delta,\sigma}\langle i|C_{\Delta, \sigma}^{\dagger} U_\Delta(g) C_{\Delta, \sigma}| i'\rangle_{\Delta,\sigma},
\label{eq:dCconj}
\end{equation}
where the notation in \eqref{eq:dCconj} emphasizes that this $C_{\Delta, \sigma}$ maps the representation of weight $\Delta$ to itself (and thus that it relates $U(g)$ to $U(g^{-1})$ within the same representation).  To be concrete, for scalar representations we may take $C_{\Delta, \sigma}$ to be the linear map defined by composing the anti-linear map $\Gamma_\nu$  of \cite{HollandsLechner2018} (evaluated at $\nu = -i \left(\Delta - \frac{d-1}{2}\right)$) with the (again anti-linear) complex-conjugation map $K$ defined by taking $K|i\rangle_{\Delta,\sigma} = |i\rangle_{\Delta,\sigma}$.  For non-scalar representations we may then use the analogous $\Gamma_\nu$ defined by composing the complex conjugation of \cite{KarateevKravchukSimmonsDuffin2019} with either their shadow transform (in the principal series) or with the identity (in the complementary series). 
If we furthermore choose our basis to be real in the sense of \cite{KarateevKravchukSimmonsDuffin2019}, then one can check the useful relations 
\begin{align}
    C^\dagger_{\Delta,\sigma}  & = C_{d-1-\Delta,\sigma} = K C_{\Delta,\sigma} K, \quad  C^T_{\Delta,\sigma}=C_{\Delta,\sigma} \ \ \ \ \ \ \ \ \ \ \ \ \ \ \ \  {\rm Principal \ Series} \ \Delta &
    \nonumber \\ \ \ \ \ \
        C_{\Delta, \sigma}  & = {\mathds 1} \quad   \ \ \ \ \ \ \ \ \ \ \  \ \ \ \ \ \ \ \  \ \ \ \ \ \ \ \   \ \ \ \ {\rm Complementary \ Series } \ \Delta\ {\rm and} \ \Delta = \frac{d-1}{2},&  \label{eq:CDelta}
\end{align}  
where $C^T_{\Delta, \sigma} = K C^\dagger_{\Delta, \sigma} K$.

We will henceforth make the above choices so that \eqref{eq:CDelta} applies.   In the principal series, $C_{\Delta, \sigma}$ is analytic in $\Delta$.  In fact, it can be extended to an operator $C_{\Delta, \sigma}^P$ that is  meromorphic in the complex $\Delta$-plane, and which is both well-defined and invertible for all $\Delta$ in the complementary series.\footnote{In particular, the results of \cite{HollandsLechner2018} show that $C^P_{\Delta,\sigma}$ is determined by the principal-series Knapp-Stein map, for which eq. (1.4) of \cite{ArendsBangJensenFrahm2026} gives an explicit formula for Weyl-invariant $\sigma$ (the case relevant for the complementary series).}  However, when continued to the complementary series one finds $C_{\Delta, \sigma}^P \neq {\mathds 1} = C_{\Delta, \sigma}$.  This feature stems from a corresponding property of the matrix elements of $U(g)$.  We will address this issue further below, but for now we simply mention that it is the $d>2$ analogue of the $d=2$ behavior associated with \eqref{eq:hDef} and \eqref{eq:hme}. In any case, the superscript $P$ is to serve as a reminder that $C_{\Delta, \sigma}^P$ is defined via analytic continuation from the principal series.

For the specific case of $N=3$, this charge-conjugation operator allows us to write the principal-series 3-particle group-averaging inner product in a useful way. In particular, by choosing an appropriate normalization for the Haar measure, we may write
\begin{align}
{}_{\vec \Delta, \vec \sigma}\langle C_{3, \vec \Delta,\vec \sigma} \vec j|C_{3, \vec \Delta,\vec \sigma}  \vec j'\rangle^{\rm GA}_{\vec \Delta, \vec \sigma} &=\sum_x \left[ \left(\widetilde C^{P,x}_{PP}\right)^{\Delta_3, \sigma_3, j_3}_{\Delta_1, \sigma_1,j_1|\Delta_2, \sigma_2, j_2}\left(\widetilde C^{P,x}_{PP}\right)^{\Delta_3, \sigma_3, j'_3\,*}_{\Delta_1, \sigma_1,j'_1|\Delta_2, \sigma_2, j'_2} \right] ,
\label{eq:PPPipd}
\end{align}
where for $\vec \Delta = (\Delta_1,\Delta_2,\Delta_3)$ and $\vec j = (j_1,j_2,j_3)$ we have defined 
\begin{equation}
    |C_{3, \vec \Delta,\vec \sigma}  \vec j\rangle : = |j_1\rangle_{\Delta_1, \sigma_1} \otimes|j_2\rangle_{\Delta_2, \sigma_2} \otimes
C_{\Delta_3,\sigma_3} |j_3\rangle_{\Delta_3, \sigma_3}.
\end{equation}

The form  \eqref{eq:PPPipd} is
analogous to \eqref{eq:PPPip} for SO$_0(2,1)$ and, in particular, the Clebsch-Gordan coefficients 
$\left(\widetilde C^{P,x}_{PP}\right)^{\Delta_3, \sigma_3, j_3}_{\Delta_1, \sigma_1,j_1|\Delta_2, \sigma_2, j_2}$ are defined in analogy with 
\eqref{eq:PD-PP-CG} (with $x$ again labeling the possible multiplicity channels), though for convenience we have absorbed any factors that depend on $j_3$ into the normalization and phase of our $\left(\widetilde C^{P,x}_{PP}\right)^{\Delta_3, \sigma_3, j_3}_{\Delta_1, \sigma_1,j_1|\Delta_2, \sigma_2, j_2}$. 

However, as discussed in footnote \ref{foot:CGchi}, since Clebsch-Gordan coefficients can be expressed in terms of integrals of matrix elements of $U(g)$, these coefficients also inherit from $U(g)$ their properties under complex conjugation.   In particular, in the principal series they satisfy an analogue of \eqref{eq:dCconj}:
\begin{equation}
\left(\widetilde C^{P,x}_{PP}\right)^{\Delta_3, \sigma_3, j_3\, *}_{\Delta_1, \sigma_1,j_1|\Delta_2, \sigma_2, j_2} =  \left(\widetilde C^{P,x}_{PP}\right)^{\Delta_3, \sigma_3, C_{\Delta_3, \sigma_3} j_3}_{\Delta_1, \sigma_1,C_{\Delta_1, \sigma_1} j_1|\Delta_2, \sigma_2, C_{\Delta_2, \sigma_2} j_2},
\label{eq:CGbetterconj}
\end{equation}
where, on the right-hand-side, we have used $C_{\Delta, \sigma} j_i$ to denote a formal linear combination of indices as determined by the action of $C_{\Delta, \sigma}$ on our basis states, 
\begin{equation}
C_{\Delta, \sigma} |j\rangle_{\Delta, \sigma} = \sum_k (C_{\Delta, \sigma})^k{}_j |k\rangle_{\Delta, \sigma}  ,
\end{equation}
and we have then extended the definition of our Clebsch-Gordan symbols to such objects by taking them to be anti-linear in each of the lower two indices and linear in the upper ($j_3$) index.  
We have also used the freedom to choose the phases (and, where several channels share the same quantum numbers, the basis) of the Clebsch-Gordan channels $x$ so that \eqref{eq:CGbetterconj} holds without additional $x$-dependent factors.\footnote{This is always possible since $\Gamma_{\Delta, \sigma} := C_{\Delta, \sigma} K$ induces an  anti-linear map on the space of Clebsch-Gordan channels that, by \eqref{eq:CDelta}, squares to the identity, and since any such map admits a basis of invariant vectors.  For $\Delta = \frac{d-1}{2}+i\nu$ this $\Gamma_{\Delta, \sigma}$ coincides with the $\Gamma_\nu$ of \cite{HollandsLechner2018}, or with the generalization to non-scalar representations discussed above \eqref{eq:CDelta}.
}

We may now also use this notation to move the additional charge-conjugation operations from the left-hand-side of \eqref{eq:PPPipd} to the right-hand-side and to thus write
\begin{align}
{}_{\vec \Delta, \vec \sigma}\langle  \vec j| \vec j'\rangle^{\rm GA}_{\vec \Delta, \vec \sigma} &=\sum_x \left(\widetilde C^{P,x}_{PP}\right)^{\Delta_3, \sigma_3, C^T_{\Delta_3,\sigma_3} j_3}_{\Delta_1, \sigma_1, j_1|\Delta_2, \sigma_2, j_2}
\left(\widetilde C^{P,x}_{PP}\right)^{\Delta_3, \sigma_3,  j'_3}_{\Delta_1, \sigma_1,C_{\Delta_1, \sigma_1} j'_1|\Delta_2, \sigma_2, C_{\Delta_2, \sigma_2} j'_2}
\nonumber \\
&=\sum_x \left(\widetilde C^{P,x}_{PP}\right)^{\Delta_3, \sigma_3, C^{P}_{\Delta_3, \sigma_3} j_3}_{\Delta_1, \sigma_1, j_1|\Delta_2, \sigma_2, j_2}
\left(\widetilde C^{P,x}_{PP}\right)^{\Delta_3, \sigma_3, j'_3}_{\Delta_1, \sigma_1,K C_{d-1-\Delta_1, \sigma_1}^{P} j'_1|\Delta_2, \sigma_2, K C_{d-1-\Delta_2, \sigma_2}^{P} j'_2},
\label{eq:PPPipd2}
\end{align}
where this equality is again valid in the principal series and we have used $C_{\Delta, \sigma}^\dagger = C_{d-1-\Delta,\sigma}$.  Note in particular the operator $C_{\Delta_3, \sigma_3}^T: = K C_{d-1-\Delta_3,\sigma_3} K$ that appears in the 1st Clebsch-Gordan coefficient on the first line due to our having removed a $C_{3,\vec \Delta, \vec \sigma}$ from the bra-vector.  Here $K$ is the complex-conjugation operator defined just above \eqref{eq:CDelta}.  The second line then follows from the first line of \eqref{eq:CDelta}, the relation $K^2= {\mathds 1}$, and the definition $K|i\rangle_{\Delta, \sigma} = |i\rangle_{\Delta, \sigma}$.
The utility of the final form is that, since the Clebsch-Gordan symbols are anti-linear in their lower indices, the factors of $K$ in the lower indices make the right-hand-side explicitly meromorphic in the $\Delta_i$.
Analyticity of the left-hand-side will be discussed in section \ref{subsec:3pgend} below.

\subsection{A framework for general  states with \texorpdfstring{$\Sigma > d-1$}{Σ > d-1}}
\label{subsec:othergend} 

We now address states that contain particles in representations outside the principal and discrete series.  In the language of \cite{Basile:2016aen} (though describing results from \cite{Schwarz:1971}), such other UIRs must be either complementary-series representations or exceptional representations.  Furthermore, the non-tempered exceptional representations can all be constructed by taking limits of complementary-series representations; see theorem 1.9, remark 1.10, and theorem 3.9 of 
\cite{ArendsBangJensenFrahm2026}. 

As a result, positivity of group averaging for general $N$-particle states satisfying  the condition \eqref{eq:ddimSigmag1} will follow if it can first be established for all such states built from both complementary-series and tempered representations. 
Furthermore, for $N$-particle states that contain at least one tempered-representation particle,  positivity can then be shown by iteratively Clebsch-Gordan-reducing the state of the other $N-1$ particles to a state with a single effective particle.  If the effective particle is described by a tempered representation, then positivity was shown in section \ref{subsec:PDgend} above.  On the other hand,  
if this effective particle does not lie in a tempered representation, then it must lie in the complementary or exceptional series and have real weight $\Delta < \frac{d-1}{2}$.  We then note that our two-particle state must again have $\Sigma > d-1$ since a complementary-series or exceptional representation of weight $\Delta$ can appear in the UIR decomposition of a tensor product of representations of weights $\Delta_1, \Delta_2$ only if $\Delta_1, \Delta_2$ are real (see again footnote \ref{foot:CwithT}) and if $\Delta \ge \Delta_1+\Delta_2$; see again theorem 2.1 of \cite{Shalom2000Rigidity}.   The first particle must thus have weight strictly greater than $\frac{d-1}{2}$, and must thus lie in the discrete series. In this case the 2-particle group-averaging norm vanishes by the same argument as for $d=2$ (see footnote \ref{foot:CD}) which wrote the result as an $L^2(G_d)$ inner product between eigenfunctions\footnote{\label{foot:notnorm}Recall that the argument did not require both ``eigenfunctions'' to be normalizable in $L^2(G_d)$.} of the self-adjoint Laplacian on $G_d$ that have distinct eigenvalues (though for $d>2$ one must also make use of the fact that this inner product also vanishes whenever the two representations are induced from distinct SO$(d-1)$ representations). As usual, these arguments involve exchanges of orders of integration justified as at the end of section \ref{sec:principal-discrete-positivity} and in the discussion surrounding \eqref{eq:PPGAIPfixs2}.

It thus remains only to consider the case where all particles lie in the complementary series (since we again note that positivity for any non-tempered exceptional-series representations will then follow by taking limits).  As noted above, the results of \cite{Shalom2000Rigidity,deLaatSiebenand2021} again guarantee that, by combining the lightest particles first, the decomposition of a tensor product into UIRs may be used to write any state with $N\ge 4$ particles satisfying ${\rm Re} \, \sum_{i=1}^N \Delta_i > d-1$ as a superposition of states involving $N-1$ effective particles in which the sum of the real parts of the effective weights again exceeds $d-1$ (by the same inequality-averaging as in section \ref{subsec:4d2}).  Iterating this procedure (and using the by-now-standard justification for exchanging orders of integration) thus reduces the proof of the desired positivity result to showing positivity of group averaging for 3-particle states with each particle lying in the complementary series and with $\Sigma = \Delta_1 + \Delta_2 + \Delta_3 > d-1$.

The rest of the analysis will now involve considering various cases.
An important fact that helps to organize this discussion is that complementary-series UIRs can  be described as being induced from UIRs $\sigma$ of SO$(d-1)$.  Indeed, specifying $\sigma$ and a weight $\Delta$ fully determines any complementary-series UIR.  Furthermore, such representations exist only (see e.g. theorem 1.9 of \cite{ArendsBangJensenFrahm2026}) for $\frac{d-1}{2} > \Delta > a_\sigma,$ where $a_\sigma\ge 0$ is the number of rows in the highest-weight Young diagram for $\sigma$.  For odd $d$ there are also certain restrictions on the representation $\sigma$ (though these will play no role in our analysis).

The scalar complementary-series representations are those induced from the trivial representation $\sigma = {\mathbf 1}$, for which $a_\sigma =0$.  As a result, they exist for all $0 < \Delta < \frac{d-1}{2}$, and this range is non-trivial for all $d$.  However, non-scalar complementary-series representations have at least $a_\sigma =1$ and so exist only for $d\ge 4$. 

For $d=2$, we saw in sections \ref{sec:principal-discrete-positivity} and \ref{subsec:3c} that, after group averaging, for each $\Delta_1,\Delta_2,\Delta_3$ with $\Sigma >1$
there were precisely two linearly-independent 3-particle states. 
It will be useful to have a similar prediction, or at least a bound, on the number of linearly-independent group-averaged 3-particle states of various types for general $d$.    Such a bound can be obtained by noting that group averaging constructs formal (not normalizable) bra-states 
\begin{equation}
\langle \Phi | := \langle \phi |\int dg \, U(g) 
\end{equation}
whose standard Fock-space inner product $\langle \Phi|\psi\rangle$ with any $|\psi\rangle$ gives the group-averaging inner product 
$\langle \phi|\psi\rangle^{\rm GA}$.  More properly, one says that group-averaging the linear functional $\langle \phi|$ defines a $G_d$-invariant linear functional $\langle \Phi|$. When $\langle \phi|$ is a three-particle state, we may call this a trilinear functional since $\langle \Phi| \psi\rangle$  is separately linear in the state of each $\psi$-particle.

The number of  such invariant trilinear forms  can then be studied
using the action of SO$_0(d,1)$ as the conformal group on the Euclidean sphere $S^{d-1}$.  In general, it is equal to the number of conformally-invariant $3$-point functions 
associated with CFT operators on this $S^{d-1}$ that transform in corresponding representations \cite{DobrevEtAl1976,ClercOrsted2011,KravchukSimmonsDuffin2018}; see also the discussion in \cite{Kitaev2018,KolchmeyerLiu2024} for the case $d=2$.  Roughly speaking, this uses the CFT state-operator correspondence to map each conformally-invariant bra-state $\langle \Phi|$ to a  corresponding 3-point function.

The number of independent such $3$-point functions was then characterized\footnote{\label{foot:tech}An important technicality is that \cite{KravchukSimmonsDuffin2018} considered only 3-point functions at non-coincident points.  However, the arguments of \cite{KolkVaradarajan1996} and \cite{ClercOrsted2011} show that additional 3-point functions (supported at coincident points) can arise only when certain weight-counting conditions are satisfied.  In particular, theorem 3.15 of \cite{KolkVaradarajan1996} requires the existence of a transverse symbol invariant under the stabilizer group. (This also follows from  corollary A.1 of \cite{ClercOrsted2011} which, as opposed to their main text, allows representations of arbitrary spin.)  This possibility can then be excluded by using the same weight counting as in steps 2 and 3 of theorem 3.1 of \cite{ClercOrsted2011} (associated with the $A$-subgroup of the stabilizer). While that theorem was stated in the context of scalar representations, the choice of inducing SO$(d-1)$ representation does not affect the relevant weights \cite{ArendsBangJensenFrahm2026} and, for all $\sigma_i$, the conditions may be read off from theorem 3.1.  For SO$_0(d,1)$, these conditions require either 
$\Delta_i+\Delta_j-\Delta_k=-n_1$ with $i,j,k$ all distinct or $\Sigma = d-1 - n_2$, where in either case $n_1,n_2$ are non-negative integers.  But  since we explicitly require $\Sigma > d-1$, using $\Delta_k < \frac{d-1}{2}$ for distinct $i,j,k$ and real $\Delta_i$ gives $\Delta_i+\Delta_j-\Delta_k = \Sigma - 2\Delta_k >0$.  Allowing the principal-series weights $\Delta = \frac{d-1}{2}+is$ for $s>0$ also fails to generate new solutions. And a shadow transformation, e.g. $\Delta_1 \rightarrow d-1-\Delta_1$, exchanges these conditions in pairs but does not create new solutions to the conditions.   The counting of \cite{KravchukSimmonsDuffin2018} is thus complete for our allowed complementary-series $\Delta_i$.}
in \cite{KravchukSimmonsDuffin2018}. For $d>2$ representations without conservation laws, it was found to be given by  the number of SO$(d-2)$-invariant states contained in the tensor product SO$(d-1)$ representation $\sigma_1\otimes \sigma_2 \otimes \sigma_3$ built from the  representations $\sigma_i$ of SO$(d-1)$ from which our SO$_0(d,1)$ complementary-series representations are induced.  That the caveat about conservation laws is not relevant to the case at hand can be seen by noting that values of $\Delta_i$ associated with conservation laws can occur only at so-called reducibility points, where the induced-representation construction becomes reducible.  But it is known that, when there is a complementary series at all, the point $\Delta  = \frac{d-1}{2}$ where it meets the principal series is not reducible \cite{Knapp2001Semisimple} and, furthermore, that the first reducibility point one reaches when further decreasing $\Delta$ is the complementary-series endpoint $\Delta = a_\sigma$; see \cite{ArendsBangJensenFrahm2026} for a modern summary of this last point citing \cite{Hirai1962a,BaldoniSilvaBarbasch1983,BransonOlafssonOrsted1996,Thieleker1973}, and see also \cite{Hirai1962b,Ottoson1968}.  The above counting thus gives an upper bound on the rank of the group-averaged Gram matrix for 3-particle states associated with representations induced from $\sigma_1,\sigma_2,\sigma_3$ for any complementary-series $\Delta_i$ with $\Sigma > d-1$.

In particular, since scalar representations are induced from the trivial representation $\sigma = {\mathbf 1}$, and since ${\mathbf 1} \otimes {\mathbf 1} \otimes {\mathbf 1}={\mathbf 1}$,  
we see immediately that for $d> 2$ there is a unique Clebsch-Gordan channel associated with any triple of scalar complementary-series representations.  This is, of course,  associated with the familiar fact that there is a unique conformally-invariant scalar $3$-point function up to normalization.   We may thus check positivity in this channel by finding any 3-particle state whose group-averaging norm is positive. Since the parity of this channel (under an orientation-reversing element of O$(d)$, such as a reflection fixing the boost axis) is even for all $d>2$, it is an analogue of the $S$-channel for $d=2$ and, in parallel with our discussion of that case, it suffices to consider the normalized  rotation-invariant one-particle states $|0\rangle_\Delta$ and their tensor product $|0,0,0\rangle_{\Delta_1, \Delta_2, \Delta_3} : = |0\rangle_{\Delta_1} \otimes |0\rangle_{\Delta_2} \otimes |0\rangle_{\Delta_3}.$  The expectation value of  $e^{i\lambda B_1}$ in $|0\rangle_{\Delta}$ is known to be \cite{HelgasonGGA}
\begin{align}
  \phi^{(d)}_\Delta(\lambda)
  \;& =\; \frac{1}{\mathrm{Vol}(S^{d-1})}\int_{S^{d-1}}\Omega(\bm{x})^{-\Delta}\, \mu_{S^{d-1}}(\bm{x}) d^{d-1} x
\nonumber \\  &=\frac{\Gamma(d/2)}{\sqrt \pi \Gamma((d-1)/2)}
       \int_0^\pi d\theta (\sin \theta)^{d-2}
       (\cosh \lambda - \sinh \lambda \cos \theta)^{-\Delta}
     \nonumber \\
       &= {}_2F_1\!\left(
\Delta,\,
d-1-\Delta;\,
\frac{d}{2};\,
-\sinh^2\frac{\lambda}{2}
\right),
\label{eq:phiint-general}
\end{align}
where we think of the boost as acting as a conformal isometry of $S^{d-1}$, the function $\Omega(\bm{x})=\cosh\lambda-\sinh\lambda\cos\theta$ is the conformal factor associated with the specified boost, $\theta$ is the angle measured from the boost axis to the relevant point $\bm{x}\in S^{d-1}$, and $\mu_{S^{d-1}} d^{d-1}x$ is the volume form on the unit round $S^{d-1}$. In particular, for real $\Delta$, the above \eqref{eq:phiint-general} is positive since
\begin{equation}
0 <   e^{-\lambda}\;\le\;\Omega(\bm{x}).
\label{eq:Omegabound-general}
\end{equation}

The product of three such functions defined by $\Delta_1,\Delta_2, \Delta_3 \in (0, \frac{d-1}{2})$ is thus positive as well. 
Since the group-averaging integral converges absolutely for $\Delta_1+\Delta_2+\Delta_3 > d-1$, this makes manifest that the  group-averaging norm of  $|0,0,0\rangle_{\Delta_1, \Delta_2, \Delta_3}$ is positive in that regime.\footnote{Though it has a simple pole of non-zero residue at $\Sigma = d-1$, so that its analytic continuation is negative for some open set with $\Sigma < d-1$.  As in section  \ref{subsec:3c}, it is possible to choose scalar 3-particle states for which the leading exponentials cancel and the divergence disappears, and where group averaging is then absolutely convergent for an open set of $\Delta_i$ with $\Sigma < d-1$ but where the group-averaged norm is then strictly negative.}

We have thus reduced our task for each $d>2$ to checking positivity of group averaging for triples of complementary-series representations such that $\Delta_1+\Delta_2+\Delta_3 > d-1$ that contain one or more non-scalar representations with $\frac{d-1}{2}> \Delta_i> a_{\sigma_i}\ge 1$.  Note that there are no such representations for $d=3$, in which case $\frac{d-1}{2}=1$ clearly cannot strictly exceed $1$ (as would be required to have a non-scalar complementary-series representation).  We have thus already proven the desired positivity result for dS$_3$ Fock states with $\Sigma > d-1=2$.

Working with ChatGPT (GPT 5.6) and Claude (Opus 5.5) suggests that it is possible to treat non-scalar cases in a similar way.  However, for such cases the number of Clebsch-Gordan channels grows with both dimension and spin so that, even in the 3-particle sector associated with a given triple of UIRs, the rank of the group-averaged Gram matrix becomes large.  In addition, we do not have a way to treat the infinite family of all such triples simultaneously via a direct computation like the one above.  We therefore follow a somewhat more abstract approach in section \ref{subsec:3pgend} below that establishes positivity simultaneously for all three-particle states with $d>3$ and $\Sigma > d-1$.

\subsection{Three complementary-series particles with \texorpdfstring{$\Sigma > d-1$}{Σ > d-1} for \texorpdfstring{$d>3$}{d > 3}}
\label{subsec:3pgend}

As noted above, for $d > 3$ it remains to address 3-particle states in which all particles lie in the complementary series. We will do so using a somewhat more abstract form of the computation reported in footnote \ref{foot:hole} that draws heavily on the existing mathematical literature.\footnote{We thank Claude (Opus 5.5) and ChatGPT (GPT 5.6) for help in assembling this argument and, in particular, for tracking down critical references.}

We begin by recalling that, as described above, Clebsch-Gordan coefficients for {\it principal}-series representations can be computed using integrals in $L^2(G_d)$ involving the product of three matrix elements of $U(g)$ in UIRs.  Such integrals converge absolutely and, 
in fact, continue to do so if we analytically continue the integrands outside the principal series so long as ${\rm Re} \, \Delta_i \le \frac{d-1}{2}$ and
\begin{equation}
\label{eq:3pb}
    \Sigma = {\rm Re} \, \left(\Delta_1+\Delta_2+ \Delta_3 \right) > d-1.
\end{equation}
Furthermore, since the $\Delta$-derivative of each matrix element has the same exponential fall-off as the undifferentiated matrix element (up to polynomial factors), this absolute convergence continues to hold if we replace one of the matrix elements by its $\Delta$-derivative.  
It follows that these Clebsch-Gordan coefficients satisfy the Cauchy-Riemann equations in each of $\Delta_1,\Delta_2,\Delta_3$ (since the individual matrix elements do by analyticity in the principal series).  Thus the integrals define analytic extensions of our principal-principal-principal Clebsch-Gordan coefficients to all $\Delta_i$ for which ${\rm Re} \, \Delta_i \le \frac{d-1}{2}$ and \eqref{eq:3pb} holds.  Below, for general $\Delta_i$, we will always use the term Clebsch-Gordan coefficients to refer to this analytic extension from the principal series.

The number of linearly-independent such Clebsch-Gordan coefficients (the analogue of the $S$ and $A$ channels for $d=2$) is dictated by the multiplicity with which a given principal-series representation appears inside a principal-principal tensor product.  For future reference, it is useful to note that theorem 16 of \cite{Martin1975} (see also theorem 3.5 of \cite{AfentoulidisAlmpanisLiu2025} for an independent derivation for Spin$_0(d,1)$) gives a formula for this multiplicity in terms of the three SO$(d-1)$ representations $\sigma_1,\sigma_2,\sigma_3$ from which our principal-series SO$_0(d,1)$ representations are induced. For $G_d = {\rm SO}_0(d,1)$ with $d\ge 3$, the desired multiplicity is the number of SO$(d-2)$-invariant states contained in the SO$(d-1)$ representation $\sigma_1\otimes \sigma_2 \otimes \sigma_3^*$  (where $\sigma_3^*$ is the dual of $\sigma_3$).

We note that this counting agrees precisely with the number of conformally-invariant three-point functions found in \cite{KravchukSimmonsDuffin2018} that can be associated with the given triple of principal-principal-principal SO$_0(d,1)$ representations in the manner discussed in section \ref{subsec:othergend} above. That a principal-principal Clebsch-Gordan channel leads to an invariant three-point function follows directly from our earlier discussion.  However, the non-trivial result implied by this counting is that {\it all} invariant three-point functions in fact define Clebsch-Gordan channels.\footnote{See also again footnote \ref{foot:tech}.}  This fact will be critical in the argument below.  

With this result in hand, we can now again proceed in analogy with our treatment for $d=2$.  In particular, recall from section \ref{sec:dS2UIRs} that, while we may consider matrix elements of $U(g)$ which are defined using a natural orthonormal basis in the principal series, their analytic continuation to the complementary series no longer gives matrix elements in an orthonormal basis.  Instead, it was necessary to insert the correction factors $h_\Delta$ defined in \eqref{eq:hDef} so that the truly analytic (and, more generally, meromorphic) quantities were those given by \eqref{eq:hme}.
We thus expect that we may choose a basis $|i\rangle_{\Delta, \sigma}$ for the SO$_0(d,1)$ representation of weight $\Delta$ induced from the SO$(d-1)$ representation $\sigma$ and a function $h_{\Delta, \sigma}(i)$ for which $h_{\Delta, \sigma}(i)$ is a pure phase in the principal series, $h_{\Delta, \sigma}(i)$ is a positive real number in the complementary series,  $h_{\Delta, \sigma}(i)=1$ for $\Delta = \frac{d-1}{2}$, and we have
$h_{(d-1-\Delta), \sigma}(i) = \frac{1}{h_{\Delta, \sigma}(i)}$ for all $\Delta$, 
and for which the combinations
\begin{equation}
\label{eq:hme2}
\sqrt{\Bigg|\frac{h_{\Delta,\sigma}(i')}{h_{\Delta,\sigma}(i)}\Bigg|}\ {}_{\Delta, \sigma}\langle i|U(g)|i'\rangle_{\Delta,\sigma}
\end{equation}
define meromorphic functions of $\Delta$; i.e., functions that, when defined by \eqref{eq:hme2} in the principal series and analytically continued to meromorphic functions of $\Delta$,  then again coincide with \eqref{eq:hme2} in the complementary series.  That this is indeed possible
was shown in theorem 3.1 of \cite{BransonOlafssonOrsted1996}; see eq. (1.4) of \cite{ArendsBangJensenFrahm2026} for explicit expressions. See also e.g. sections 1 and 7 of chapter 7 in \cite{Knapp2001Semisimple}.  A final useful property (which is manifest in the explicit expressions in \cite{ArendsBangJensenFrahm2026}) is that, if our basis is adapted to an SO$(d)$ subgroup of SO$_0(d,1)$, then $h_{\Delta, \sigma}$ is proportional to the identity on each SO$(d)$ representation.  Such bases are manifestly compatible with the conditions we imposed earlier in our discussion of charge-conjugation around equation \eqref{eq:dCconj}.
We refer to the $h_{\Delta, \sigma}(i)$ as Knapp-Stein factors, as they are the concrete expressions for the so-called Knapp-Stein intertwiner \cite{KnappStein1971} that relates complementary-series representations of weights $\Delta$ and $d-1-\Delta$.

Let us now return to \eqref{eq:PPPipd2} and note that the second line gives an expression for the principal-series group-averaging inner product that involves no complex-conjugated Clebsch-Gordan symbols. Furthermore, any $\Delta_i$'s in the indices (e.g., associated with the action of $C^P_{\Delta_i, \sigma_i}$ on an index) also appear without complex conjugation (due to the convention that Clebsch-Gordan symbols are anti-linear in their lower indices when combined with the explicit factors of $K$). Thus, just as in section \ref{subsec:3c}, we use the fact that $h_{\Delta, \sigma}(j)$ is a pure phase in the principal series to multiply the left-hand-side (but not the right-hand-side) of \eqref{eq:PPPipd2} by
\begin{equation}
\label{eq:gendhfactor}
\sqrt{\Bigg|\frac{h_{\Delta_1,\sigma_1}(j'_1)h_{\Delta_2,\sigma_2}(j'_2) h_{\Delta_3,\sigma_3}(j'_3)}{h_{\Delta_1,\sigma_1}(j_1) h_{\Delta_2,\sigma_2}(j_2)h_{\Delta_3,\sigma_3}(j_3)} \Bigg|}\, . 
\end{equation}

Finally, we note that i) 
the  formula for $h_{\Delta, \sigma}(j)$ from 
\cite{ArendsBangJensenFrahm2026} depends only on the representation of the SO$(d)$ subgroup to which the state $|j\rangle_{\Delta, \sigma}$ belongs and ii) the action of $C^P_{\Delta, \sigma}$ does not change this SO$(d)$ representation.  As a result, our multiplication of the left-hand-side of \eqref{eq:PPPipd2} by \eqref{eq:gendhfactor} yields an object which can be extended to a meromorphic function of the $\Delta_i$, and which must thus also agree with the above form of the right-hand-side for all complex $\Delta_i$.  In the complementary series (where $h_{\Delta, \sigma}$ is positive), for states $|\phi\rangle = \sum_{j_1,j_2,j_3} \phi_{j_1,j_2,j_3}|{j_1,j_2,j_3}\rangle_{\vec \Delta, \vec \sigma}$ and $|\psi\rangle = \sum_{j'_1,j'_2,j'_3} \psi_{j'_1,j'_2,j'_3}|{j'_1,j'_2,j'_3}\rangle_{\vec \Delta, \vec \sigma}$,  we thus find
\begin{align}
\langle \phi |\psi \rangle^{\rm GA}&=\sum_x \Bigg[ \left( \sum_{j'_1,j'_2,j'_3} \psi_{j'_1,j'_2,j'_3}\frac{
\left(\widetilde C^{P,x}_{PP}\right)^{\Delta_3, \sigma_3, j'_3}_{\Delta_1, \sigma_1,K C_{d-1-\Delta_1, \sigma_1}^{P} j'_1|\Delta_2, \sigma_2, K C_{d-1-\Delta_2, \sigma_2}^{P} j'_2}}{\sqrt{h_{\Delta_1,\sigma_1}(j'_1) h_{\Delta_2,\sigma_2}(j'_2)h_{\Delta_3,\sigma_3}(j'_3)}}\right)
\nonumber \\
&\times \left( \sum_{j_1,j_2,j_3} \phi^*_{j_1,j_2,j_3} \sqrt{h_{\Delta_1,\sigma_1}(j_1) h_{\Delta_2,\sigma_2}(j_2)h_{\Delta_3,\sigma_3}(j_3)}  \left(\widetilde C^{P,x}_{PP}\right)^{\Delta_3, \sigma_3, C^{P}_{\Delta_3,\sigma_3} j_3}_{\Delta_1, \sigma_1, j_1|\Delta_2, \sigma_2, j_2}\right) \Bigg].
\label{eq:PPPipd5}
\end{align}

Recall that the invariance of the Haar measure under $g\rightarrow g^{-1}$ makes the group-averaging inner product explicitly Hermitian for unitary representations.  As a result, up to complex conjugation and when evaluated in the complementary series, the two parenthetical factors on the right-hand-side of \eqref{eq:PPPipd5} must be related by a Hermitian matrix; i.e., for complementary-series weights $\Delta_i$ with $\Sigma > d-1$, since $h_{\Delta, \sigma}(i)$ is real we must have
\begin{align}
\frac{
\left( \left( \widetilde C^{P,x}_{PP}\right)^{\Delta_3, \sigma_3, j_3}_{\Delta_1, \sigma_1,KC_{d-1-\Delta_1, \sigma_1}^{P} j_1|\Delta_2, \sigma_2,KC_{d-1-\Delta_2, \sigma_2}^{P} j_2}\right)^*} {\sqrt{h_{\Delta_1,\sigma_1}(j_1) h_{\Delta_2,\sigma_2}(j_2)h_{\Delta_3,\sigma_3}(j_3)}} \quad \quad \quad \quad \quad \quad \quad \quad \quad \quad \quad \quad \quad \quad \quad
 \nonumber \\=\sum_{y}M^{x}{}_{y}(\vec \Delta, \vec \sigma)\,
 \sqrt{h_{\Delta_1,\sigma_1}(j_1) h_{\Delta_2,\sigma_2}(j_2)h_{\Delta_3,\sigma_3}(j_3)}  
 \left( \widetilde C^{P,y}_{PP}\right)^{\Delta_3, \sigma_3, C^{P}_{\Delta_3,\sigma_3} j_3}_{\Delta_1, \sigma_1, j_1|\Delta_2, \sigma_2, j_2} 
 \label{eq:Meq}
\end{align}
for some $M^{x}{}_{y}(\vec \Delta, \vec \sigma) = \left(M^{y}{}_{x}(\vec \Delta, \vec \sigma)\right)^*$.  We see that the group-averaging inner product is non-negative for 3-particle states in representations labeled by $(\vec \Delta, \vec \sigma)$ if and only if the matrix $M^{x}{}_{y}(\vec \Delta, \vec \sigma)$ is positive semi-definite.

We will now show this to be true by first establishing that $M^{x}{}_{y}$ is the identity when $\Delta_1= \Delta_2=\Delta_3 = \frac{d-1}{2}$ (for all choices of $\sigma_1,\sigma_2, \sigma_3$) and then arguing that $M^{x}{}_{y}$ is both continuous in the $\Delta_i$ and invertible everywhere in the complementary series where $\Sigma > d-1$.  Since $M^{x}{}_{y}$ is Hermitian, its eigenvalues are real.  But since $M^{x}{}_{y}$ is invertible, none of its eigenvalues can vanish.  Continuity then implies that they remain positive throughout the complementary series.

To show that $M^{x}{}_{y}$ is the identity 
for $\Delta_1= \Delta_2=\Delta_3 = \frac{d-1}{2}$, note that here we have $h_{\Delta_i,\sigma_i}(j)=1$ for all $i,j$ and $C_{\Delta_i,\sigma_i} = {\mathds 1}$ for all $i$.  Using \eqref{eq:CGbetterconj} with the above then also requires the Clebsch-Gordan coefficients to be real at this point, whence \eqref{eq:Meq} gives $M^{x}{}_{y} = \delta^x{}_y$.

To verify the rest of the above claims, it is useful to better understand the matrix $M^{x}{}_{y}$ from an abstract point of view.  In particular, for $\Sigma > d-1$ the absolute convergence of the group-averaging integral \eqref{eq:GAIP} requires \eqref{eq:PPPipd5} to be separately invariant under arbitrary $G_d$ transformations of $|\psi\rangle$ and $|\phi\rangle$.  Thus for any choice of $\phi_{j_1,j_2,j_3}$ we obtain an SO$_0(d,1)$-invariant functional on an appropriate space of states $|\psi\rangle$.

Similarly, for any choice of $\psi_{j_1,j_2,j_3}$ we obtain an SO$_0(d,1)$-invariant {\it anti-linear} functional on an appropriate space of states $|\phi\rangle$.  However, due to the fact that (with our conventions) $C_{\Delta,\sigma} = {\mathds 1}$ in the complementary series (see \eqref{eq:CDelta}), composing this map with $K$ gives an invariant linear functional defined by precisely the same coefficients.\footnote{In particular,  acting with $U(g)$ on $\phi_{j_1,j_2,j_3}$ is equivalent in this context to acting with $U(g^{-1})$ on $\phi^*_{j_1,j_2,j_3}$.}

To understand the implications of this observation, recall from the beginning of this subsection that \cite{Martin1975} counted the number of linearly-independent Clebsch-Gordan channels (labeled by our index $x$) in the principal series, and that the result depends only on the SO$(d-1)$ representations $\sigma_i$ and is thus explicitly independent of $\Delta$.  In particular, the proof of Theorem 16 in \cite{Martin1975} shows that the channels correspond to a basis of the $\Delta$-independent SO$(d-2)$-invariant tensors described above on $\sigma_1 \otimes \sigma_2 \otimes \sigma_3^*$ which, following \cite{KarateevKravchukSimmonsDuffin2019}, we may use to define the $x$-label of our Clebsch-Gordan coefficients.  In the language of section \ref{subsec:othergend}, the Clebsch-Gordan coefficients for channel $x$ are (up to the factors of $h_{\Delta,\sigma}$ and $C^P_{\Delta,\sigma}$ discussed below) the SO$(d)$-harmonic coefficients of a conformally-covariant 3-point function on $(S^{d-1})^3$ whose tensor structure is specified by the invariant tensor $x$ and does not depend on the $\Delta_i$, and whose dependence on the $\Delta_i$ is entirely through a scalar prefactor $\prod_{i<j} |\bm{x}_{ij}|^{e_{ij}}$.  For the 3-point functions relevant to \eqref{eq:linf1} below, the exponents are $e_{12} = \Sigma - 2(d-1)$, $e_{13} = -\delta_1$, and $e_{23}=-\delta_2$, where $\delta_i := \Sigma - 2\Delta_i$; for those relevant to \eqref{eq:linf2}, they are $e_{12} = (d-1) - \Sigma$, $e_{13} = \delta_1 - (d-1)$, and $e_{23} = \delta_2-(d-1)$.  Since $0 < \delta_i < d-1$ for complementary-series weights with $\Sigma > d-1$, in both cases the exponents exceed $-(d-1)$ at each pairwise coincidence and their sum exceeds $-2(d-1)$ at the triple coincidence, so that these 3-point functions are locally integrable on $(S^{d-1})^3$ throughout the region of interest.\footnote{For \eqref{eq:linf1}, integrability at pairwise coincidences is precisely the condition $\Sigma > d-1$; at the triple coincidence it is $\delta_3 > 0$ (which we have seen follows from $\Sigma > d-1$ and $\Delta_i \le \frac{d-1}{2}$).}  Their harmonic coefficients are thus given by absolutely convergent integrals that depend holomorphically on the $\Delta_i$ and so coincide with the analytic continuations of the principal-series Clebsch-Gordan coefficients discussed above.  
Moreover, the harmonic coefficients of a locally integrable function determine it uniquely (almost everywhere).  Thus, suppose that some linear combination $\sum_x c_x \widetilde C^{P,x}_{PP}$ of channels vanished as a functional on all states.  The corresponding linear combination of 3-point functions would then vanish on all configurations of three distinct points, and dividing by the non-vanishing scalar prefactor would give a vanishing linear combination $\sum_x c_x t_x$ of the $\Delta$-independent invariant tensors $t_x$.  Since the $t_x$ are linearly independent, this requires all $c_x$ to vanish.  We conclude that the analytic continuations of the $\widetilde C^{P,x}_{PP}$ to the complementary series with $\Sigma > d-1$ are again linearly independent functionals of their arguments $j_1,j_2,j_3$.


Furthermore, we may also note that the counting of \cite{Martin1975} used only the behavior of states under the ${\rm SO}(d-2) \subset {\rm SO}(d-1)$ subgroup of SO$_0(d,1)$.  As a result, since the $h_{\Delta, \sigma}(j)$ factors are independent of $j$ for labels associated with basis states of a given SO$(d)$ representation, and since (as seen explicitly from eq. (1.4) of \cite{ArendsBangJensenFrahm2026}) they are well-defined positive (non-zero) functions in the complementary series, these factors again cannot affect linear independence of the channels.  

Since the various factors of $C_{\Delta, \sigma}^P, C_{d-1-\Delta, \sigma}^{P}$ that appear in  \eqref{eq:PPPipd5} are invertible in both the principal and complementary series (see again the comments below \eqref{eq:CDelta}), they  again cannot affect linear independence of the Clebsch-Gordan channels.  
 It thus follows that, if we consider the set of SO$_0(d,1)$-invariant linear functionals 
 \begin{align} \sum_{j_1,j_2,j_3} \phi_{j_1,j_2,j_3} \sqrt{h_{\Delta_1,\sigma_1}(j_1) h_{\Delta_2,\sigma_2}(j_2)h_{\Delta_3,\sigma_3}(j_3)}  \left(\widetilde C^{P,x}_{PP}\right)^{\Delta_3, \sigma_3, C^{P}_{\Delta_3,\sigma_3} j_3}_{\Delta_1, \sigma_1, j_1|\Delta_2, \sigma_2, j_2} \label{eq:linf1} \end{align}
defined by the final line of \eqref{eq:PPPipd5} after replacing $|\phi\rangle$ by $K|\phi\rangle$, the functionals defined by various choices of $x$ are linearly independent for all $\Delta_i$ in the complementary series with $\Sigma > d-1$.  The same argument also applies to the SO$_0(d,1)$-invariant functionals 
\begin{align} \sum_{j'_1,j'_2,j'_3} \frac{ \psi_{j'_1,j'_2,j'_3} }{\sqrt{h_{\Delta_1,\sigma_1}(j'_1) h_{\Delta_2,\sigma_2}(j'_2)h_{\Delta_3,\sigma_3}(j'_3)}}\left(\widetilde C^{P,x}_{PP}\right)^{\Delta_3, \sigma_3, j'_3}_{\Delta_1, \sigma_1,KC_{d-1-\Delta_1, \sigma_1}^{P} j'_1|\Delta_2, \sigma_2, KC_{d-1-\Delta_2, \sigma_2}^{P} j'_2} \label{eq:linf2} \end{align}
defined by the first line of \eqref{eq:PPPipd5}.  The key point is then that, as observed before, any such invariant functional must in fact define a conformally-invariant 3-point function on $S^{d-1}$, and that, for the corresponding triple of SO$(d-1)$ representations $\vec \sigma$ from which our SO$_0(d,1)$ representations are induced, the number of such functionals \cite{KravchukSimmonsDuffin2018} (see also eq. (2.19) of \cite{KarateevKravchukSimmonsDuffin2019}) agrees precisely, as noted above, with the number of Clebsch-Gordan channels from \cite{Martin1975} (see also footnote \ref{foot:tech}, which shows that this counting remains complete for the complementary-series $\Delta_i$ of interest).
As a result, each of the two sets of linear functionals \eqref{eq:linf1} and \eqref{eq:linf2} must span the same linear space at each complementary-series $\vec \Delta$ with $\Sigma > d-1$.  It then follows that the matrix $M^x{}_y$ relating them must be continuous and invertible at all such $\vec \Delta$.

Note that the above argument proceeds separately for each triple $\vec \sigma$.  As a result, if e.g. the parity properties of $\vec \sigma$ imply that, for all finite linear combinations of states in our chosen basis, group averaging remains absolutely convergent on a larger domain (e.g., for $\Sigma > d-2$), then the same argument implies positivity on that extended domain.  In particular, taking into account parity further restricts the weight-counting argument of footnote \ref{foot:tech} to rule out 3-point functions with singular support in precisely the sectors in which parity removes poles in the group-averaging norm.

On the other hand, for general states in representations induced from general choices of $\vec \sigma$ there will be a pole at $\Sigma = d-1$ of non-zero residue. We then expect to find that, while we can find particular families of states for which this pole is absent and for which group averaging continues to converge at smaller $\Sigma$, in that case the group-averaging norm of some such state will pass through zero and will become negative for $\Sigma < d-1$.    This would then give a general-$d$ analogue of the phenomenon discussed for even-parity states in section \ref{subsec:3c}.

\section{Discussion}
\label{sec:disc}

Our work above studied the group-averaging inner product for Fock-space states of bosonic (integer-spin) particles in dS$_d$ with $d\ge 2.$ The main result was to establish positivity of group averaging for {\it all} such states (with each particle in a finite superposition of standard global dS modes\footnote{Or in the dS-invariant closure of the space so defined under the group algebra $L^1(G_d)$ as in section IIIA of \cite{Giulini_1999}.}) with
\begin{equation}
\label{eq:DiscSigma}
    \Sigma := {\rm Re} \, \sum_{i=1}^N\Delta_i > d-1.
\end{equation}
Here $\Delta_i$ is the weight associated with the $i$th particle using the convention $\Delta_i < \frac{d-1}{2}$ in the complementary series and $\Delta_i > \frac{d-1}{2}$ in the discrete series.  In particular, the above conclusion applies to states of arbitrary (integer) spin.  

However, we also found that group averaging can continue to converge absolutely for some states that violate \eqref{eq:DiscSigma}.  In some cases this improved convergence was guaranteed by a parity-symmetry argument.  For $d=2$, in those cases the group-averaging norm remained absolutely-convergent and positive for all $\Sigma > d-2$, and we expect this to be true for higher $d$ as well.\footnote{In particular, this will be the case if the parity argument also removes the potential poles at $\Sigma-2\Delta_i=0$, and if the required exchange of orders of integration remains justified for $d>2$.  This seems likely, though we have not checked it in detail.}  This was in particular so on $N$-particle states in the subspace ${\cal H}^N_{\text{3EP-RIPO}}$ which can be expressed using Clebsch-Gordan reduction as an RIPO state of three effective particles.  In contrast, in cases where we obtained improved convergence by fine-tuning coefficients in the state to cancel the potential pole at $\Sigma = d-1$, the group-averaging norm then vanished at $\Sigma= d-1$ and became negative for an open set of $\Delta_i$ with $\Sigma < d-1.$  The latter behavior was displayed explicitly only for $d=2$, but the arguments of section \ref{subsec:3c} show that it will necessarily also occur for $d>2$ in any sector where one can use the large-boost asymptotic expansion of the matrix elements to show that the potential pole at $\Sigma=d-1$ has non-zero residue.  In particular, for any $d > 2$ this is the case for generic scalar-scalar-scalar three-particle states; see e.g. \eqref{eq:phiint-general} for the explicit one-particle boost matrix elements in the complementary series.  

Our violation of positivity is similar to that found previously in \cite{Louko:2003cn} which involved absolutely-convergent group-averaging of the covering group SL$(2, {\mathbb R})$ of the de Sitter group for dS$_2$.  However, we now see that similar behavior can arise naturally in perturbative  quantum gravity on dS${}_d$. In both cases, the absolute convergence implies that the relevant domain is an $L^1$ subspace in the sense of \cite{Giulini_1999}, so that the uniqueness theorem of that reference applies.   The theorem considers the $*$-algebra of de Sitter invariant operators that preserve a domain $\Phi$ in the original Fock space on which this convergence holds for all pairs of states (i.e., $\langle\phi_1|U(g)|\phi_2\rangle\in L^1(G_d)$ for all $\phi_1,\phi_2\in\Phi$), and where $\Phi$ is invariant under the group-algebra defined by the de Sitter group $G_d$.   It then shows that, if one wishes to find $*$-representations of this algebra acting on invariant functionals in the (anti-linear) algebraic dual $\Phi^*$ of $\Phi$, the inner product on the image of the rigging map in $\Phi^*$ must agree with group averaging up to a single overall multiplicative constant.  As a result, there can be no inner product that achieves this goal when $\Phi$ contains both states with $\Sigma > d-1$ and finely-tuned states with $\Sigma < d-1$ with finite-but-negative group-averaging norms.

Our current suggestion is thus to simply assume that physically-meaningful states of perturbative quantum gravity about dS$_d$ arise only from group-averaging Fock states with $\Sigma > d-1$ (or, in appropriate symmetry-protected sectors, with appropriately smaller values of $\Delta_i$); see also below for comments on non-Fock states.\footnote{We can of course also include the dS-invariant Fock vacuum, to which, since it always lies in a different superselection sector from states with $\Sigma >d-1$, we can assign an arbitrary norm in the Hilbert space of invariant states; see e.g. the discussion in section 4 of \cite{Marolf_2009} (drawing on ideas from \cite{Marolf:1995cn,GomberoffMarolf1999}). This is consistent with the uniqueness theorem of \cite{Giulini_1999} discussed above since, as $\langle 0|U(g)|0\rangle=1$, the vacuum does not lie in any $L^1$ domain $\Phi$.}  However, it would be interesting to return in the future to the question of whether this suggestion can be further justified.  

Let us now recall that our work above focused on implementing the linearization-stability constraints, and that one might describe the effect of such constraints as `gauging the connected component SO$_0(d,1)$ of the dS$_d$ isometry group'.  However, as discussed in e.g. \cite{Harlow:2023hjb} (see also \cite{DineLeighMacIntire1992,ChoiKaplanNelson1993} for CP), one expects a theory of quantum gravity to gauge parity as well.\footnote{Ref. \cite{Harlow:2023hjb} also advocates gauging time-reversal symmetry in quantum gravity, though see e.g. \cite{Witten:2025ayw,VincentCRT} for contrasting comments.}  This is easily implemented by simply projecting out the parity-odd states.  Interestingly, in sectors with parity-odd Clebsch-Gordan channels, this can still leave a non-trivial subspace of ${\cal H}^N_{\text{3EP-RIPO}}$ on which group averaging remains both absolutely-convergent and positive for smaller values of $\Sigma$ (and likely for $\Sigma > d-2$).

Returning now to the analysis of positivity in the main text, our treatment was straightforward for general $d$ for particles in tempered UIRs, which are essentially the so-called principal-series and discrete-series representations of SO$_0(d,1)$.  Subtleties arose only when considering non-tempered representations, which then required more careful discussion   of the complementary series.

The extension to include fermions (i.e., spinorial representations, or representations with  half-integer spin) turns out to be essentially trivial simply because all spinorial (half-integer spin) representations of the covering group Spin$_0(d,1)$ are tempered.  In particular, there are simply no spinorial complementary-series representations of Spin$_0(d,1)$; see e.g. appendix B.2 of \cite{PethybridgeSchaub2022} or \cite{Schaub2024} for $d>2$ (the unitary fermionic representations that \cite{Schaub2024} calls `exceptional' occur only for $d=4$, where they belong to the discrete series) and \cite{Bargmann1947} for $d=2$.  Positivity for states satisfying \eqref{eq:DiscSigma} that include fermions thus follows by the rather simple general arguments in sections \ref{subsec:PDgend} and \ref{subsec:othergend}.
Here, for even $d$, the tempered spinorial representations also include limits of discrete series (e.g., $D^\pm_{1/2}$ for $d=2$); since their matrix elements are again bounded by a polynomial times $e^{-\frac{d-1}{2}\lambda}$, the $L^2$ argument of sections~\ref{subsec:PDgend} and~\ref{subsec:othergend} applies to them as well; see again footnotes \ref{foot:CwithT} and \ref{foot:notnorm}.

An immediate implication is that group averaging can be used to define a consistent theory of perturbative quantum gravity about  dS$_d$ (with $d>2$) for states with $\Sigma > d-1$ in any theory whose linearized states define Fock states as above.  In particular, since any given positive-norm state has some definite norm that exceeds zero by a finite amount, perturbative corrections cannot violate positivity.  Thus positivity of the resulting Hilbert space must persist to all orders in perturbation theory.\footnote{One could imagine that e.g. first-order corrections to the inner product might grow with either frequency $\omega$ or particle number $N$.  But even if the sign of the correction were negative, that need not indicate failure of positivity at large $\omega, N$.  Instead, it would more naturally be interpreted as indicating that the regime of validity for such perturbation theory decreases as $\omega, N$ become large.}

Now, as discussed in section \ref{sec:intro}, $d=2$ gravity theories typically require a dilaton that breaks the dS$_2$ symmetry and reduces any required group averaging to a one-dimensional integral that is easily shown to be positive.  But if one were interested in group averaging of general dS${}_2$ Fock states, one would consider anyonic fields in addition to the bosons and fermions discussed above. By this we mean that one would consider general UIRs of the universal cover $\widetilde {\rm SL}(2,{\mathbb R})$ of SO$_0(2,1)$.  This universal cover effectively unwraps the circle of dS${}_2$ into a line.  There are then associated IR divergences that one must regulate, e.g. by considering $\widetilde {\rm SL} (2,{\mathbb R})$ as the $n\rightarrow \infty$ limit of the $n$-fold cover of SO$_0(2,1)$, group-averaging over each $n$-fold cover, and then taking the $n\rightarrow \infty$ limit of the result.  While the details would be a significant digression, we mention that Claude (Opus 5.5) reports that the analogue of the matrix $M^x{}_y$ in \eqref{eq:Meq} can be computed directly and is positive definite.\footnote{While we do not claim to have independently verified  this calculation, doing so should be straightforward for an interested reader.}

While for $d>2$ our work addressed general Fock-space states for dS$_d$ quantum field theory, one should recall that fields with $M^2\le0$ can have modes that are not described by Fock states at all.  These include the zero-mode of a massless scalar field and the unstable modes of tachyons.  For a dS${}_2$ massless scalar, group averaging was shown to be positive semi-definite in \cite{Marolf:2008it} for states satisfying an analogue of \eqref{eq:DiscSigma} that included an appropriate contribution from the zero-mode.  It thus seems likely that a corresponding result will hold for other cases with non-Fock modes, though we save a full investigation for future work.

\acknowledgments
    We thank Jiuci Xu for useful discussions.  We also thank Xi Dong for motivating conversations over several years. This work was supported by NSF grants PHY-2408110
    and PHY-2107939, and by funds from the University of California. X.Y. was also supported by the National Science Foundation Graduate Research Fellowship under Grant No. 2139319. Any opinions, findings, and conclusions or recommendations expressed in this material are those of the author(s) and do not necessarily reflect the views of the National Science Foundation.

\appendix

\section{Absolute convergence for RIPO reductions with
\texorpdfstring{$\Sigma \le 1$}{Sigma <= 1}}
\label{app:RIPOconv}

The iterative arguments of section \ref{subsec:4d2} exchange the group-averaging integral with the integrals over the weights $s$ of effective principal-series particles.  For $\Sigma>1$ this exchange was justified at the end of section \ref{sec:principal-discrete-positivity} by absolute convergence of the full integral.  For $\Sigma \le 1$ the convergence of group averaging for RIPO (rotation-invariant, parity-odd) states relies on a cancellation of the leading large-$\lambda$ terms, so that bounds on the full matrix elements (such as item 2 of theorem 2.1 of \cite{Shalom2000Rigidity}) no longer suffice.  Here we supply the required estimate for $d=2$.\footnote{We thank Claude (Fable 5.1) for help in assembling this argument.}

Consider a fixed set of original representations and a fixed finite sequence of Clebsch-Gordan reductions, applied to a finite superposition of basis states.  Let $\vec s$ denote the collection of principal-series labels of the effective particles produced along the way.  Since $U(g)$ preserves these labels, the reduced state is a direct integral over $\vec s$ of 3-particle states $|\psi\rangle_{\vec s}$, each a finite superposition of basis states $|m_1,m_2,m_3\rangle$ with angular momenta drawn from a finite set that, by angular momentum conservation, is independent of $\vec s$.  We normalize the $|\psi\rangle_{\vec s}$ to unit norm at each fixed $\vec s$ (i.e., each $|\psi\rangle_{\vec s}$ lives in a Hilbert space with fixed $\vec s$), absorb the spectral measure into the coefficient function $f(\vec s)$, and project each $|\psi\rangle_{\vec s}$ onto its RIPO part (which preserves all of the above properties).  The group-averaging inner product is then $\int d\vec s\,|f(\vec s)|^2\int_G dg\,{}_{\vec s}\langle\psi|U(g)|\psi\rangle_{\vec s}$, and the exchange of the two integrals is justified by Fubini's theorem if
\begin{equation}
\label{eq:Fubinicond}
\int d\vec s\,|f(\vec s)|^2\int_G dg\,\big|{}_{\vec s}\langle\psi|U(g)|\psi\rangle_{\vec s}\big|<\infty .
\end{equation}
Since RIPO states have $J=0$, the rotations in \eqref{eq:PD-Haar-measure} act trivially, and $\int_G dg$ reduces to $4\pi^2\int_0^\infty d\lambda\,\sinh\lambda$ of the boost matrix element.

For such states, we have  \begin{equation} \label{eq:RIPO-spectral-moments} \int d\vec s\,(1+|\vec s|)^p|f(\vec s)|^2<\infty \qquad(p\ge0). \end{equation} The bounds \eqref{eq:RIPO-spectral-moments} follow by again using (at each reduction) the fact that Clebsch-Gordan coefficients decrease faster than any power of $s$ (item (iv) in the discussion at the end of section \ref{sec:principal-discrete-positivity}).  We will combine \eqref{eq:RIPO-spectral-moments} with further bounds derived in the Lemmas below.

\paragraph{Lemma A.1 (uniform asymptotics).}  Let $\Delta$ be either a principal-series weight $\frac12+is$ or a complementary-series weight in $(0,\frac12)$, and let $\kappa_\Delta := 4^\Delta\, \Gamma(1-2\Delta)/(\Gamma(\Delta)\Gamma(1-\Delta))$.  Then for all $m,m'$ and $\lambda \ge 0$, the matrix elements between unit-normalized basis states satisfy
\begin{equation}
\label{eq:LRsplit}
{}_\Delta\langle m | e^{i\lambda B_1}|m'\rangle_\Delta = L^\Delta_{mm'}(\lambda) + R^\Delta_{mm'}(\lambda),
\end{equation}
where the leading part $L^\Delta_{mm'}(\lambda)$ is a sum of two terms.  In the principal series, we find
\begin{equation}
\label{eq:Lterm}
L^\Delta_{mm'}(\lambda) = i^{m+m'}\Big[\kappa_\Delta\, h_\Delta(m')\, e^{-\Delta\lambda} + \kappa_{1-\Delta}\, h_{1-\Delta}(m)\, e^{-(1-\Delta)\lambda}\Big].
\end{equation}
On the complementary series, where the basis states are normalized by \eqref{eq:hDef}, one should make the replacements $h_\Delta(m') \to \sqrt{h_\Delta(m)h_\Delta(m')}$ and $h_{1-\Delta}(m) \to \sqrt{h_{1-\Delta}(m)h_{1-\Delta}(m')}$.  Moreover, there are a constant $C$ and a polynomial $P$ such that, uniformly in real $s$ and in $m,m'$,
\begin{equation}
\label{eq:LRbounds}
|L^\Delta_{mm'}(\lambda)| \le C P(s,m,m')\, (1+\lambda)\, e^{-\lambda/2}, \qquad
|R^\Delta_{mm'}(\lambda)| \le C P(s,m,m')\, (1+\lambda)\, e^{-3\lambda/2}
\end{equation}
in the principal series, while in the complementary series (for fixed $\Delta$ and finitely many $m,m'$) there is a constant $C$ such that
\begin{equation}
\label{eq:LRboundscomp}
|L^\Delta_{mm'}(\lambda)| \le C e^{-\Delta\lambda}, \qquad |R^\Delta_{mm'}(\lambda)| \le C e^{-(\Delta+1)\lambda} .
\end{equation}

\paragraph{Proof.}  Recall from \eqref{eq:SO21explicit} that $t = \tanh\frac\lambda2$, $c=\cosh\frac\lambda2$ and $r = m-m'$.  Consider first $m \ge m'$ and set $z=t^2$, so that
\begin{equation}
1-z = {\rm sech}^2\tfrac{\lambda}{2} = \frac{4e^{-\lambda}}{(1+e^{-\lambda})^2}, \qquad c^{-2\Delta}=(1-z)^\Delta .
\end{equation}
We may thus expand around $z=1$ using the connection formula
\begin{align}
\label{eq:connectionformula}
{}_2F_1(a,b;c;z) =& \frac{\Gamma(c)\Gamma(c-a-b)}{\Gamma(c-a)\Gamma(c-b)}\,{}_2F_1(a,b;a+b-c+1;1-z) \nonumber \\ +& (1-z)^{c-a-b}\frac{\Gamma(c)\Gamma(a+b-c)}{\Gamma(a)\Gamma(b)}\,{}_2F_1(c-a,c-b;c-a-b+1;1-z),
\end{align}
valid for $c-a-b\notin{\mathbb Z}$.  For the hypergeometric function in \eqref{eq:SO21explicit} we have $a=\Delta-m'$, $b=\Delta+m$ and $c=r+1$, so that $c-a-b = 1-2\Delta$, $c-a = m+1-\Delta$ and $c-b = 1-\Delta-m'$; the condition $c-a-b\notin{\mathbb Z}$ holds for $s\ne0$ in the principal series and throughout the complementary series.  The resulting Gamma-function prefactors combine with the Pochhammer symbol in \eqref{eq:SO21explicit} into expressions involving only $h_\Delta$:  using $(m'+1-\Delta)_r = \Gamma(m+1-\Delta)/\Gamma(m'+1-\Delta)$ together with the reflection formula $\Gamma(x)\Gamma(1-x)=\pi/\sin\pi x$, one finds
\begin{align}
\frac{(m'+1-\Delta)_r\,\Gamma(1-2\Delta)}{\Gamma(m+1-\Delta)\,\Gamma(1-\Delta-m')} &= (-1)^{m'}\,\frac{\Gamma(1-2\Delta)}{\Gamma(\Delta)\Gamma(1-\Delta)}\,h_\Delta(m') = (-1)^{m'}\,\frac{\kappa_\Delta}{4^\Delta}\, h_\Delta(m'),
\label{eq:gammaid1}\\
\frac{(m'+1-\Delta)_r\,\Gamma(2\Delta-1)}{\Gamma(\Delta-m')\,\Gamma(\Delta+m)} &= (-1)^{m'}\,\frac{\Gamma(2\Delta-1)\sin\pi\Delta}{\pi}\,h_{1-\Delta}(m) \nonumber \\ &= (-1)^{m'}\,\frac{\kappa_{1-\Delta}}{4^{1-\Delta}}\, h_{1-\Delta}(m),
\label{eq:gammaid2}
\end{align}
where in the last step of each line we have used the definition of $\kappa_\Delta$ and, in \eqref{eq:gammaid2}, the identity $\Gamma(\Delta)\Gamma(1-\Delta) = \pi/\sin\pi\Delta$.  Inserting \eqref{eq:connectionformula}--\eqref{eq:gammaid2} into \eqref{eq:SO21explicit} and writing $(it)^r(-1)^{m'} = t^r\, i^{m-m'}\, i^{2m'} = t^r\, i^{m+m'}$, the right-hand side of \eqref{eq:SO21explicit} becomes
\begin{equation}
\label{eq:connected}
t^{r}\, i^{m+m'}\Big[\kappa_\Delta\, h_\Delta(m') \Big(\tfrac{1-z}{4}\Big)^{\Delta} G_+(1-z) + \kappa_{1-\Delta}\, h_{1-\Delta}(m) \Big(\tfrac{1-z}{4}\Big)^{1-\Delta} G_-(1-z)\Big],
\end{equation}
where $G_+ := {}_2F_1(\Delta-m',\Delta+m;2\Delta;1-z)$ and $G_- := {}_2F_1(m+1-\Delta, 1-\Delta-m';2-2\Delta;1-z)$ are hypergeometric series in $1-z$ with $G_\pm(0)=1$.  Since $t^r = 1 + O(r e^{-\lambda})$, $(\frac{1-z}{4})^{\Delta} = e^{-\Delta\lambda}(1+ O(e^{-\lambda}))$, and $G_\pm = 1 + O(e^{-\lambda})$, the leading terms of \eqref{eq:connected} are precisely those in \eqref{eq:Lterm}.  In the principal series, $|h_\Delta|=1$ and \eqref{eq:SO21explicit} gives the matrix element directly, while in the complementary series the matrix element is obtained by multiplying \eqref{eq:connected} by the real factor $\sqrt{h_\Delta(m)/h_\Delta(m')}$, which converts $h_\Delta(m')$ and $h_{1-\Delta}(m)$ into the symmetric square roots stated in the lemma.

For $m<m'$, we use the symmetry properties of the matrix elements.  In the principal series, \eqref{eq:Pshadow} and \eqref{eq:Prel} give ${}_\Delta\langle m | e^{i\lambda B_1}|m'\rangle_\Delta = \frac{h_\Delta(m')}{h_\Delta(m)}\,{}_\Delta\langle m' | e^{i\lambda B_1}|m\rangle_\Delta$, and applying this to \eqref{eq:connected} with $m \leftrightarrow m'$ reproduces \eqref{eq:Lterm} unchanged.  In the complementary series, \eqref{eq:SO21explicit} together with the unitarity relation noted below it shows that the normalized matrix elements are symmetric under $m\leftrightarrow m'$, as is the symmetric form of \eqref{eq:Lterm}.  Thus \eqref{eq:LRsplit}--\eqref{eq:Lterm} hold for either ordering of $m$ and $m'$.

For the bounds, note first that the coefficients of the series $G_\pm$ are rational in $\Delta, m, m'$ with poles only at $2\Delta \in -{\mathbb Z}_{\ge 0}$ or $2-2\Delta\in -{\mathbb Z}_{\ge 0}$, which are not in the ranges considered, and the same is true of their derivatives with respect to $s$.  Thus for $\lambda \ge \lambda_0(s,m,m') := \log P_1(s,m,m')$, with $P_1$ a suitable polynomial satisfying $P_1 \ge e$, the series converge and
\begin{equation}
\label{eq:Gbounds}
|G_\pm - 1| \le P_2(s,m,m')\, e^{-\lambda}, \qquad |\partial_s G_\pm| \le P_2(s,m,m')\, e^{-\lambda},
\end{equation}
for a polynomial $P_2$.  Since $|h_\Delta|=1$ in the principal series, the only potential obstruction to uniformity in $s$ is the factor $\kappa_{\frac12\pm is}\propto\Gamma(\mp2is)$, which has a simple pole at $s=0$.

Consider first $|s| \ge \frac14$, where $|\kappa_{\frac12\pm is}|$ is bounded by a constant.  Then \eqref{eq:Lterm} gives $|L|\le Ce^{-\lambda/2}$ directly.  For $\lambda\ge\lambda_0$, the remainder is $R = \sum_\pm (\text{term}_\pm)\times\big[t^r(1+e^{-\lambda})^{-2\Delta_\pm}G_\pm - 1\big]$ with $\Delta_\pm \in \{\Delta, 1-\Delta\}$, and \eqref{eq:Gbounds} together with $t^r = 1+O(re^{-\lambda})$ gives $|R|\le CP e^{-3\lambda/2}$.  For $\lambda<\lambda_0$ we instead use unitarity, $|{}_\Delta\langle m | e^{i\lambda B_1}|m'\rangle_\Delta| \le 1$, so that $|R| \le 1 + |L|$; since $e^{-3\lambda/2} \ge P_1^{-3/2}$ on this range, this is again of the form \eqref{eq:LRbounds} once the polynomial $P$ is taken large enough.

For $|s|<\frac14$ we must control the cancellation between the two terms of \eqref{eq:Lterm}.  To this end, define for real $u$ and for $x\in\{m,m'\}$
\begin{equation}
\ell_x(u) := i^{m+m'}\,\kappa_{\frac12+iu}\, h_{\frac12+iu}(x)\, e^{-(\frac12+iu)\lambda},
\end{equation}
so that $L^\Delta_{mm'} = \ell_{m'}(s) + \ell_{m}(-s)$ (the second term of \eqref{eq:Lterm} being obtained from the first by $\Delta\to1-\Delta$, i.e.\ $s\to -s$, and $m'\to m$).  Each $\ell_x$ has a simple pole at $u=0$, but since $h_{\frac12}(x)=1$ for all $x$ the residue $\rho(\lambda) = \lim_{u\to0}u\,\ell_x(u)$ of $\ell_x$ is independent of $x$, and the poles cancel in $L$.  Writing $F_x(u) := u\,\ell_x(u)$, which is analytic in $u$ with $F_x(0)=\rho(\lambda)$, we have
\begin{equation}L^\Delta_{mm'} = \frac{F_{m'}(s)-F_{m'}(0)}{s} + \frac{F_{m}(-s)-F_{m}(0)}{-s},
\end{equation}
so that by the fundamental theorem of calculus $|L^\Delta_{mm'}| \le \sup_{|v|\le\frac14}\big(|F'_{m'}(v)| + |F'_{m}(v)|\big)$ (where the choice of the constant $\frac{1}{4}$ was arbitrary).  Differentiating $F_x(v) = i^{m+m'}\,[v\kappa_{\frac12+iv}]\, h_{\frac12+iv}(x)\, e^{-(\frac12+iv)\lambda}$, the factor $v\kappa_{\frac12+iv}$ and its derivative are bounded on $|v|\le\frac14$, $|h_{\frac12+iv}(x)|=1$ while $|\partial_v h_{\frac12+iv}(x)| = |\psi(x+\tfrac12+iv)+\psi(x+\tfrac12-iv)|$ (with $\psi$ the digamma function) grows at most logarithmically in $|x|$, and $\partial_v e^{-(\frac12+iv)\lambda} = -i\lambda\, e^{-(\frac12+iv)\lambda}$.  Thus $|L| \le CP(1+\lambda)e^{-\lambda/2}$, which is the first bound in \eqref{eq:LRbounds}; the factor of $(1+\lambda)$ is the familiar logarithmic enhancement of principal-series matrix elements at $s=0$.  The remainder $R$ is treated in the same way: for $\lambda\ge\lambda_0$ we write $R = r_{m'}(s) + r_m(-s)$ with $r_x(u) := \ell_x(u)\,B_x(u)$ and $B_x(u) := t^r(1+e^{-\lambda})^{-(1+2iu)}G_x - 1$, 
where $G_{m'}(u) := G_+\big|_{\Delta=\frac12+iu}$ and $G_m(u) := G_-\big|_{\Delta=\frac12-iu}$, so that $r_{m'}(s)$ and $r_m(-s)$ are exactly the two terms of \eqref{eq:connected} with their leading parts removed.  The residues of $r_{m'}$ and $r_m$ at $u=0$ again agree, since $B_{m'}(0)=B_m(0)$ (at $\Delta=\frac12$ the series $G_\pm$ coincide), and the same  argument applies verbatim to $u\,r_x(u)$: by \eqref{eq:Gbounds} the factor $B_x$ and its $u$-derivative are both $O(Pe^{-\lambda})$, so $|R|\le CP(1+\lambda)e^{-3\lambda/2}$.  For $\lambda<\lambda_0$ we use $|R|\le 1+|L|$ as before.  This establishes \eqref{eq:LRbounds}.

The complementary-series statement follows in the same way, with no $s$-dependence to control. 

\paragraph{Lemma A.2 (parity cancellation).} Let $|\psi\rangle$ and $|\psi'\rangle$ be RIPO 3-particle states in the same tensor product of principal- or complementary-series representations.  Then the product of leading terms $\prod_{i=1}^3 L^{\Delta_i}_{m_im_i'}(\lambda)$, summed against the coefficients of $\psi$ and $\psi'$, vanishes identically:
\begin{equation}
\sum_{\vec m, \vec m'} \overline{\psi_{\vec m}}\, \psi'_{\vec m'} \prod_{i=1}^3 L^{\Delta_i}_{m_im_i'}(\lambda) = 0 .
\end{equation}

\paragraph{Proof.}  Expanding the product of the two-term expressions \eqref{eq:Lterm} gives eight terms, each of which is a $\lambda$-dependent constant times $\prod_i i^{m_i+m_i'}\, \alpha_i(m_i)\beta_i(m_i')$ with $\alpha_i, \beta_i$ each equal to $1$, to $h_{\Delta_i}$, or to $h_{1-\Delta_i}$ evaluated at the indicated argument (or to the symmetric square roots in the complementary case).  Each such term therefore factorizes into a functional of $\psi$ times a functional of $\psi'$, and it suffices to show that $\sum_{\vec m} \psi_{\vec m} \prod_i i^{m_i}\alpha_i(m_i)$ vanishes for any RIPO $\psi$ (though the same will be true of the corresponding bra-factor).  Since $\psi$ has $J=0$, $\sum_i m_i = 0$ and the factor $\prod_i i^{m_i} = 1$; since $h_\Delta(-m) = h_\Delta(m)$ by \eqref{eq:hDef} and the reflection formula, the remaining coefficient $\prod_i\alpha_i(m_i)$ is even under $\vec m \to -\vec m$; and since $\psi$ is parity-odd, $\psi_{-\vec m} = -\psi_{\vec m}$.  The sum thus vanishes.  

\paragraph{Proposition A.3.} For the states considered in this appendix, \eqref{eq:Fubinicond} holds, including cases where the original weights satisfy $0<\Sigma\le1$. 

\paragraph{Proof.} We now fix the remaining representation labels and define $\Sigma_{\mathrm{eff}}:=\sum_{i=1}^3{\rm Re}\,\Delta_{\mathrm{eff},i}>0$, which is independent of $\vec s$.  For triples without discrete-series factors, Lemma A.2 removes the $LLL$ term from the expansion of the matrix element in terms of \eqref{eq:LRsplit}, so that every remaining term contains at least one factor of $R$.  Together with the finite angular-momentum support, the bounds \eqref{eq:LRbounds} and \eqref{eq:LRboundscomp} then give
\begin{equation}
\big|{}_{\vec s}\langle\psi|e^{i\lambda B_1} |\psi\rangle_{\vec s}\big| \le C P(\vec s)\, (1+\lambda)^3\, e^{-(\Sigma_{\mathrm{eff}}+1)\lambda} ,
\end{equation}
where $P$ is a polynomial.  Since $\Sigma_{\mathrm{eff}}>0$, this is integrable against $\sinh\lambda$.

For triples containing a discrete-series factor we necessarily have $\Sigma_{\mathrm{eff}}>1$, since the discrete-series weight is at least 1 and the other two weights are positive.  Here no cancellation is needed: matrix elements of $U(e^{i\lambda B_1})$ between fixed basis states of $D^\pm_\Delta$ are bounded by $Ce^{-\Delta\lambda}$ (see section \ref{sec:dS2UIRs}), and \eqref{eq:LRbounds}--\eqref{eq:LRboundscomp} bound the remaining factors, so the full matrix element is bounded by $C P(\vec s)(1+\lambda)^3 e^{-\Sigma_{\mathrm{eff}}\lambda}$, which is again integrable against $\sinh\lambda$.

In both cases the polynomial factor $P(\vec s)$ is controlled by \eqref{eq:RIPO-spectral-moments}, and the finite sums over angular momenta therefore establish \eqref{eq:Fubinicond} and the required interchange.

\bibliographystyle{JHEP}
\bibliography{pos2}

\end{document}